\documentclass[preprint,12pt,authoryear,fleqn]{elsarticle}
\usepackage{bbm}
\usepackage{amssymb}
\usepackage{hyperref}
\usepackage{amsmath}
\usepackage[document]{ragged2e}
\usepackage[utf8]{inputenc}
\usepackage{ucs}
\usepackage{textcomp}
\usepackage{float}
\usepackage{tikz}

\usepackage{caption}
\usepackage{algorithm}
\usepackage{algpseudocode}
\usepackage{subfig}
\usepackage{graphicx}
\usepackage{color}
\renewcommand\footnote{}

\def\initialI{Model I}
\def\initialII{Model II}

\journal{\ }

\begin{document}

\begin{frontmatter}
\title{Group-Representative Functional Network Estimation from Multi-Subject fMRI Data via MRF-based Image Segmentation}
\author[a]{Aditi Iyer} \ead{iyer13@purdue.edu} \author[b]{Bingjing Tang} \ead{tang272@purdue.edu} \author[b]{Vinayak Rao} \ead{varao@purdue.edu} \author[c]{Nan Kong\corref{cor1}} \ead{nkong@purdue.edu}
\cortext[cor1]{Corresponding author:  Nan Kong}
\address[a]{School of Electrical and Computer Engineering, Purdue University, West Lafayette, Indiana, USA}
\address[b]{Department of Statistics, Purdue University, West Lafayette, Indiana, USA}
\address[c]{Weldon School of Biomedical Engineering, Purdue University, West Lafayette, Indiana, USA}
 
\begin{abstract}
We propose a novel two-phase approach to functional network estimation 
of multi-subject functional Magnetic Resonance Imaging (fMRI) data, which 
applies model-based image segmentation to determine a group-representative 
connectivity map. In our approach, we first improve clustering-based 
Independent Component Analysis (ICA) to generate maps of components 
occurring consistently across subjects, and then estimate the 
group-representative map through MAP-MRF (Maximum a priori - Markov 
random field) labeling. For the latter, we provide a novel and efficient 
variational Bayes algorithm. We study the performance of the proposed 
method using synthesized data following a theoretical model, and 
demonstrate its viability in blind extraction of group-representative 
functional networks using simulated fMRI data. We anticipate the proposed 
method will be applied in identifying common neuronal characteristics in 
a population, and could be further extended to real-world clinical 
diagnosis.
\end{abstract}

\end{frontmatter}

\footnote{\textit{Abbreviations}: CPV, Cumulative Percent Variance; DTW, Dynamic Time Warping; FDR, False Discovery Rate; ICA, Independent Component Analysis; ICM, Iterated Conditional Modes; IMED, Image Euclidean Distance; MAP, Maximum a priori; MCMC, Markov Chain Monte Carlo; ML, Maximum Likelihood; MRF, Markov Random Field; MSE, Mean Squared Error; simTB, Simulation Toolbox; SNR, Signal-to-Noise Ratio.}

\section{Introduction}\label{sec1}
fMRI systems capture neuronal activity by imaging the accompanying changes 
in blood flow. Increased activity demands greater energy, which triggers 
increased local blood flow and oxygenation \citep{42}. Further, 
differences in magnetic properties of oxyhaemoglobin (diamagnetic) and 
deoxyhaemoglobin (paramagnetic) enable us to record the level of neuronal 
activity under an applied magnetic field, as a function of the 
deoxyhaemoglobin content of blood \citep{43}. fMRI systems are 
noninvasive with high spatial and temporal resolution, leading to their 
popularity as tools for identification of brain regions that are active 
in tandem, referred to as functionally connected networks.

\

There has been growing interest in using functional connectivity patterns, 
determined from fMRI data, to characterize groups of individuals 
exhibiting common traits. Applications include identifying individuals 
with neurological and psychiatric diseases on the basis of observed abnormalities in functional connectivity \citep{1}. Recent work analyzing clinical data of patients with neurological diseases, such as Alzheimer's \citep{3}, epilepsy \citep{4}, and multiple sclerosis \citep{5}, demonstrates the strong potential of fMRI as a diagnostic tool in clinical practice. However, the present challenge lies in efficient and accurate identification of distinct functional connectivity patterns observed consistently across multiple subjects \citep{44}.

\

Functional networks of {\em individual} subjects have been successfully 
uncovered from their fMRI data through 
independent components analysis \citep[ICA,][]{7}. ICA is a technique 
that separates individual source components from a linear mixture, under 
assumptions of independence and non-Gaussianity \citep{6}. For the 
identification of connectivity patterns 
{\em across multiple subjects}, several extensions of single-subject 
ICA have been proposed, primarily involving data 
aggregation through across-subject averaging \citep{8}, data 
concatenation \citep{9}, and clustering-based schemes \citep{11}. While 
across-subject averaging reduces computation, it requires 
perfect registration across subjects; leading to loss of sensitivity due 
to suppression of unique and minority sources, as well as loss of 
resolution \citep{11}. Concatenation-based methods make the imposition of 
a common observation space (for temporal concatenation) or common time 
course (for spatial concatenation) \citep{12}. While these methods are 
intended to prevent overfitting \citep{13}, only a single spatial map or 
time course is attainable, applying to all subjects.

\

The third ICA scheme is the so-called self-organized clustering-based 
ICA (SOG-ICA, \cite{11}). This two-stage procedure applies spatial ICA to 
individual subject data, followed by across-subject clustering to identify 
independent components that occur consistently. Cluster centroids are used 
to form the group-representative map. This has the advantage of allowing 
for differences between individual subject (spatial) maps and time 
courses, and for incorporation of both spatial and temporal similarities 
in  measuring across-group consistency. 
However, the similarity measures 
recommended in \cite{11} limit dimensionality reduction during 
pre-processing, and do not make allowance for minor relative shifts 
which may be reasonably expected between maps of different subjects. 
Additionally, the use of cluster centroids may yield incorrect results, 
as the averaged series may not belong to the valid space of fMRI 
signals \citep{13}.

\

In this paper, we propose an alternative framework that redefines the 
problem of estimating the group map as an image segmentation problem. 
We first employ an improved clustering-based ICA scheme, incorporating 
spatial and temporal similarity measures that accommodate for minor shifts 
between subjects, to determine consistent components underlying individual 
subject maps.  Our main contribution is then an MAP-MRF framework that 
models this data, as well as a novel and efficient variational Bayes 
algorithm~\cite{WaiJor2003} to identify distinct functional networks across the 
subjects. Our framework exploits spatial information underlying the 
connectivity maps, and accounts for uncertainty in the estimation process, 
overcoming limitations in more traditional schemes. 

\

The remainder of this paper is organized as follows. In Section \ref{sec:preproc}, 
we describe our problem in more detail, and the pre-processing steps. 
In Section~\ref{sec:model}, we specify the 
various components of our modeling framework, as well as the details of 
our proposed
computational framework. In Section \ref{sec:expt}, we specify the 
experimental parameters used to test the proposed framework, along with a 
discussion of the results obtained. We conclude the paper with 
Section \ref{sec:conc}, in which we detail our inferences and suggest 
avenues for future research.


\section{Data and pre-processing}\label{sec:preproc}

We represent the fMRI image sequence of each subject $i = 1,\ldots,N$ with a 
$t_i \times V$ matrix $D_{i}$, where $t_i$ represents the number of scanning 
time points, and $V$ the number of voxels. The $t$\textsuperscript{th} 
row of such a matrix contains all voxels imaged at 
time-point $t$, and the $v$\textsuperscript{th} column contains the time 
course of the corresponding voxel.
Through a series of data processing steps, we convert this collection of 
$\sum_{i=1}^Nt_i$ images into a smaller set of $M$ images. We write these as $Y_i$ 
for $i = 1,\ldots,M$. We summarize the pre-processing steps below, all 
details are provided in the appendix:
\begin{enumerate}
  \item Run spatial independent components analysis 
(ICA) \citep{24} to decompose each image sequence as $D_{i} = A_i B_i$, 
where, assuming $P$ independent components, $B_i$ is a $P \times V$ matrix 
denoting the spatial maps of the $P$ components, and $A_i$ is 
a $t_i \times P$ matrix denoting the time course of the component 
proportions. 
\item Cluster the $PN$ components to obtain $M$ images. To do this, for 
  each pair of components, we calculate spatial similarity (between voxel 
  values) and temporal similarity (between time-courses), with the overall 
inter-component distance the average of the two. 
Components are clustered based on these distances using the Average-Link 
method of hierarchical clustering \citep{41}.
These $M$ resulting spatial patterns are then rescaled to \textit{z}-scores and 
thresholded using FDR-corrected \textit{p}-values, to highlight the voxels that are active under each of the components. 

\end{enumerate}

\section{Estimation of Group-Representative Activation Map}\label{sec:model} 
For fruitful comparison of functional network patterns across subjects, 
the analysis of activation at the regional-level, rather than at locations 
of specific significance has been recommended; see e.g., \cite{14}. 
Accordingly, we formulate this task as an image segmentation problem, in 
which functionally homogeneous regions (i.e., sets of voxels active under 
the same component) identified across subjects are accorded distinct 
labels. These labels specify the group-representative activation map.

\

We take a model-based approach, characterizing the image segmentation 
problem as the solution to a {\em maximum a posteriori} Markov random 
field (MAP-MRF) inverse optimization problem. Towards this, we first 
establish a forward (generative) model, which models the unknown 
group-representative activation map, and describes the generation 
of the fMRI data from it. Together with the measured data, 
this model determines a posterior distribution over the latent activation 
map.  Our estimate is then the maximizer of this posterior distribution. 
Towards solving this efficiently, we develop a novel variational 
Bayes algorithm~\citep{WaiJor2003}. As we will see, our algorithm 
estimates the unknown global activation map while maintaining uncertainty 
at the level of individual maps, allowing robustness to model 
misspecification and noise, as well as relative insensitivity to local 
optima in the optimization landscape.



\subsection{Forward Model}\label{sec211}
As stated in Section~\ref{sec:preproc}, we write the observed subject 
maps as $Y_i$, with $i=1,\ldots,M$; these are the outputs of the 
pre-processing stage. We wish to estimate the common group-representative map $X$, whose element 
$X(s)$ gives the true label at voxel $s$. Each of the voxels in $X$ and 
the $Y_i$'s can take integer values between $0$ and $K-1$. {Inter-subject 
variation is incorporated through subject-specific binary masking matrices 
$H_i$. If element $H_i(s)$ equals $0$, then the group label $X(s)$ at 
voxel $s$ is propagated to $Y_i(s)$, otherwise $Y_i(s)$ takes on a random 
value $N_i(s)$, that equals $k$ with probability $\pi_k$, for $k=0,\ldots,
K-1$. When the label at voxel $s$ is propagated onwards, we also model 
measurement noise, allowing random mislabeling~\citep{17} through a random 
variable $Z_i(s)$. This equals $0$ (no error) with 
probability $1-\epsilon$, and takes values from $1$ to $K-1$ with 
probability $\epsilon/(K-1)$. Effectively, if propagated on, $Y_i(s)$ 
equals $X(s)$ with probability $1-\epsilon$, and takes any other value 
with probability $\epsilon/K$. The overall process can be written 
compactly as
\begin{align}
  Y_i(s) &= 
     \begin{cases}
       \left[X(s)+Z_i(s)\right]\!\!\!\!\!\mod K & \quad \text{if } H_i(s) = 0\\
       N_i(s) & \quad \text{otherwise.}
     \end{cases}
     \label{eq:gen_model1}
\end{align}
Given measurements $Y_i$, estimating the objects $X, Z_i$ and $N_i$ 
is clearly an ill-posed problem. We regularize the problem above, 
and allow identifiability by specifying {\em prior} probability 
distributions over $X, Y_i, H_i, Z_i, N_i$ and $\pi$. The priors also help 
incorporate domain knowledge about the unknown quantities. 
In particular, we expect both group labels $X$ and the individual masks 
$H_i$ to exhibit spatial structure, and capture this by modeling them with 
Markov random fields \citep{elson2012markov} (MRFs). For the binary $H_i$, 
this becomes the Ising model, where the conditional probability of voxel 
$s$ given all other voxels equals the conditional probability 
given just its neighbors $\partial s$, and satisfies
\begin{align}
  P(H_i(s)&|H_i(\neg s)) = P(H_i(s)|H_i(\partial s)) \nonumber \\ 
          &\propto 
\exp\{-\beta_{H} \ \Sigma_{r{\in}{\partial}s}  
V_{s,r}(H_i(s),H_i(r))\}. \label{h_dist} 
\end{align}
Here $s$ and $r$ are voxels on the 2D lattice $S$ on which the 
group-representative map is defined, $\neg s$ is the set 
of voxels excluding $s$, and ${\partial s}$ is the set of all 
neighbors of voxel $s$. 
In this work, we use a 8-neighbor system on the 2D lattice, with voxels at the boundaries of each slice having fewer neighbors.
The potential function, $V_{s,r}(H_i(s),H_i(r))$ equals $0$ if its 
arguments are equal, else it equals $1$. This induces a penalty $\beta_H$
whenever two neighboring voxels disagree. $\beta_{H}$ is the 
{\em inverse temperature}, determining the degree of spatial cohesion of 
mask $H_i$ \citep{19}. The overall log probability over $H_i$ for each 
subject is the sum over all neighboring pairs $C = \{(r,s)\}$, 
which defines the prior distribution as
\begin{align}
  P(H_i) \propto  \exp\{ -\beta_{H}\Sigma_{ \{r,s\} \in C} V_{s,r}(H_i(s),H_i(r)) \}\ i=1\ldots M.  \nonumber
\end{align}
The unknown group-representative map $X$ is similarly 
modeled, now with a $K$-level Pott's distribution \citep{45}:
\begin{align*}
  P(X(s)&|X(\neg s)) = P(X(s)|X({\partial s})) \\ 
        & \propto \exp\{ -\beta_X \ \Sigma_{ r\in \partial s}  V_{s,r}(X(s),X(r))\}. \label{Potts}
\end{align*}
Again, $\beta_{X}$ is the inverse temperature, and the potential function $V_{s,r}(\cdot,\cdot)$ is defined the same way as in \eqref{h_dist}.

\

The measurement errors $Z_i(s)$ are assumed to be independent and identically 
distributed with discrete distribution 
$P(Z_i(s)=0)=1-\epsilon$, $P(Z_i(s)=1) =, ..., =P(Z_i(s)=K-1)=\epsilon/(K-1)$. Finally, the 
individual label at voxel $s$ (if the group-label voxel is masked 
out), denoted by $N_i(s)$, is assumed to be independent and 
identically distributed with discrete distribution of 
$P(N_i(s) = k) = \pi_k$ for $k = 0, 1, \ldots, K-1$. }. We place a  
Dirichlet prior on the vector $(\pi_0,\ldots, \pi_{K-1})$, and a 
Beta prior on $\epsilon$, and learn these from the data. The overall 
prior distribution is then:
  \vspace{-.1in}
\begin{equation}
 \renewcommand{\arraystretch}{1}
  \vspace{-.05in}
  \begin{tabular}{ll}
  $\pi$  $\sim \text{Dirichlet}(1), \  \epsilon\sim \text{Beta}(1,10),$ \\
  $N_i(s)$  $\sim \pi, \ \  \qquad Z_i(s) \sim 
  (1-\epsilon, \frac{\epsilon}{(K-1)}, \ldots, \frac{\epsilon}{(K-1)}), \forall i,s $ \\
  $X $ $\sim \text{Ising}(\beta_X), \ \  
  H_i \sim \text{Potts}(\beta_H), \quad i = 1\ldots M $
\end{tabular}
     \label{eq:gen_model2}
\end{equation}
We write $\theta$ for the variables $(\pi,\epsilon,\beta_X,\beta_H)$.

\subsubsection{ MAP Estimation}\label{sec213}
The forward model defines a joint probability $p(Y,X,H,\theta)$. Given 
recordings $Y$, this then specifies a Bayesian posterior 
distribution $P(X, H, \theta|Y)$. A natural estimate of the latent 
group-representative map $X^*$, and one that estimates the masking 
matrices $H_i$ as well, is the {\em maximum a posteriori} (MAP) 
solution $(X^*,H^*,\theta^*)$: 
\begin{equation}
\{{H}^\ast,{X}^\ast, \theta^*\}= \arg\max_{H,X,\theta} P(X,H,\theta|Y). \label{mapcrit}
\end{equation}
A practical algorithm to maximize equation~\eqref{mapcrit} is 
coordinate-ascent, alternately maximizing with respect to X given $(H,\theta)$, $H$ 
given $(X,\theta)$ and $\theta$ given $(H,X)$ 
\citep{17}: 
\begin{itemize}
  \item $\hat{H}^{(n+1)} = \arg\max_{H} P(H|Y,\hat{X}^{(n)},\theta^{(n)})$ 
  \item $\hat{X}^{(n+1)} = \arg\max_X P(X|Y,\hat{H}^{(n+1)},\theta^{(n)}))$ 
  \item $\hat{\theta}^{(n+1)} = \arg\max_\theta P(\theta|Y,\hat{X}^{(n+1)},\hat{H}^{(n+1)})$.
\end{itemize}
As we will see in our experiments, coupling between $X$ and $H$ 
can cause 
severe practical problems with local optima, resulting in sensitivity 
to initialization and poor performance. 
In particular, any initialization $\hat{X}^{(0)}$, along with the prior 
and likelihood on $H$ strongly constrains $\hat{H}_i^{(1)}$. This 
in turn will strongly constrain  $\hat{X}^{(1)}$, resulting in 
$\hat{X}^{(1)} \approx \hat{X}^{(0)}$, and preventing the algorithm from 
escaping from its initial value.
To overcome this sensitivity to initialization, we propose a variational 
Bayes algorithm~\citep{WaiJor2003}, which optimizes over $X$ directly, while 
marginalizing out the individual masks $H=\{H_i\}$. Recall, that of primary interest to us 
is the group-representative map $X$, and an estimate of it can be 
obtained by directly optimizing $P(X,\theta|Y)$:
\begin{equation}
  ({X}^\ast,\theta^*)= \arg\max_{X,\theta} P(X,\theta|Y) = \arg\max_{X,\theta}\sum_{H} P(X, H,\theta|Y). \label{mapmargcrit}
\end{equation}
Evaluating this objective requires summing over exponentially 
many configurations of each of the $H_i$'s, which is 
intractable. The idea behind variational Bayes is to 
optimize a tractable lower bound to this quantity. Recognizing that log 
is a concave function and using Jensen's inequality~\citep{cover2006elements}, we have, 
for any probability distribution $q(H)$:
\begin{align}
  \log P&(X,\theta|Y) 
          = \log\sum \frac{P(X, H, \theta|Y)q(H)}{q(H)}  \nonumber\\
          &\geq\sum q(H)\log \frac{P(X, H, \theta|Y)}{q(H)} 
  := F(X,\theta,q). \label{eq:vb}
\end{align}
Variational Bayes now alternately optimizes this lower bound $F$ with 
respect to $X, \theta$ and $q$. Without any additional constraints on 
$q$, for any $(X,\theta)$, there exists a $q^*$ such that the bound is 
tight (i.e.\ $\log P(X,\theta|Y) = F(X,\theta,q^*)$), and variational 
Bayes reduces to solving the original intractable problem via the 
so-called EM algorithm~\citep{EM}. However by restricting $q$ to simpler 
class of probability distributions $\mathcal{Q}$, evaluating $F$ can 
be made tractable, and an approximate solution $(X^*,\theta^*)$ can be 
found to original MAP problem. Two choices of $\mathcal{Q}$ suggest themselves:\\
\tikz\draw[black,fill=black] (0,0) circle (.3ex);   {\em $\mathcal{Q}$ is the family of delta functions.} Here, $q$ 
    supports only one value for each $H_i$, and the 
    summation in equation~\eqref{eq:vb} reduces to an optimization over 
    $H$, recovering the earlier coordinate-ascent algorithm.  As we 
    mentioned before, the resulting tractability comes at the price of 
    poor convergence properties, easily getting trapped in local optima.  
    Additionally, the restriction to delta functions discards uncertainty 
    about the unobserved $H$ by $q$, and the resulting $F(X,\theta,q)$ 
    can be a poor approximation to $\log P(X,\theta|Y)$.\\
\tikz\draw[black,fill=black] (0,0) circle (.3ex);   {\em $\mathcal{Q}$ is the family of mean-field approximations.} 
    Here, under any element of $\mathcal{Q}$,
each voxel takes values independently:
$\mathcal{Q} = \{q(H) \text{ s.t. }q(H) = \prod q_i(H_i), 
q_i(H_i) = \prod_s q_{is}(H_i(s))\}$. Now, optimizing over $q(H)$ involves optimizing 
the components $q_{is}$, each of which is a number between $0$ and $1$ giving 
the probability that voxel $s$ in mask $i$ is on. 
This is a relaxation of the original coordinate-ascent algorithm, 
where each component of $H$ was set to either $0$ or $1$. 
As we will see, this is just a moment matching problem, where for each 
voxel, to set $q(H)$, we only need to calculate the marginal probability 
that it equals 1 from $P(H|X,Y)$. 

\subsection{Mean-field variational Bayes algorithm}
In this section, we outline the details of the mean-field variational Bayes 
algorithm. At a high level, this is an iterative process that starts with 
initial values $X^{(0)},q^{(0)}(H),\theta^{(0)}$, 
and then updates $q(H), X$ and $\theta$ in turns. 
For compactness, we drop dependence on $\theta$ and write $H_{is}$ for 
$H_i(s)$ (and similarly for $X$ and $Y$).
We first note that
\begin{align*}
  & \log P(Y_i|H,X) = \sum_s \large[ \mathbbm{1}_0(H_{is}) 
  \{\mathbbm{1}_{X_s}({Y_{is}})\log(1-\epsilon)+ \nonumber \\
  & (1\!-\!\mathbbm{1}_{X_s}({Y_{is}}))\log \frac{\epsilon}{K\!-\!1} \}\!+\! 
    \mathbbm{1}_1(H_{is})\! \sum_{k=1}^K \mathbbm{1}_k(Y_{is}) \log \pi_k] \\ \nonumber
    &:= \sum_s \mathbbm{1}_0(H_{is})A_s + \mathbbm{1}_1(H_{is}) B_s
\end{align*}
This, equations \eqref{eq:gen_model1} and \eqref{eq:gen_model2},
and the factorial assumption on $q(H)$ allows the easy calculation 
of $F$ from equation \eqref{eq:vb}. 
To update $q_{is}$, component $s$ of $q_i$, we set 
$\frac{dF}{dq_{is}}=0$, giving 
$q_{is}=\frac{\exp(B_s)}{\exp(A_s)+\exp(B_s)}$. 
We update $X$ one voxel at a time, with the update rule for $X_s$ given by
\begin{align*}
  & X^{(n+1)}_s=\arg\max_{X}\Sigma_{ H }q^{(n+1)}(H)\log P(Y,X|H) \\
  & =\arg\max_{X}\!\sum_{i=1}^{M}\!\sum_{s'=1}^{N}(1-q^{(n+1)}_{is'})
  \{\mathbbm{1}_{X_{s'}}({Y_{is'}})\log(1-\epsilon) + \\ 
  &\quad (1-\mathbbm{1}_{X_{s'}}({Y_{is'}})\log(\frac{\epsilon}{K-1})\} 
    -\beta_X \Sigma_{ \{r,s'\}} V_{s',r}(X_{s'},X_r) \\
    &=\arg\max_{X_s}\sum_{i=1}^{M}(1-q^{(n+1)}_{is})\{\mathbbm{1}_{X_s}({Y_{is}})\log(1-\epsilon)+ (1- \\
&\quad\mathbbm{1}_{X_s}({Y_{is}})\log(\frac{\epsilon}{K-1})\}
            -{\mathtt{\beta_X}}\Sigma_{r \in {\partial}s} V_{s,r}(X_s,X^{(n)}_r).
\end{align*}
The last step is to maximize $F$ with respect to the parameters $\theta$, 
this can be carried out easily using standard MRF estimation 
techniques~\cite{Baddeley00}.
We repeat these steps until convergence 
(which is guaranteed by the fact that $F$ is a lower-bound to 
$\log p(X,\theta|Y)$, and that every variational Bayes step increases $F$. 

\

\section{Experiments and Results}\label{sec:expt} 
\begin{figure*}
  \centerline{
  \subfloat[]{
  \includegraphics[width=0.34\textwidth]{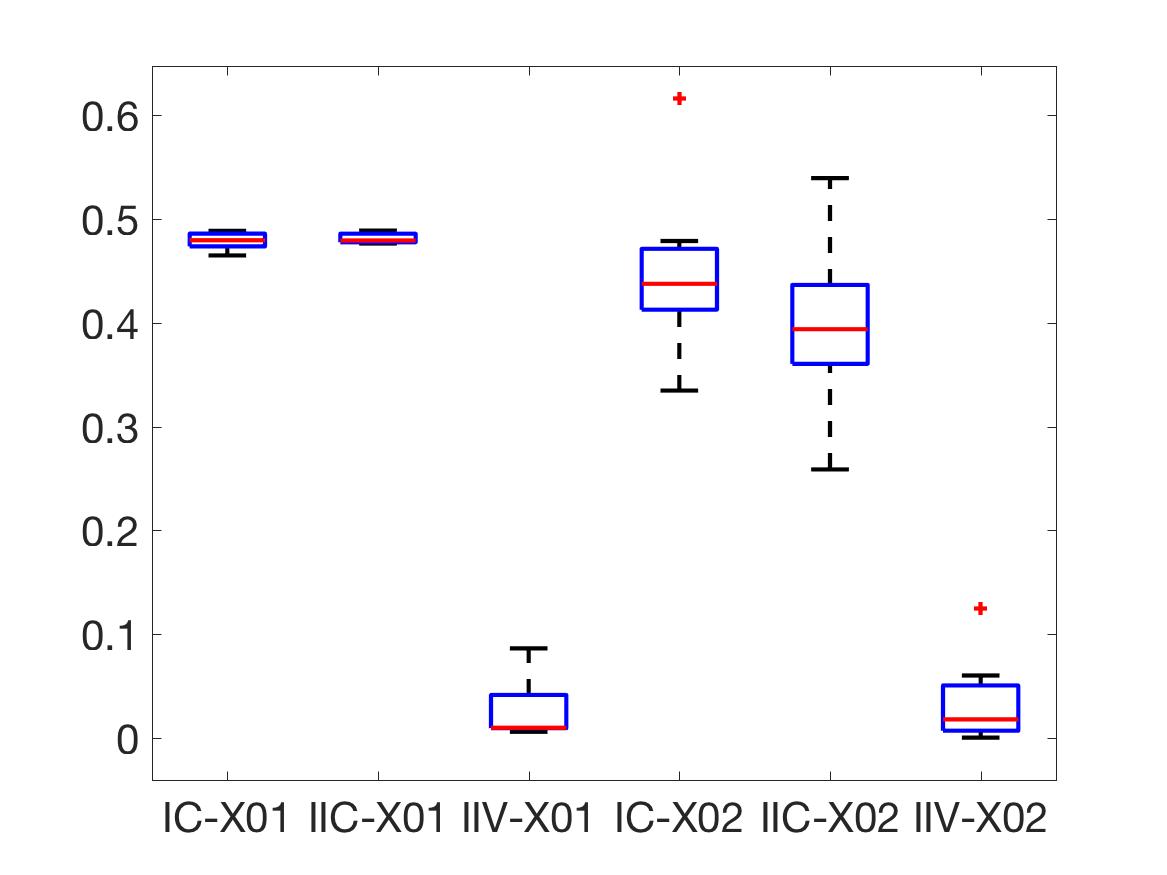}
  \label{fig:M10K2}
  }   
  \subfloat[]{
  \includegraphics[width=0.34\textwidth]{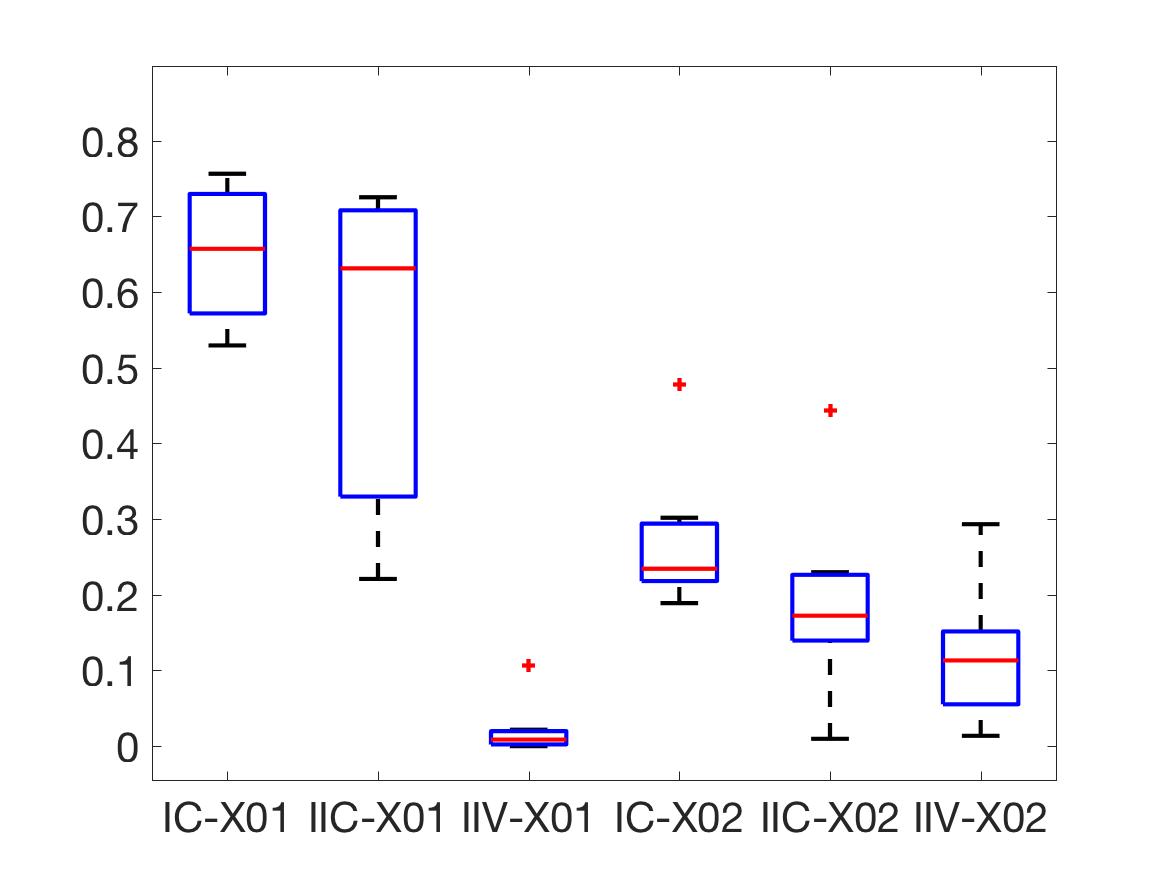}
  \label{fig:M10K5}
  }
  \subfloat[]{
  \includegraphics[width=0.34\textwidth]{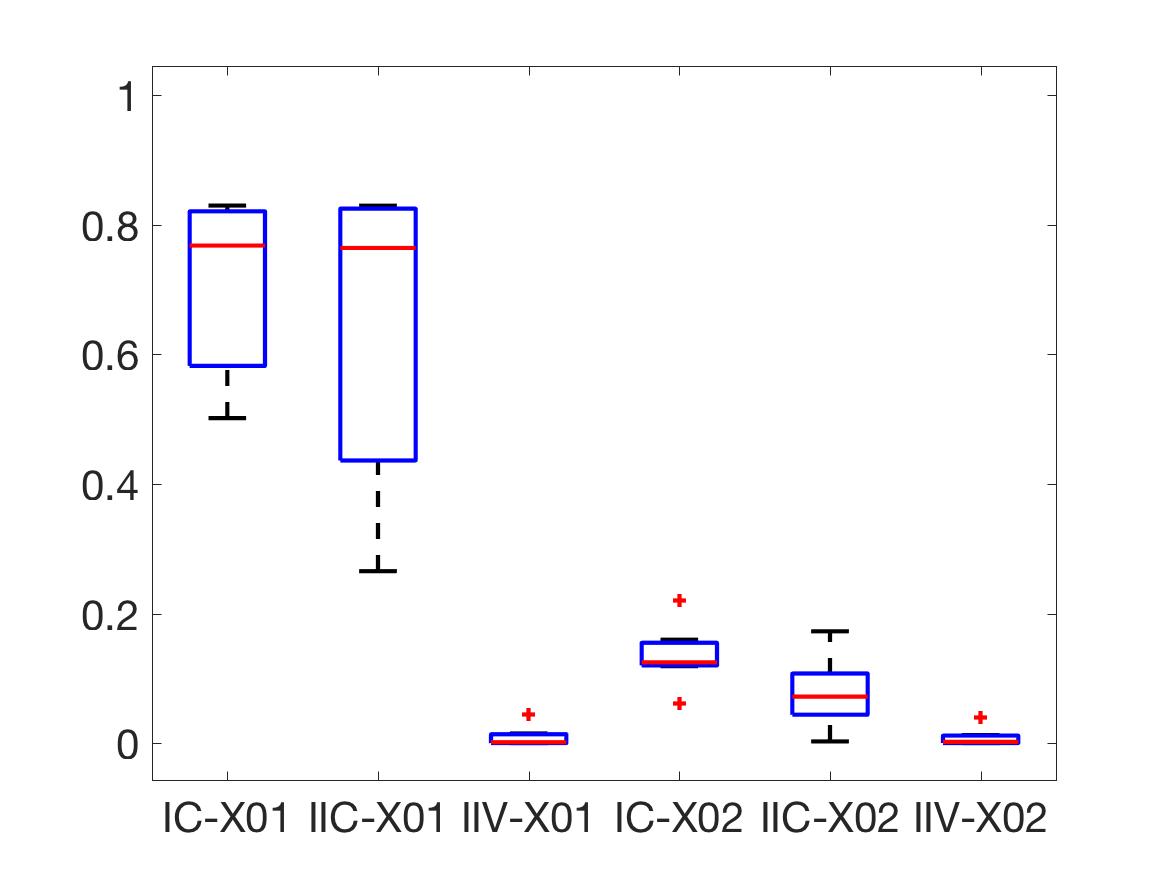}
  \label{fig:M10K10}
  }
  }
  \centerline{
  \subfloat[]{
  \includegraphics[width=0.34\textwidth]{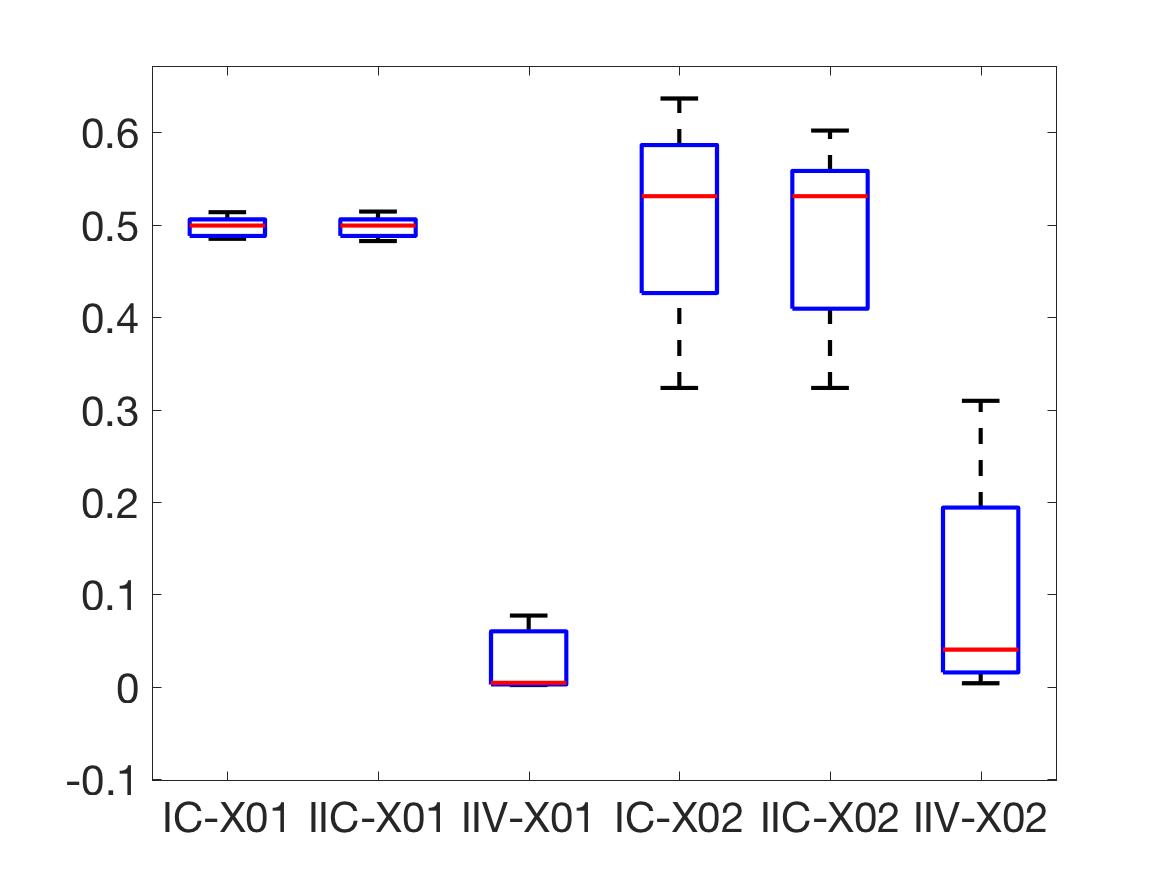}
  \label{fig:M20K2}
  }  
  \subfloat[]{
  \includegraphics[width=0.34\textwidth]{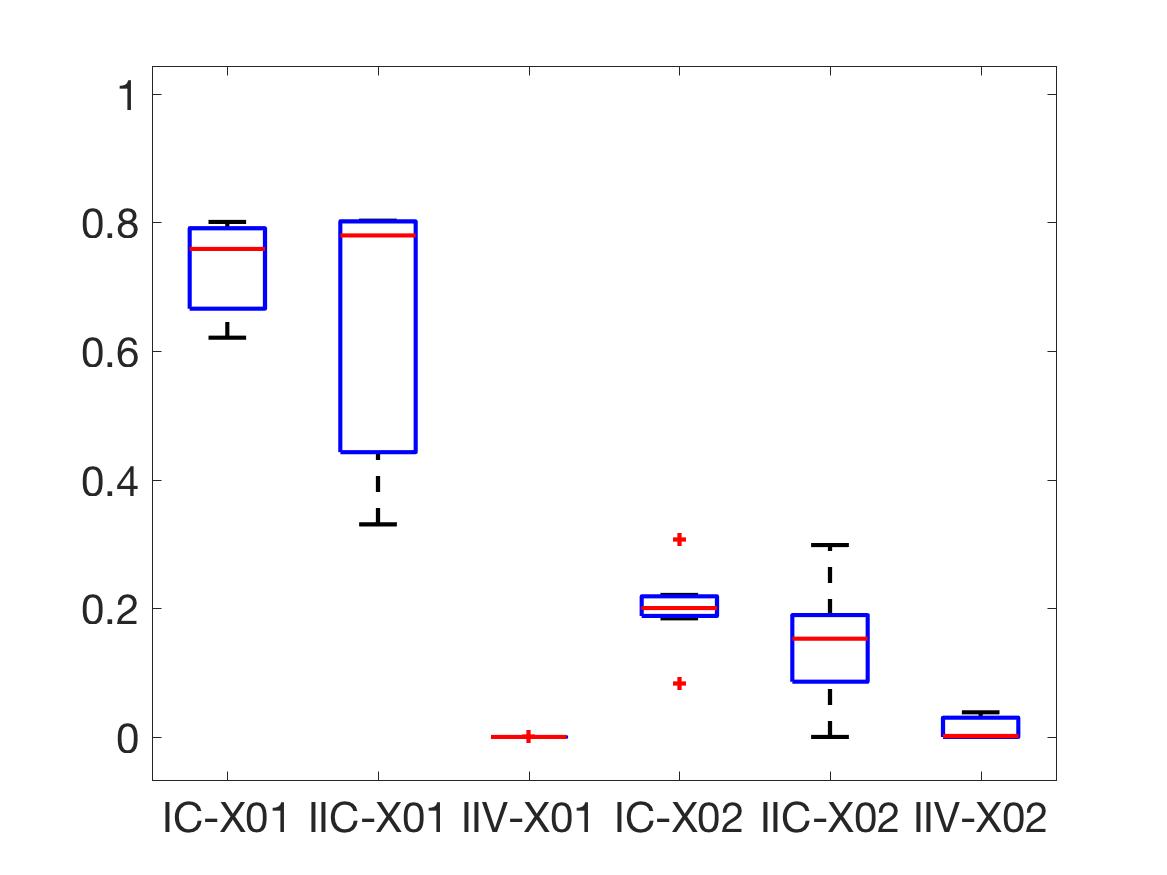}
  \label{fig:M20K5}
  }
  \subfloat[]{
  \includegraphics[width=0.34\textwidth]{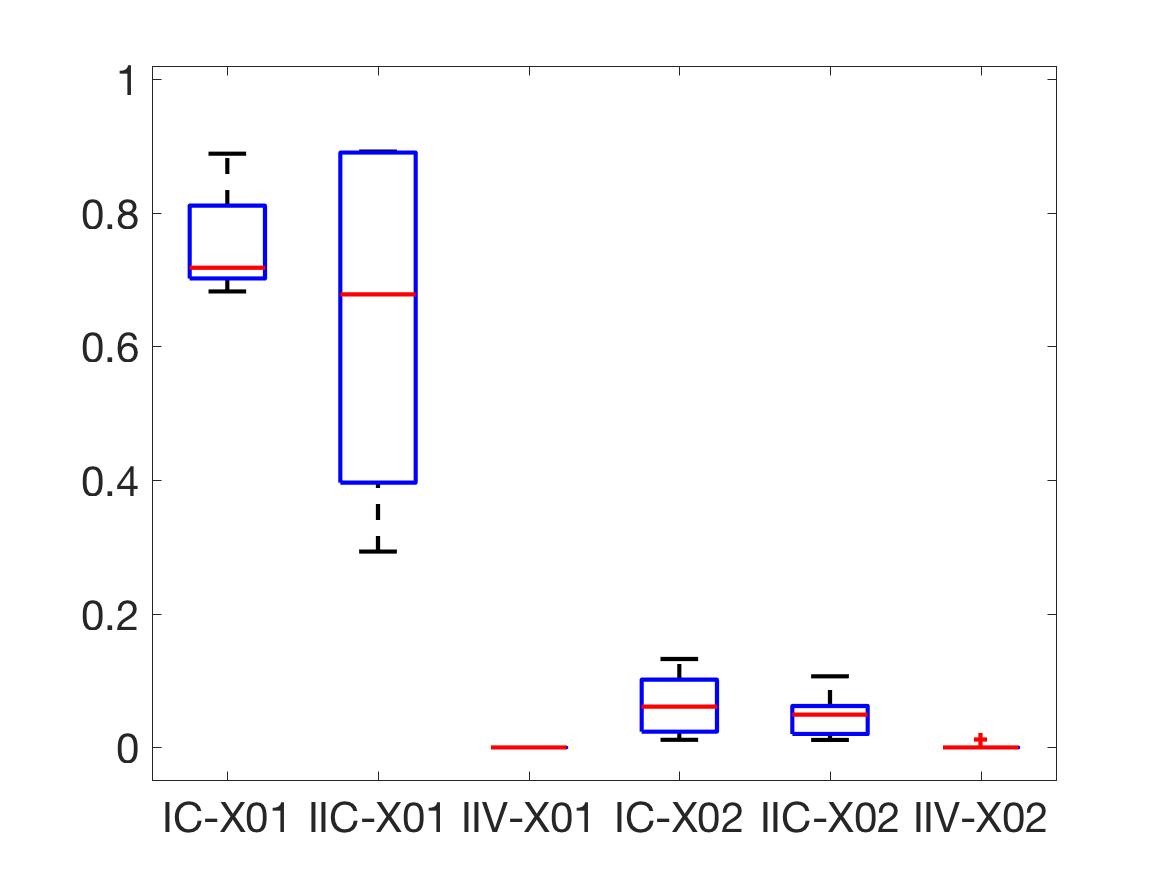}
  \label{fig:M20K10}
  }
  }
  
  \centerline{
  \subfloat[]{
  \includegraphics[width=0.34\textwidth]{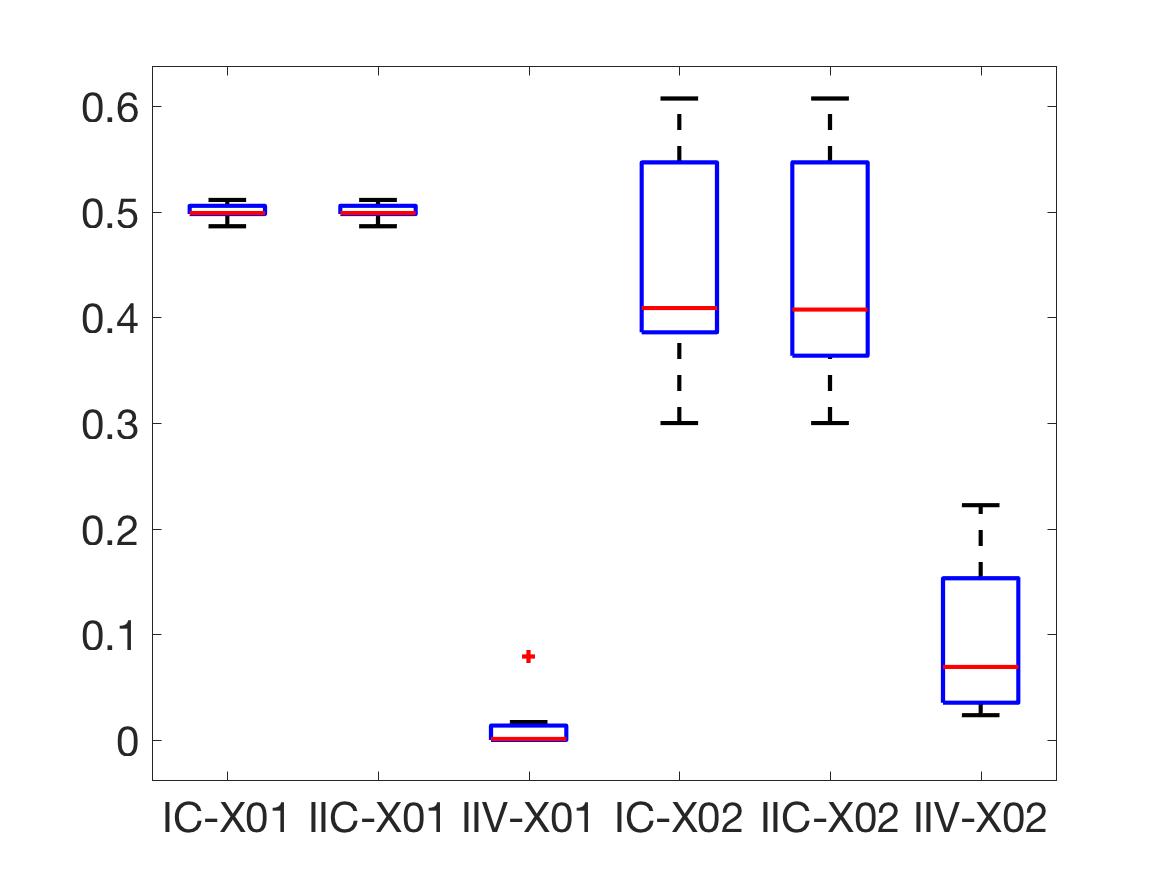}
  \label{fig:M40K2}
  }   
  \subfloat[]{
  \includegraphics[width=0.34\textwidth]{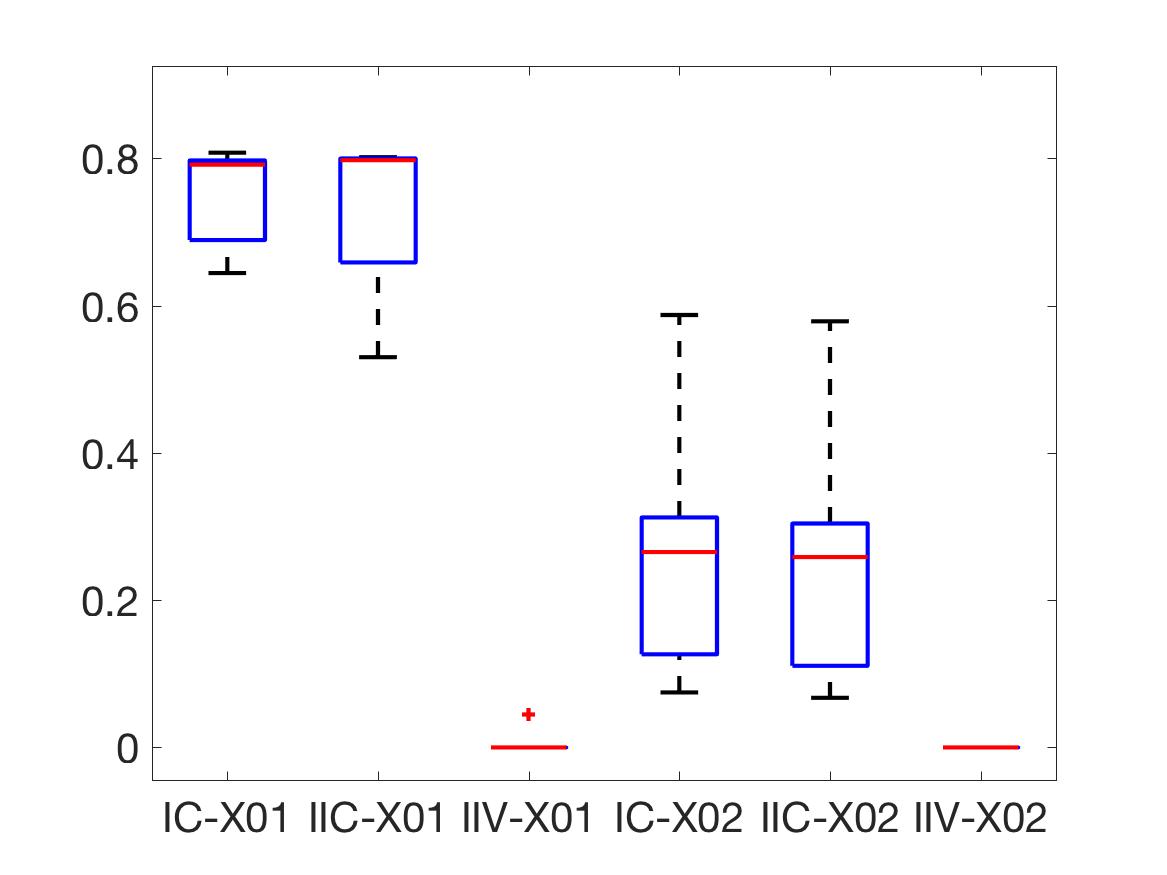}
  \label{fig:M40K5}
  }
  \subfloat[]{
  \includegraphics[width=0.34\textwidth]{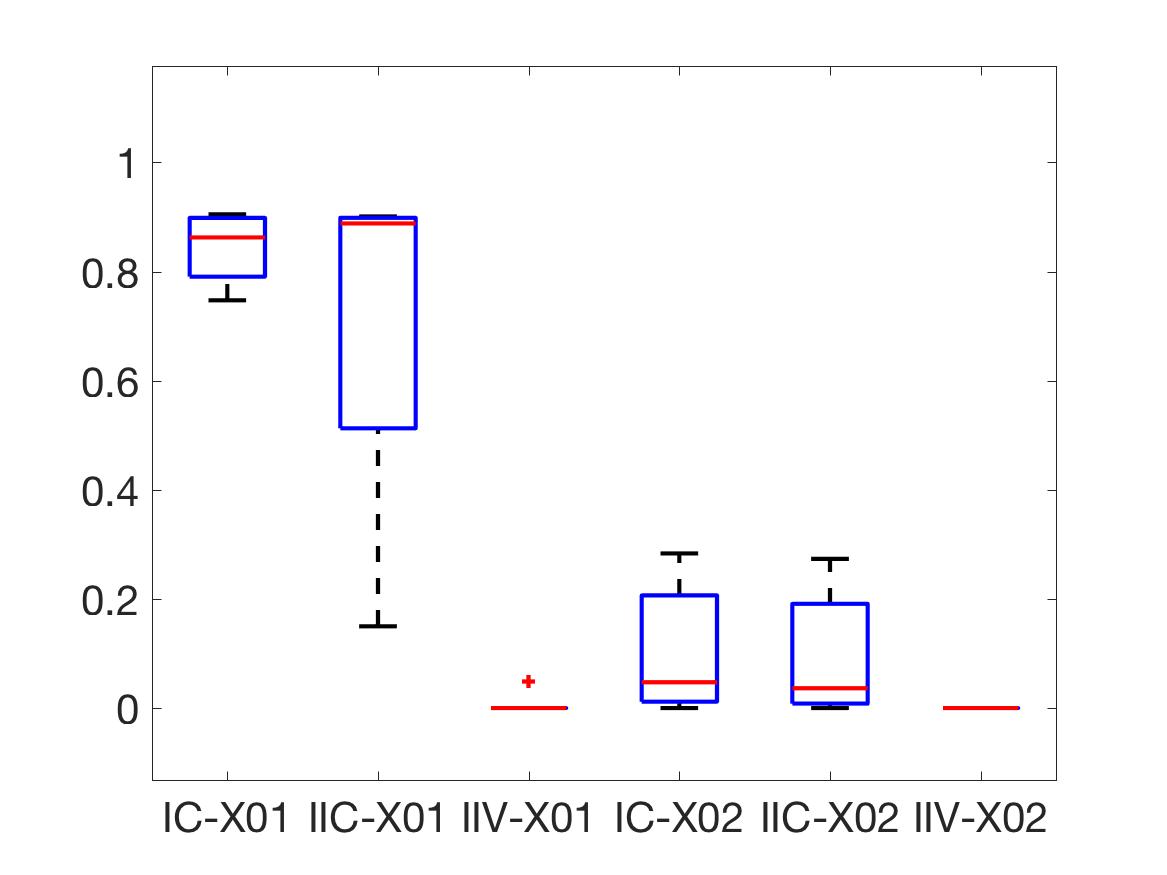}
  \label{fig:M40K10}
  }
  }

  \caption{Results for datasets generated from forward model \initialI. 
    Columns correspond to different settings of $K$ (2, 5 and 10 
    respectively).  
    Rows correspond to different settings of $M$ (10, 20 and 40 
    respectively).  
    Boxplots within each subplot represent misclassification rates for 
    10 repeats for (from left to right), \initialI\ with 
    coordinate-ascent ($IC$), \initialII\space with coordinate-ascent ($IIC$), 
    \initialII\space with variational Bayes ($IIV$). These are repeated 
    twice, with random initializations ($X_{01})$, and with a greedy 
    initialization that
    contains all the nonzero components from the $M$ images
    ($X_{02}$). 
}
\label{fig:fig1}
\end{figure*}
In this section, we first validate the proposed MAP-MRF framework 
using synthetically generated maps from two observation processes: 
the model specified in equations~\eqref{eq:gen_model1}-\eqref{eq:gen_model2},
as well as a simplified version without measurement noise (i.e.\ $Z(s)=0$).  
Table 1 summarizes these two models. We also compare two algorithms: a 
coordinate-ascent optimization algorithm and our variational Bayes algorithm.
Across different settings, we compare the estimated group map to the 
synthetically generated ground truth group map, allowing us to assess the 
viability of the proposed framework as well as its robustness to modeling 
error.

\begin{table}[H]
\small
\begin{center}
 \begin{tabular}
{
  @{\kern-.5\arrayrulewidth}
  |p{\dimexpr2.2cm-2\tabcolsep-.5\arrayrulewidth}
  |p{\dimexpr3.4cm-2\tabcolsep-.5\arrayrulewidth}
  |p{\dimexpr4.0cm-2\tabcolsep-.5\arrayrulewidth}
  |p{\dimexpr3.2cm-2\tabcolsep-.5\arrayrulewidth}
  |@{\kern-.5\arrayrulewidth}
}

 \hline
 Generative Model & Forward Model at Voxel s& Prior Model \\ [0.5ex] 
 \hline\hline
 \initialI \qquad & $Y_i(s)=H_i(s) \odot X(s)+N_i(s)$ &$H_i, X$ $\sim$ MRF, 
 \
 
$N_i(s)$ $\overset{\text{i.i.d.}}{\sim} P , P \sim$ Dir(K,1) \\ 
 \hline
 \initialII \qquad (proposed) &$Y_i(s)=H_i(s) \odot (X(s)+Z_i(s)) + N_i(s)$  & $H_i, X$ $\sim$ MRF, 
 \
 
 $N_i(s)$ $\overset{\text{i.i.d.}}{\sim} P, P \sim$ Dir(K,1) \\ 
 [1ex] 
 \hline
\end{tabular}
\end{center}
\caption{Details of both generative models. For \initialII, we have 
both coordinate-ascent as well as variational Bayes.}
\label{table:1}
\end{table}
To generate the group map $X$, we simulate a 
$K$-level Potts model. We also generate binary masks $H_i$,  
$i=1,2,\ldots,M$, following an Ising model. For both we use random 
temperature parameters drawn uniformly between 0 and 1. Next, 
individual 
subject maps $Y_i$ are produced from $X$ and $H_i$, according to 
\initialI\ and \initialII.
In section~\ref{411}, 
we use datasets from \initialI\ to evaluate our two algorithms:
coordinate-ascent, as well as our proposed variational 
Bayes algorithm, by comparing misclassification rates.
In section~\ref{412}, we repeat this, now with synthetic datasets generated 
from \initialII.
Finally, we analyze the efficacy of our proposed MAP-MRF 
framework, i.e., \initialII\space with variational Bayes, on a simulated 
fMRI dataset. In section~\ref{413}, we present the estimation results, 
and show the robustness of the proposed MAP-MRF 
framework. 
\begin{figure*}
\begin{minipage}{.7\textwidth}
  \centerline{
  \subfloat[]{
  \includegraphics[width=0.3\textwidth]{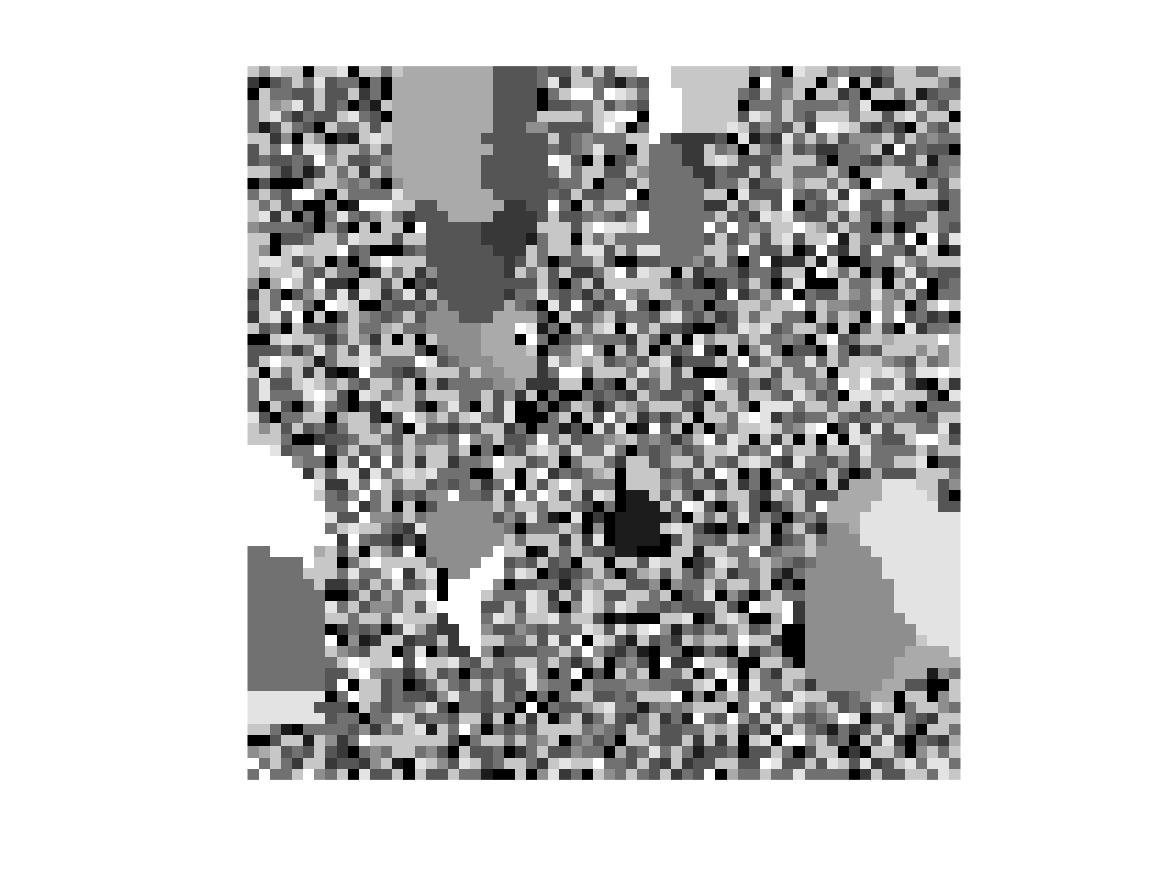}
  \label{fig:Y1}
  }   
  \subfloat[]{
  \includegraphics[width=0.3\textwidth]{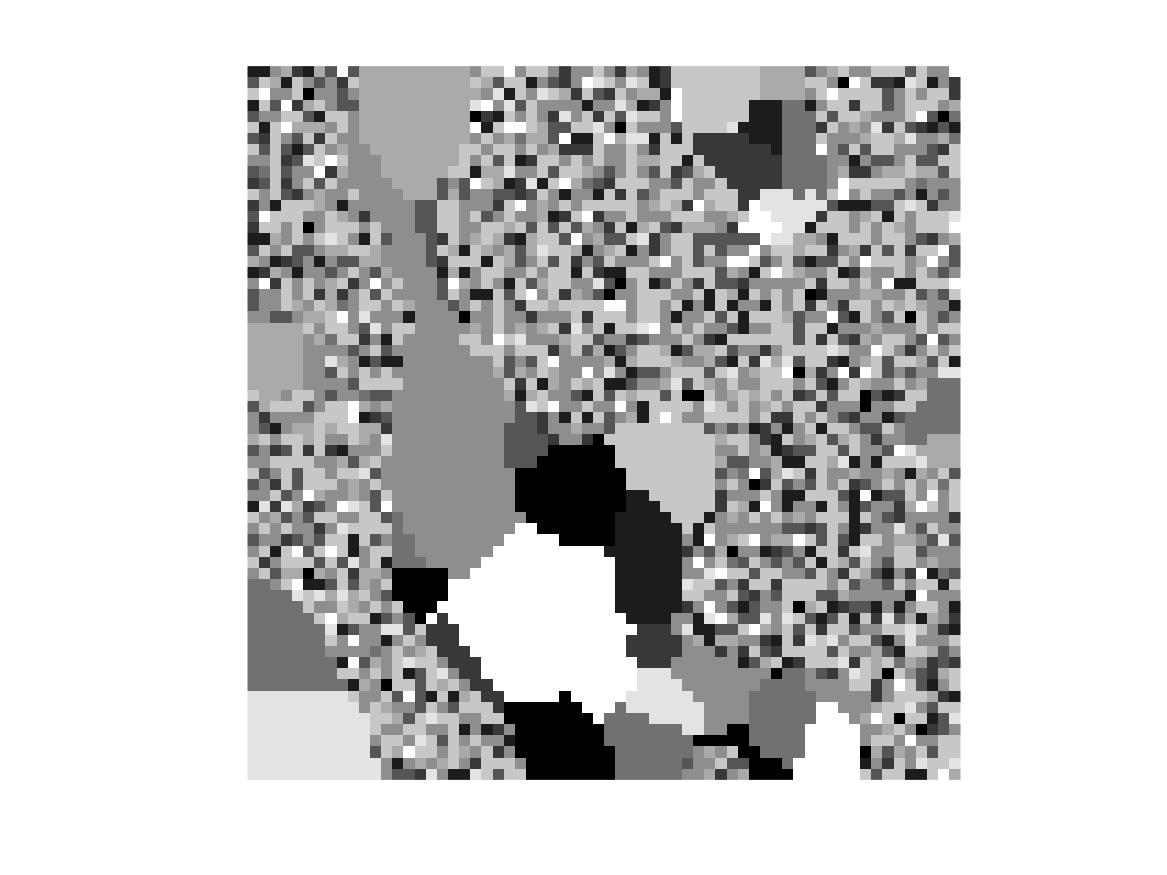}
  \label{fig:Y2}
  }
  \subfloat[]{
  \includegraphics[width=0.3\textwidth]{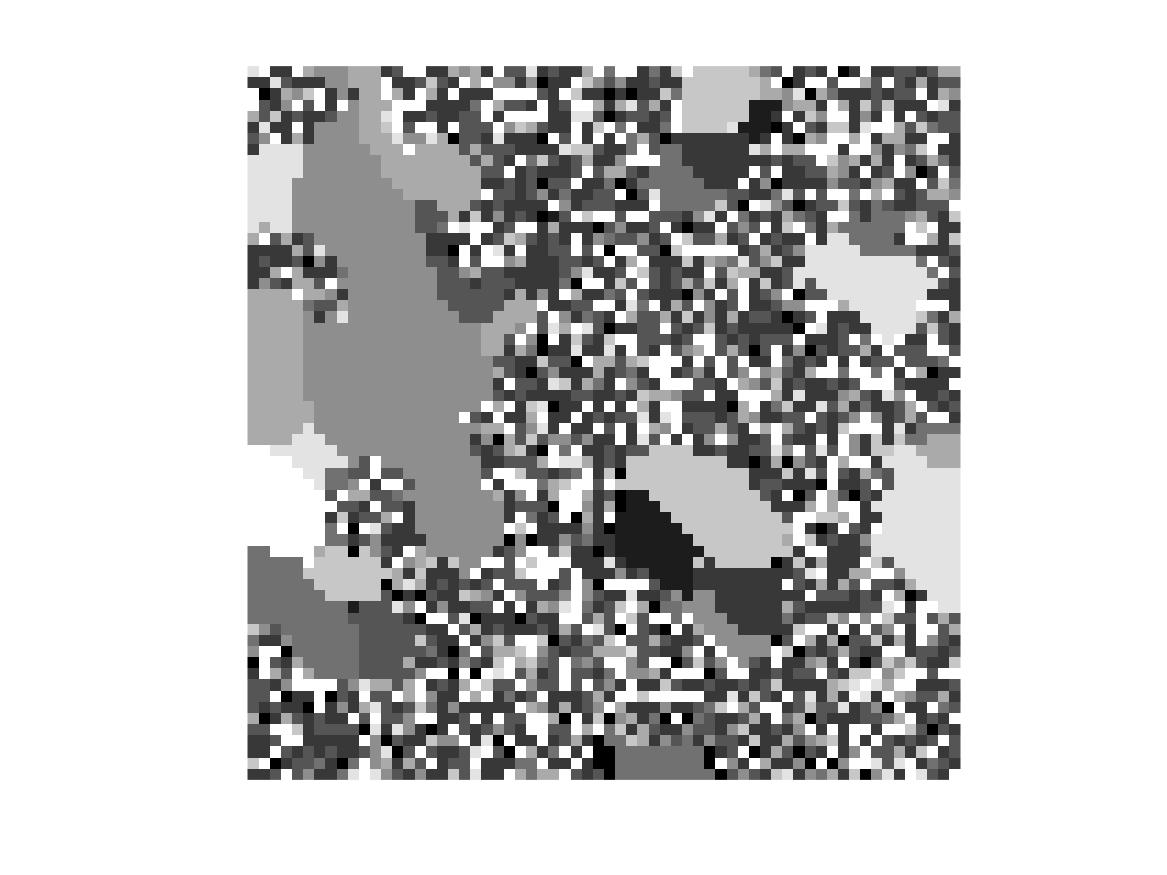}
  \label{fig:Y3}
  }
  }
\end{minipage}
\begin{minipage}{.25\textwidth}
  \caption{Figures \protect\subref{fig:Y1}-\protect\subref{fig:Y3} show individual subject maps, $Y_1$, $Y_2$, $Y_3$, respectively.
}
\label{fig:fig2}
\end{minipage}

\begin{minipage}{.7\textwidth}
   
  \centerline{
  \subfloat[]{
  \includegraphics[width=0.3\textwidth]{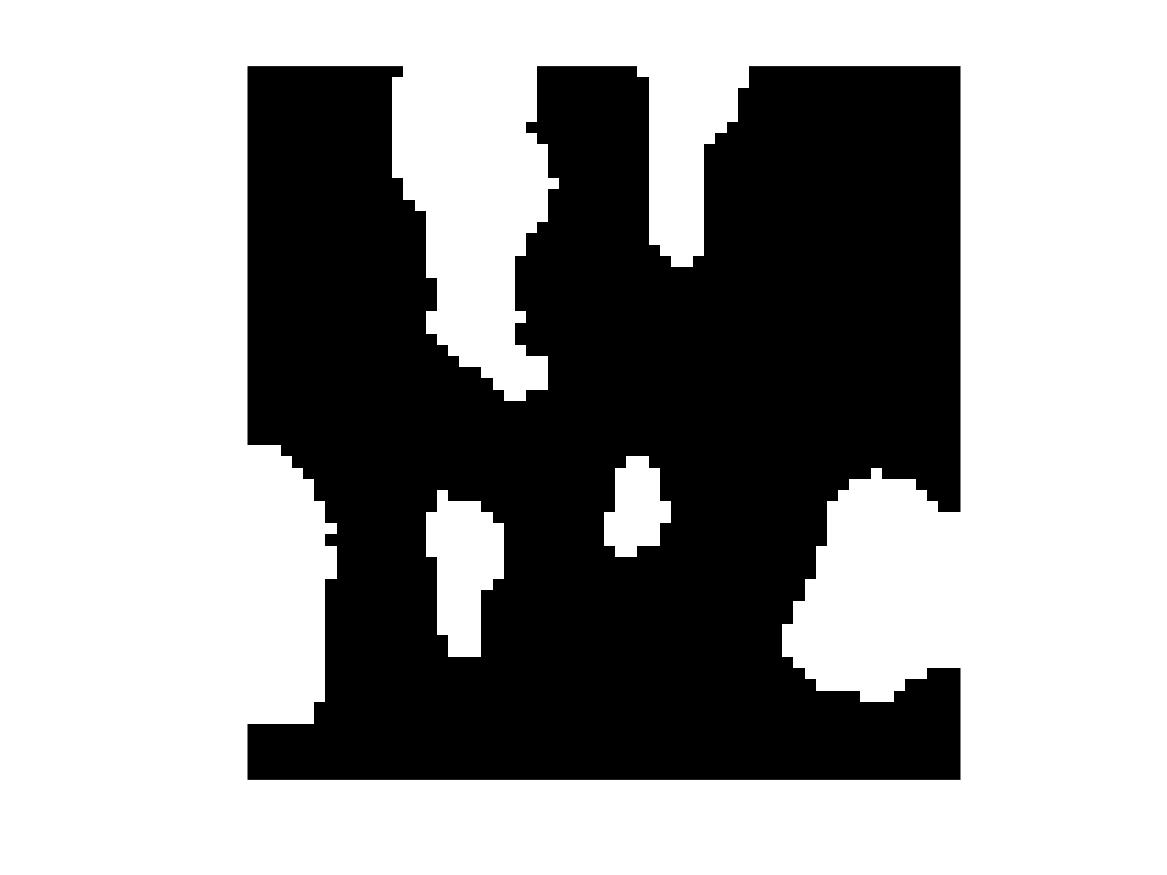}
  \label{fig:H1_gt}
  }   
  \subfloat[]{
  \includegraphics[width=0.3\textwidth]{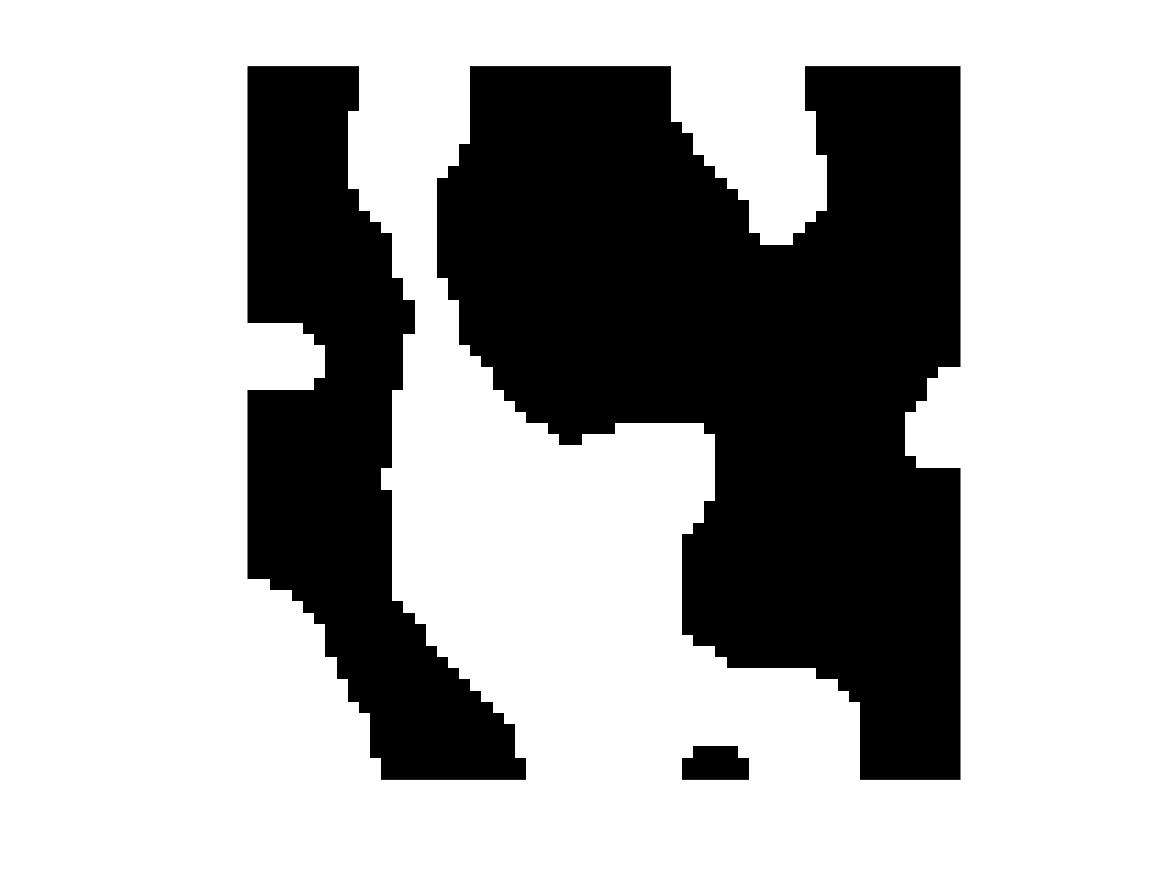}
  \label{fig:H2_gt}
  }
  \subfloat[]{
  \includegraphics[width=0.3\textwidth]{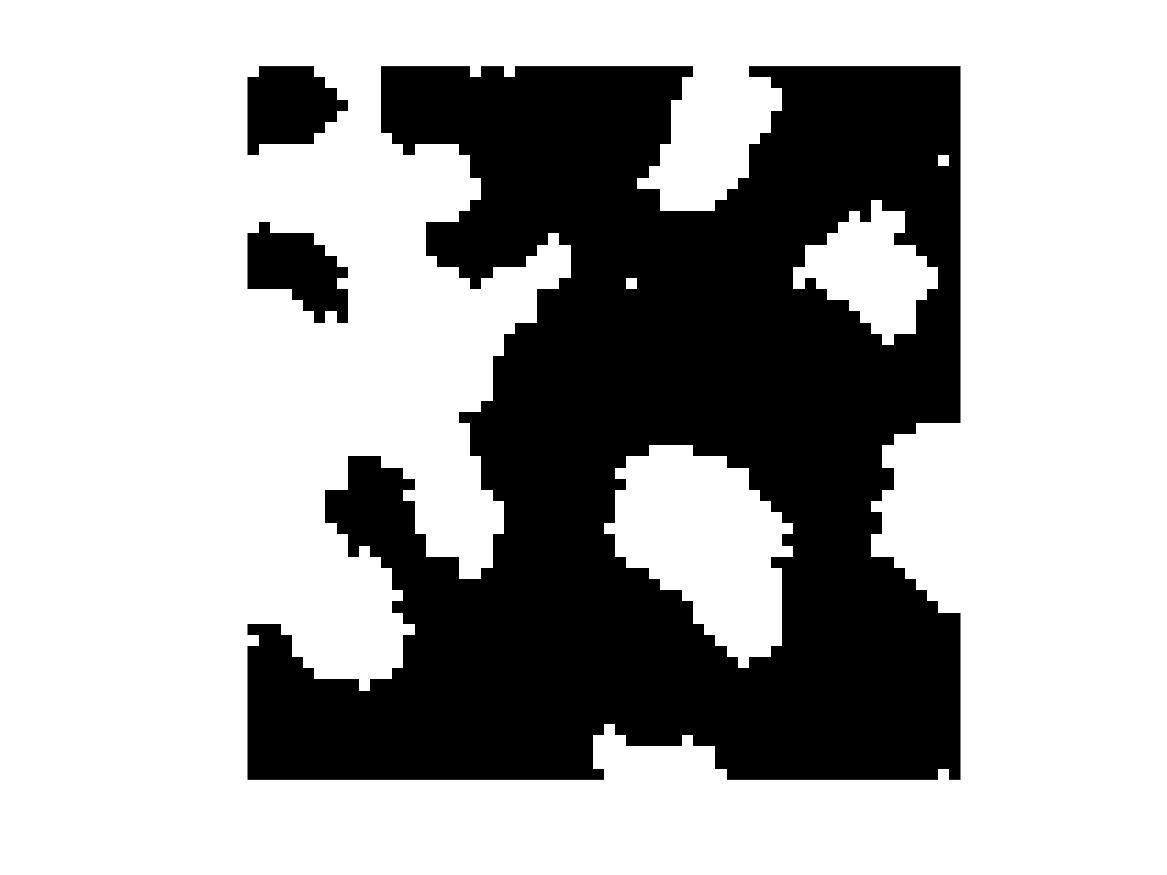}
  \label{fig:H3_gt}
  }
  }
 
\centerline{
\subfloat[]{
\includegraphics[width=0.3\textwidth]{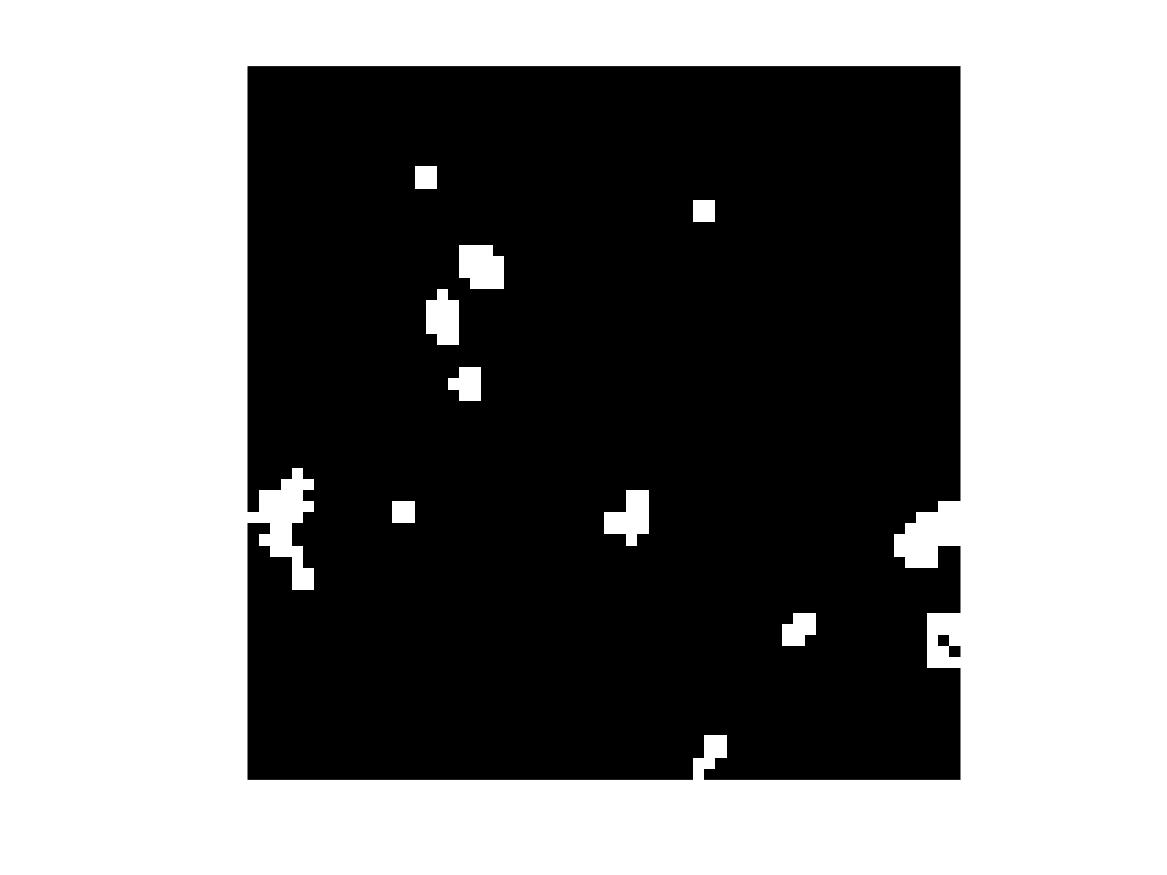}
\label{fig:Hout1Irandom}
}   
\subfloat[]{
\includegraphics[width=0.3\textwidth]{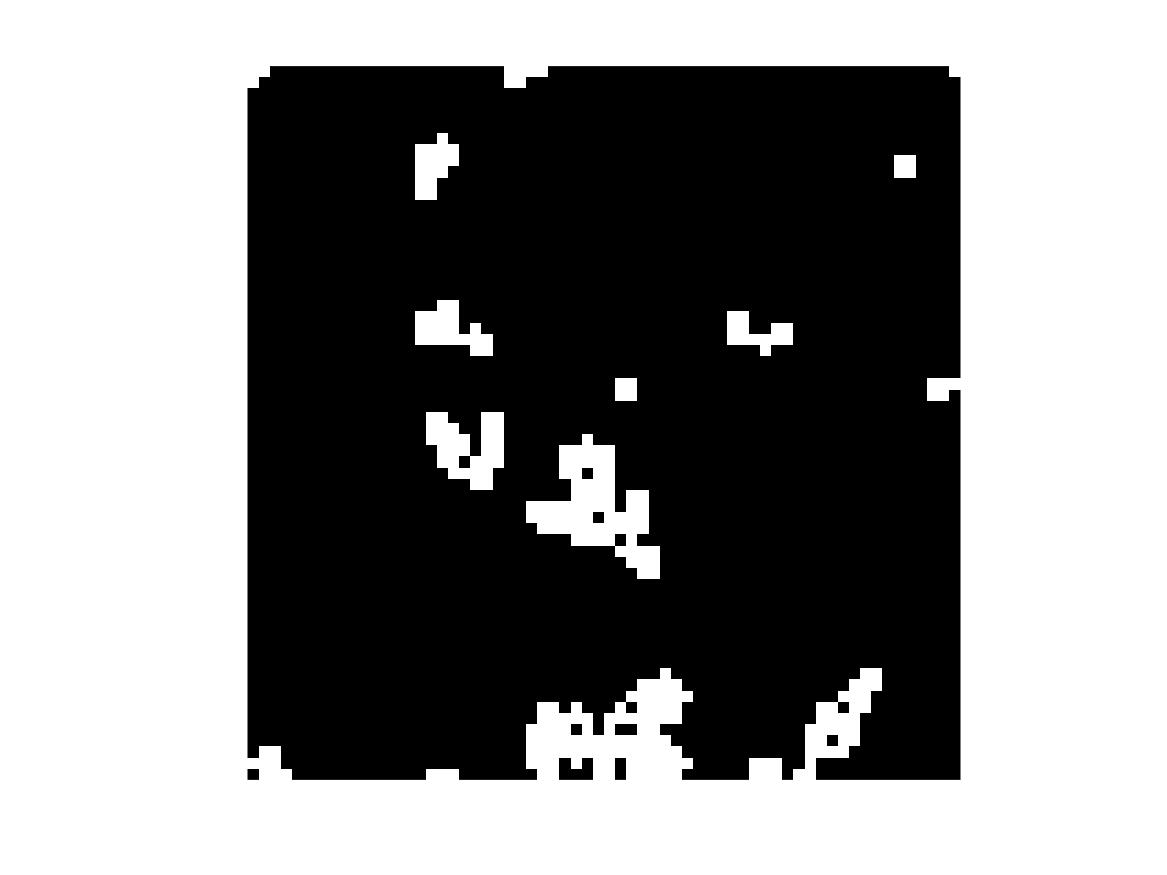}
\label{fig:Hout2Irandom}
}
\subfloat[]{
\includegraphics[width=0.3\textwidth]{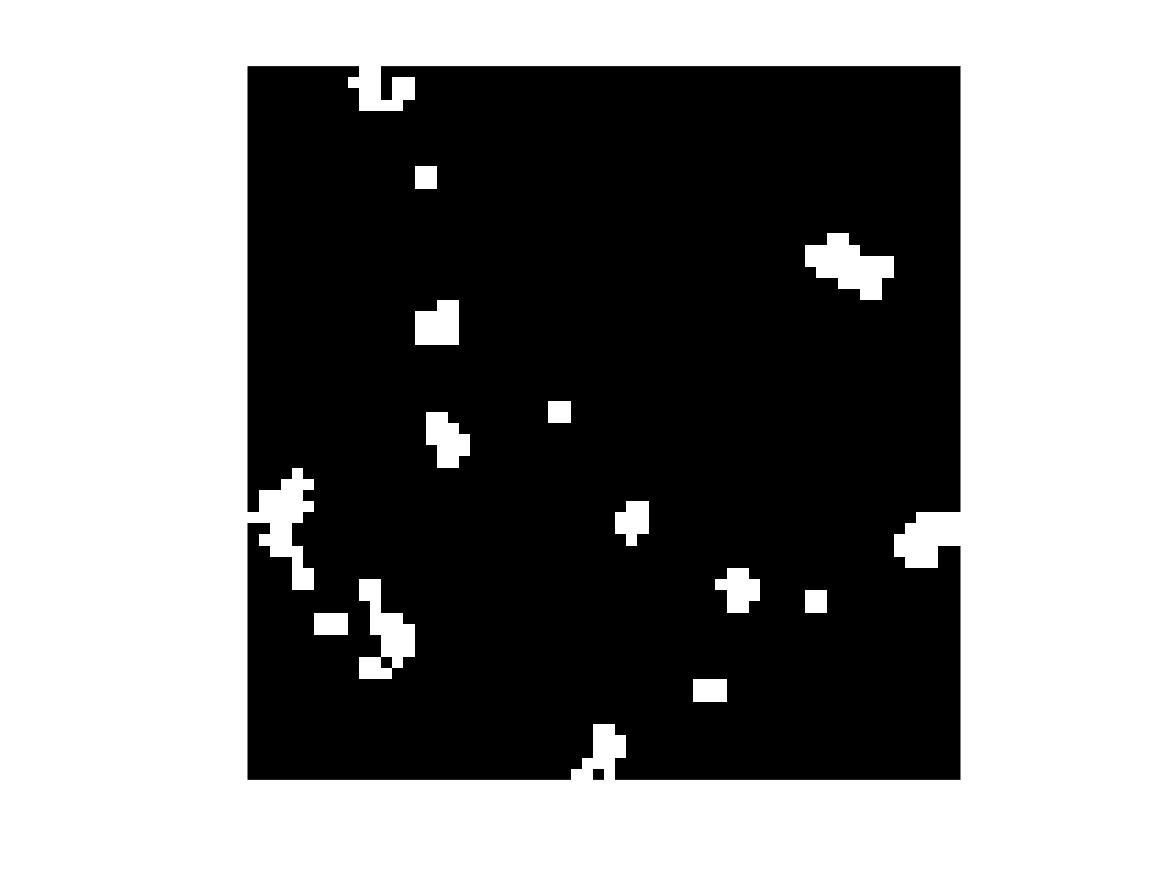}
\label{fig:Hout3Irandom}
}
}
  \centerline{
  \subfloat[]{
  \includegraphics[width=0.3\textwidth]{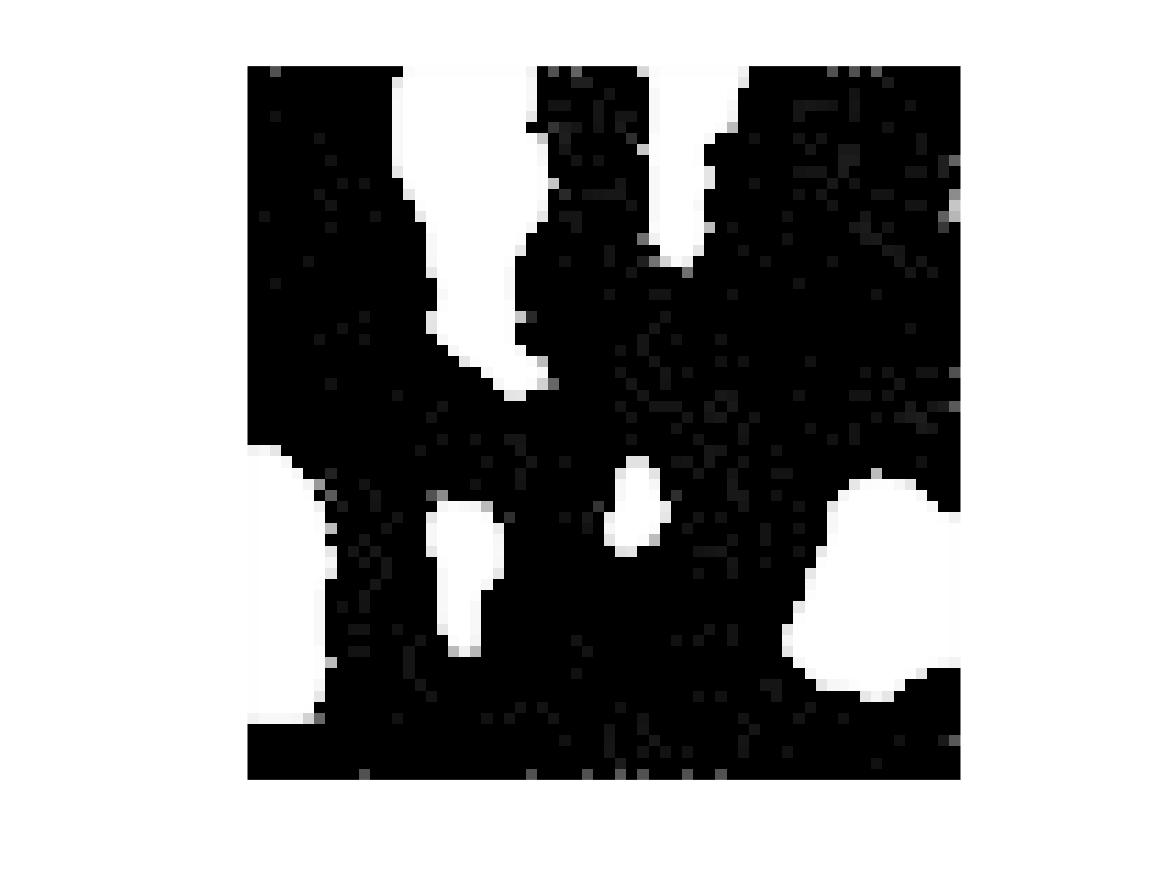}
  \label{fig:Qout1random}
  }   
  \subfloat[]{
  \includegraphics[width=0.3\textwidth]{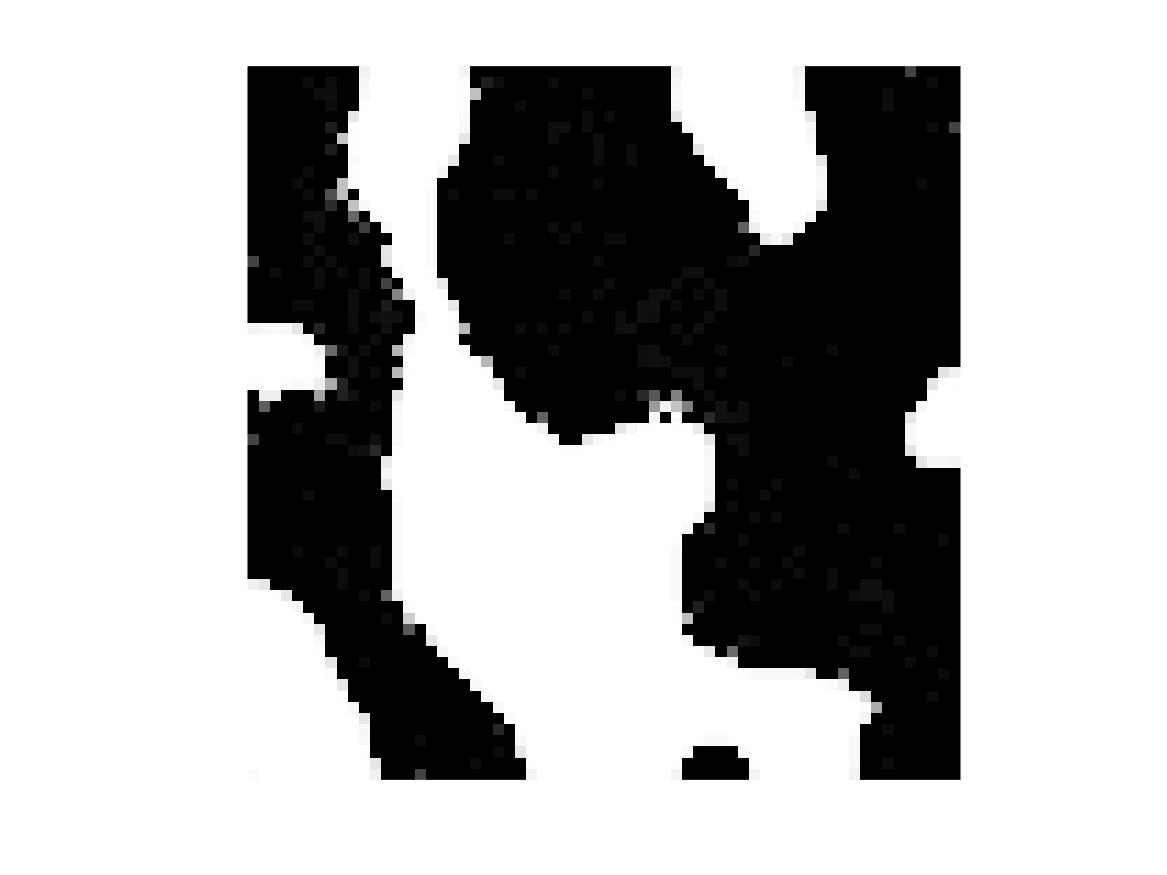}
  \label{fig:Qout2random}
  }
  \subfloat[]{
  \includegraphics[width=0.3\textwidth]{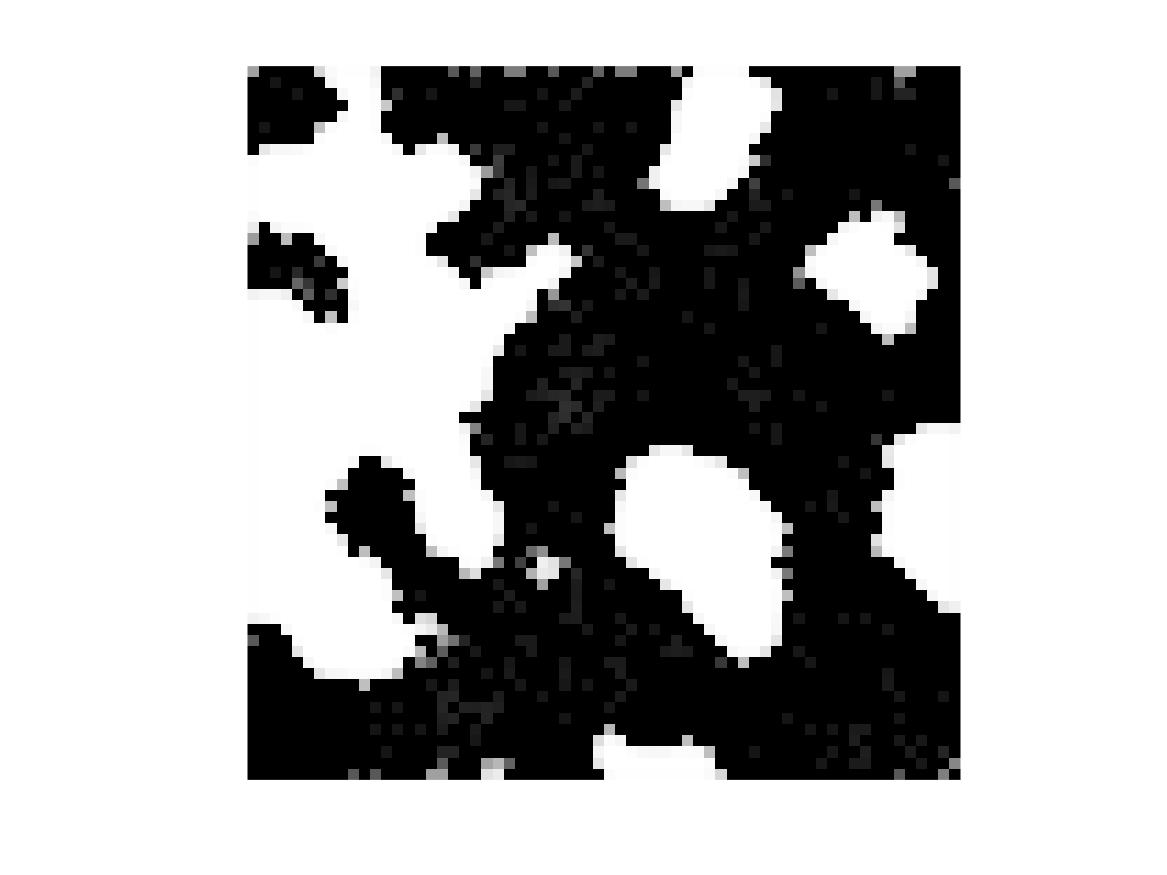}
  \label{fig:Qout3random}
  }
  }
  
%
\end{minipage}
\begin{minipage}{.25\textwidth}
 
  \caption{Figures\protect\subref{fig:H1_gt}-\protect\subref{fig:H3_gt} 
    show the ground truth binary masks, $H_1$, $H_2$, $H_3$, respectively. 
    Figures \protect\subref{fig:Hout1Irandom}-\protect\subref{fig:Hout3Irandom} show the 
    corresponding final 
    estimates of the three ground truth binary masks based on \initialII\space with 
    coordinate-ascent and $X_{01}$  (results for \initialI\ are similar). 
    Figures \protect\subref{fig:Qout1random} -\protect\subref{fig:Qout3random} the final estimated 
    probability matrices of the three ground truth binary masks based 
    on \initialI\space with variational Bayes and $X_{01}$. 
}
\label{fig:fig3}
\end{minipage}
\end{figure*}
\subsection{Synthetic data from \initialI }\label{411} 
Here, we generate synthetic datasets with different numbers of labels and 
individuals, 
setting $K=2, 5, 10$, and $M=10, 20, 40$, resulting in 9 combinations.
Figure~\ref{fig:fig1} shows results by applying \initialI\ (the true model), and 
\initialII\space (our proposed model) to the synthetic data. Both these 
models are fit using coordinate-ascent. We also fit \initialII\ using our 
proposed variational Bayes algorithm. In the figure, we report 
misclassification rates, viz. the proportion of labels in the true $X$ 
incorrectly labeled under the estimated $X$. We see that our model with variational Bayes outperforms 
other competitors, even under model-misspecification.  Using variational 
Bayes offers a significant 
improvement in performance over coordinate-ascent, with almost no 
additional computational overhead.
\bigbreak

\begin{figure*}
\begin{minipage}{.3\textwidth}
  \centerline{
  \subfloat[]{
  \includegraphics[width=0.8\textwidth]{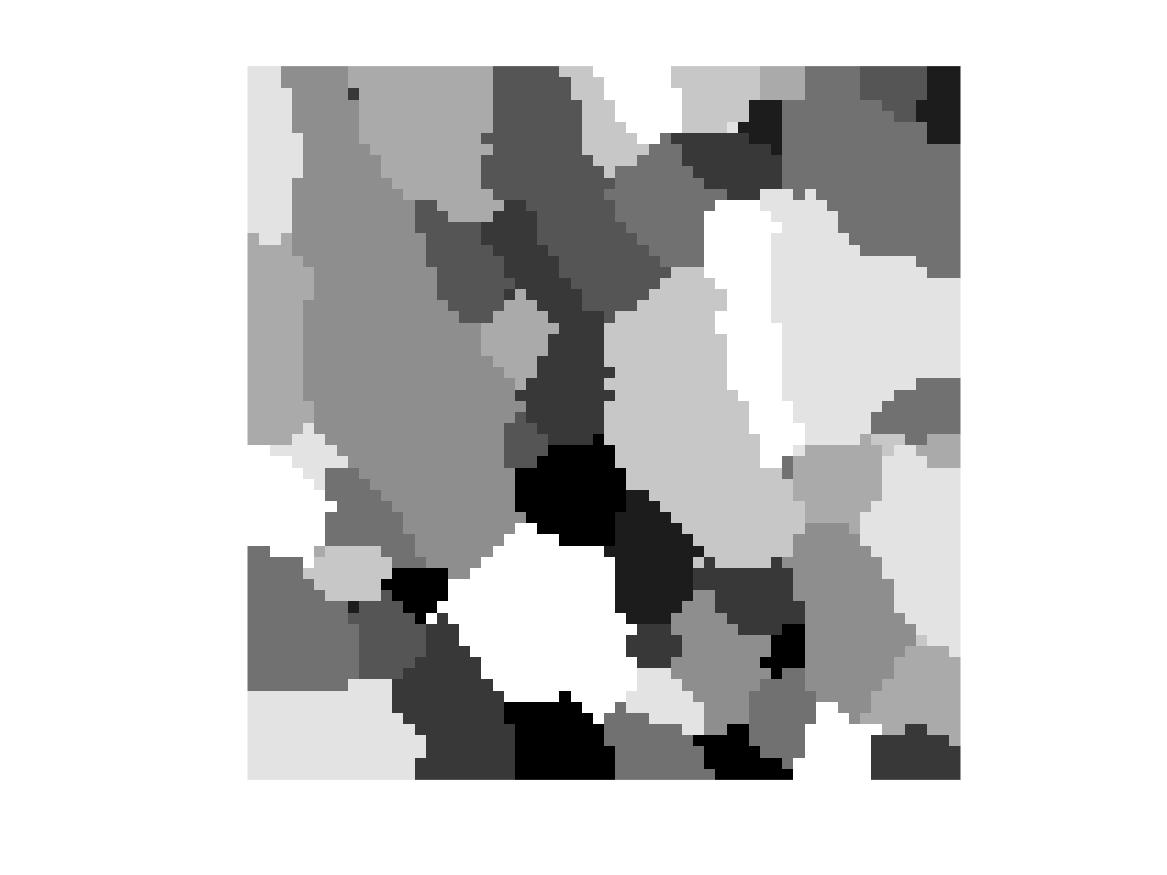}
  \label{fig:Ground truth group-representative map X}
  }}
\end{minipage}
\begin{minipage}{.5\textwidth}
  \caption{\protect\subref{fig:Ground truth group-representative map X} the true group-representative map $X$. 
    The leftmost column in the two rows below show two different 
    initializations, random and greedy. The remaining columns show corresponding estimates 
    of the group-representative map produced by (from left to right) 
\initialI\space with coordinate-ascent, \initialII\space with coordinate-ascent, and 
 \initialII\space with variational Bayes.  
} 
\label{fig:fig4}
\end{minipage}
\begin{minipage}{.98\textwidth}
  \centerline{
  \subfloat[]{
  \includegraphics[width=0.23\textwidth]{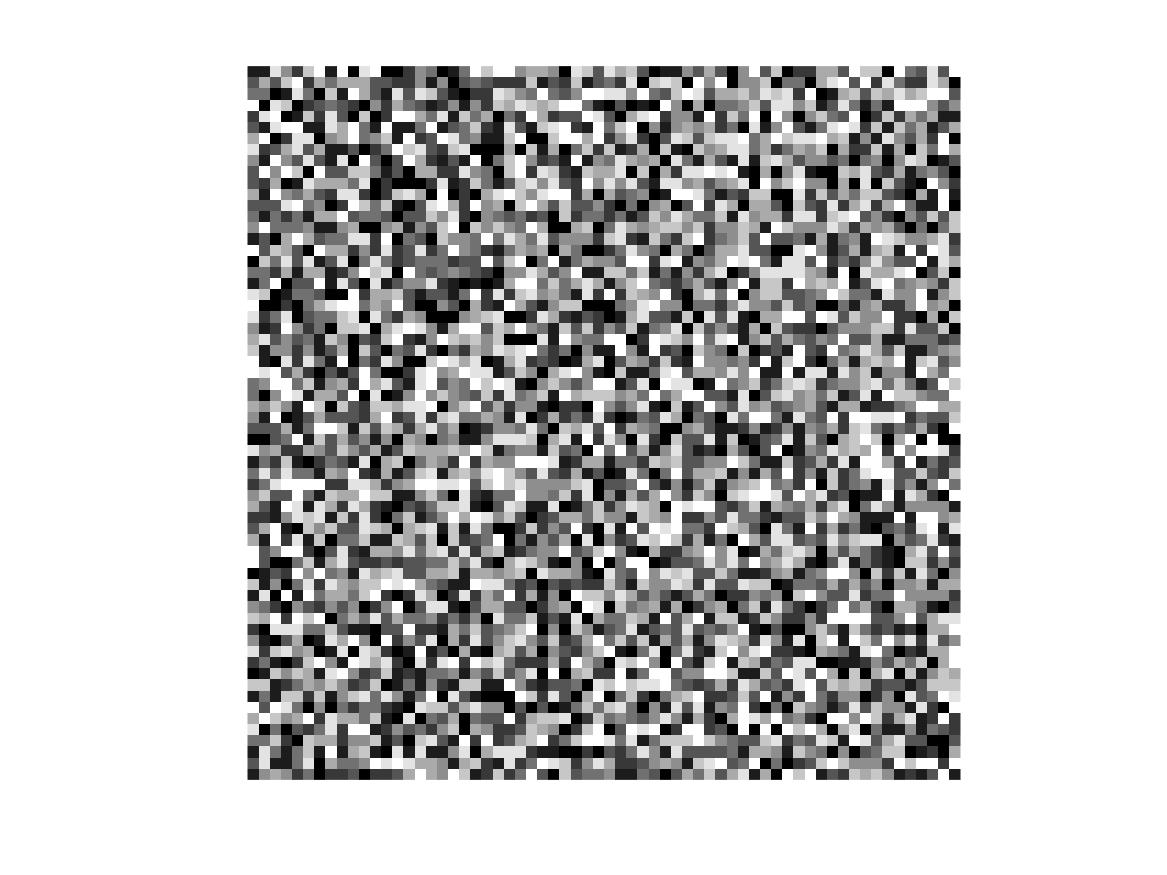}
  \label{fig:First Initial I Estimate X01}
  }   
  \subfloat[]{
  \includegraphics[width=0.23\textwidth]{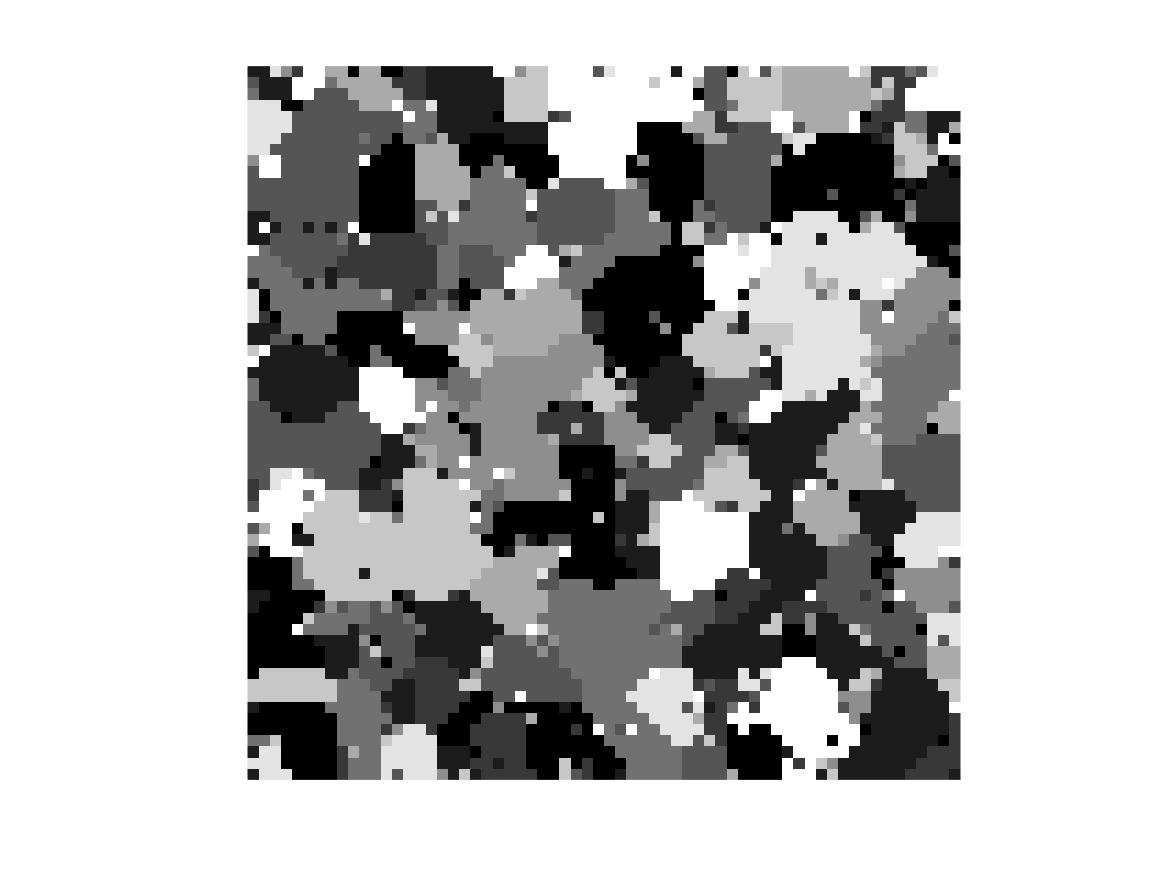}
  \label{fig:First Final I Estimate X1}
  }
  \subfloat[]{
  \includegraphics[width=0.23\textwidth]{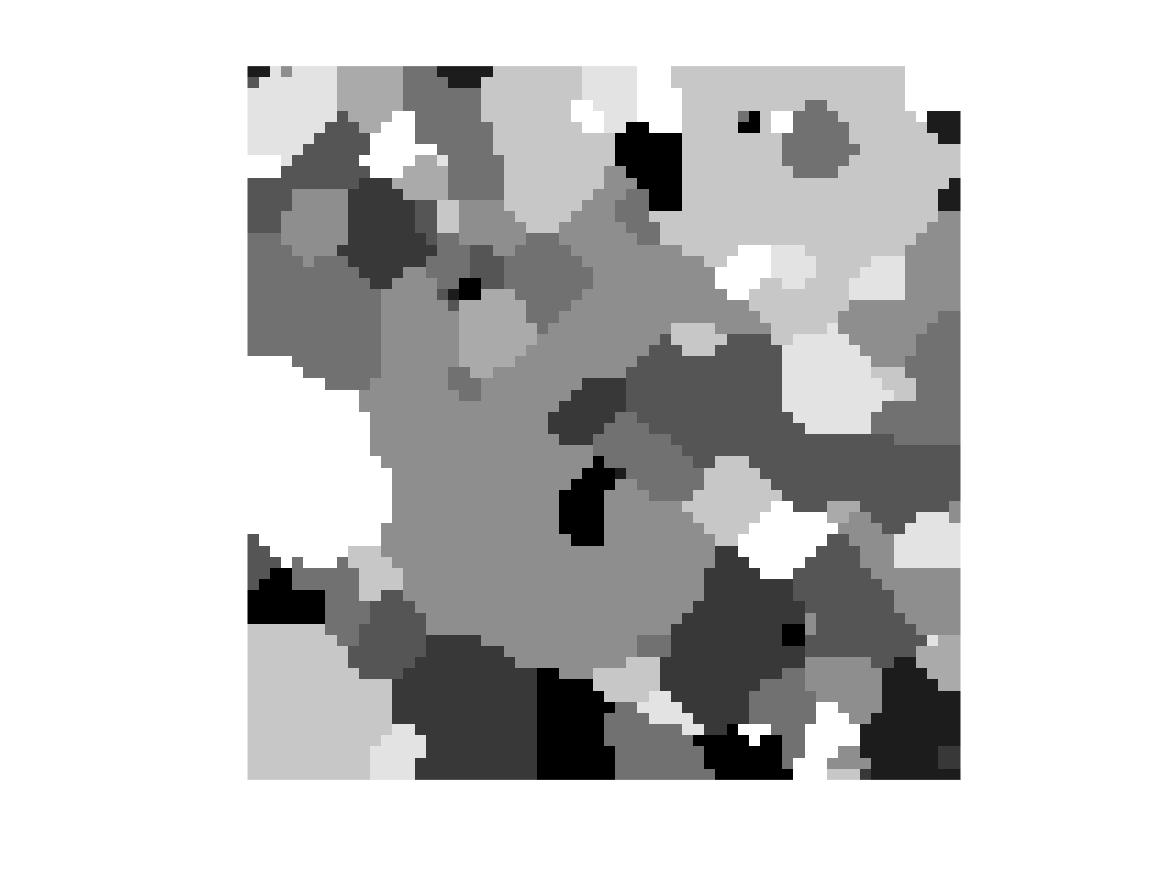}
  \label{fig:First Final II Estimate X2}
  }
  \subfloat[]{
  \includegraphics[width=0.23\textwidth]{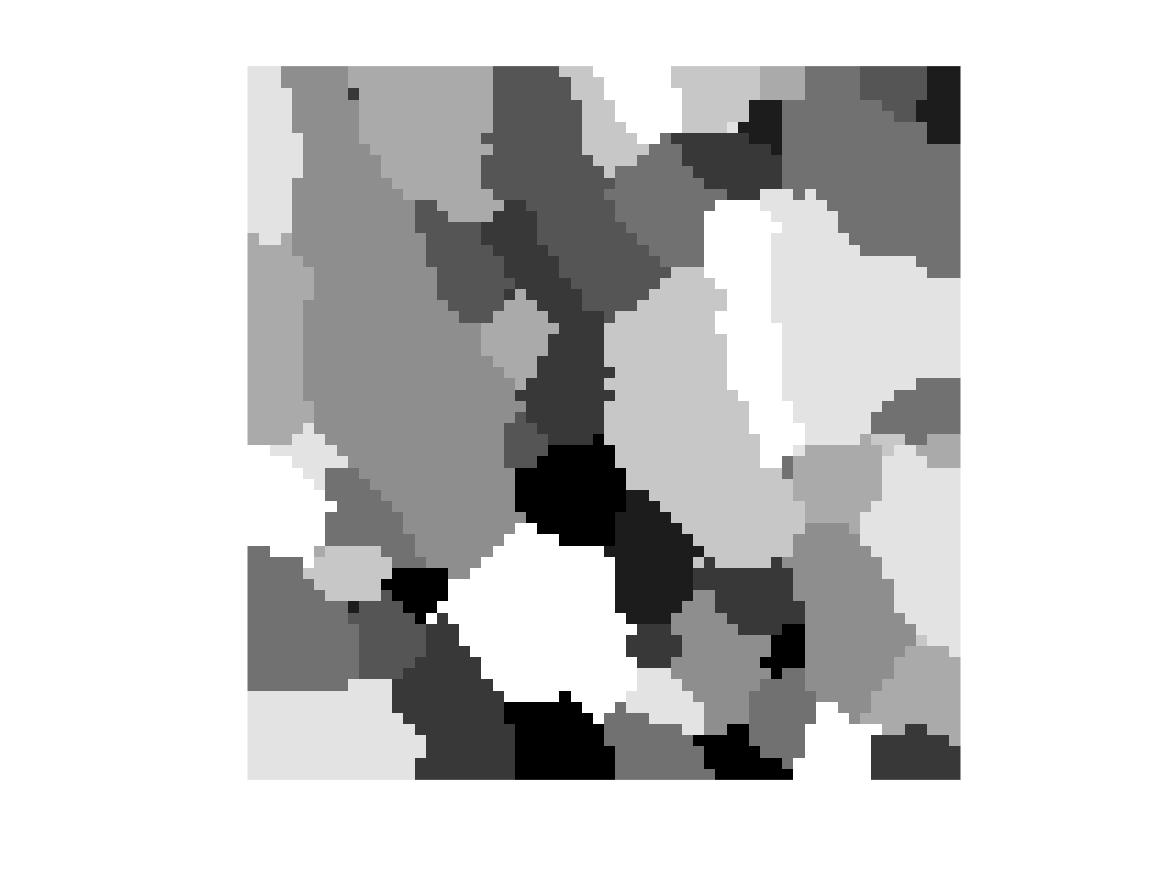}
  \label{fig:First Final VEM Estimate X3}
  }}  
  \centerline{
   \subfloat[]{
  \includegraphics[width=0.23\textwidth]{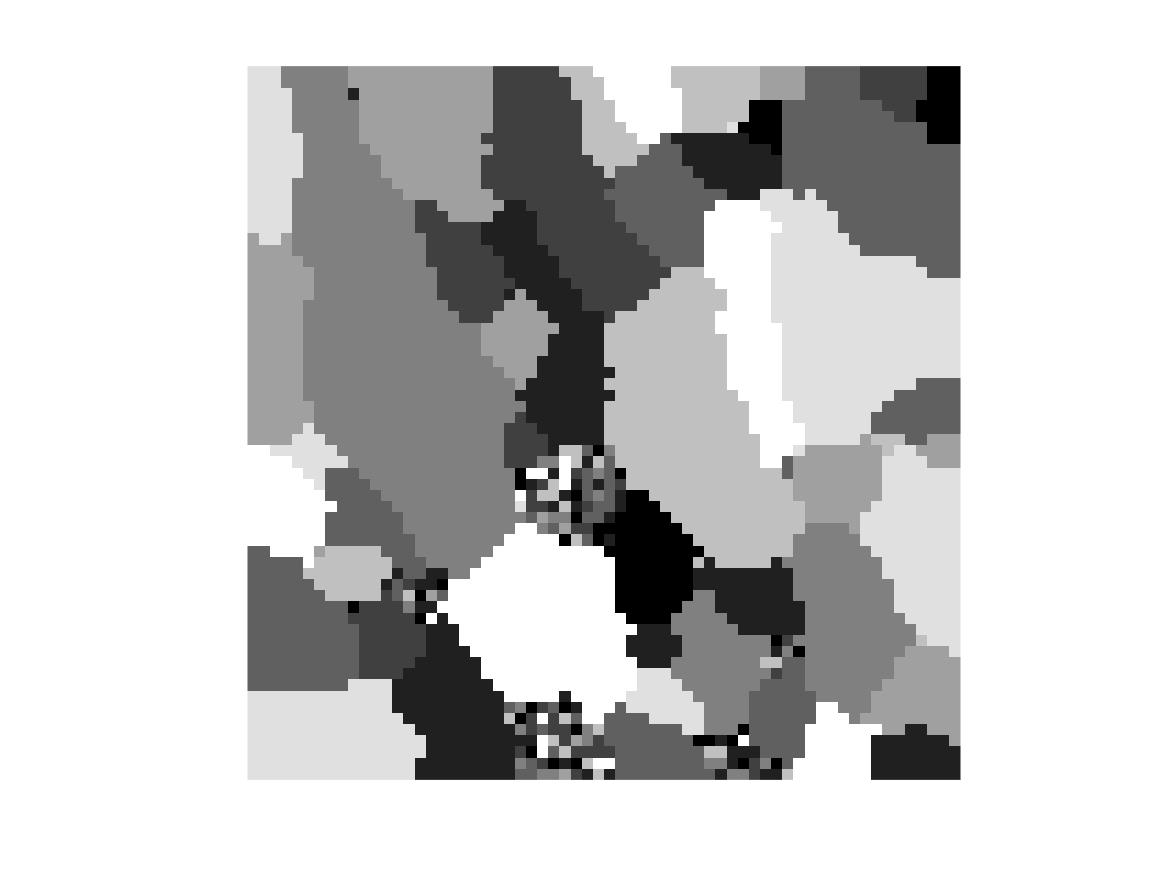}
  \label{fig:Second Initial I Estimate X02}
  }   
  \subfloat[]{
  \includegraphics[width=0.23\textwidth]{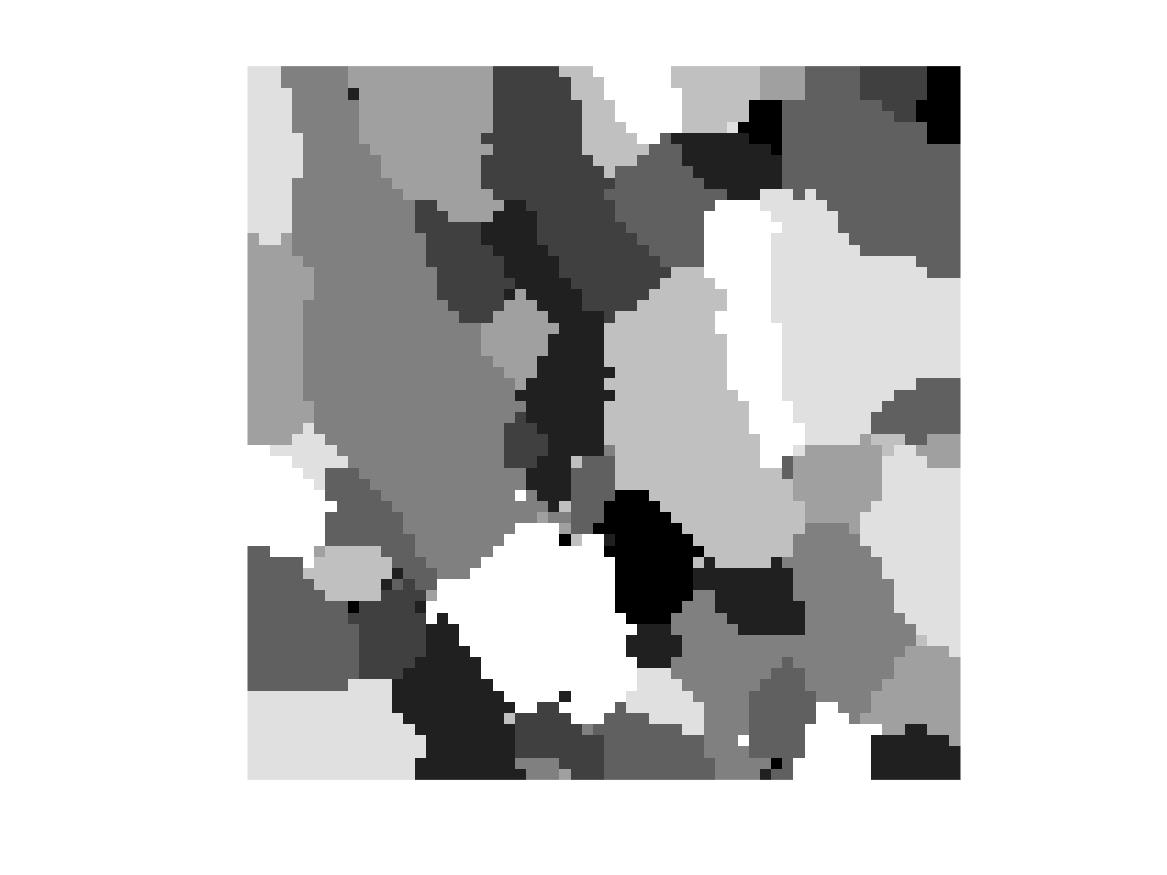}
  \label{fig:Second Final I Estimate X4}
  }
  \subfloat[]{
  \includegraphics[width=0.23\textwidth]{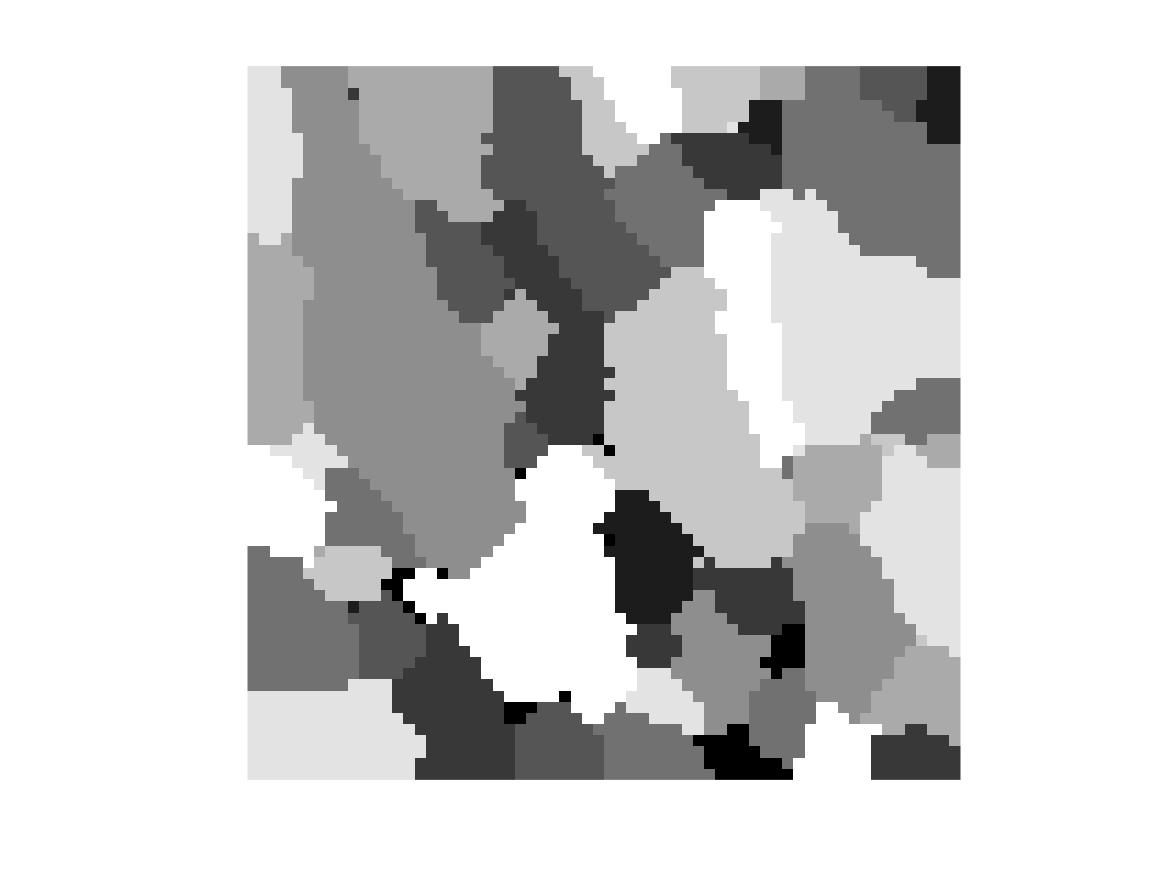}
  \label{fig:Second Final II Estimate X5}
  }
  \subfloat[]{
  \includegraphics[width=0.23\textwidth]{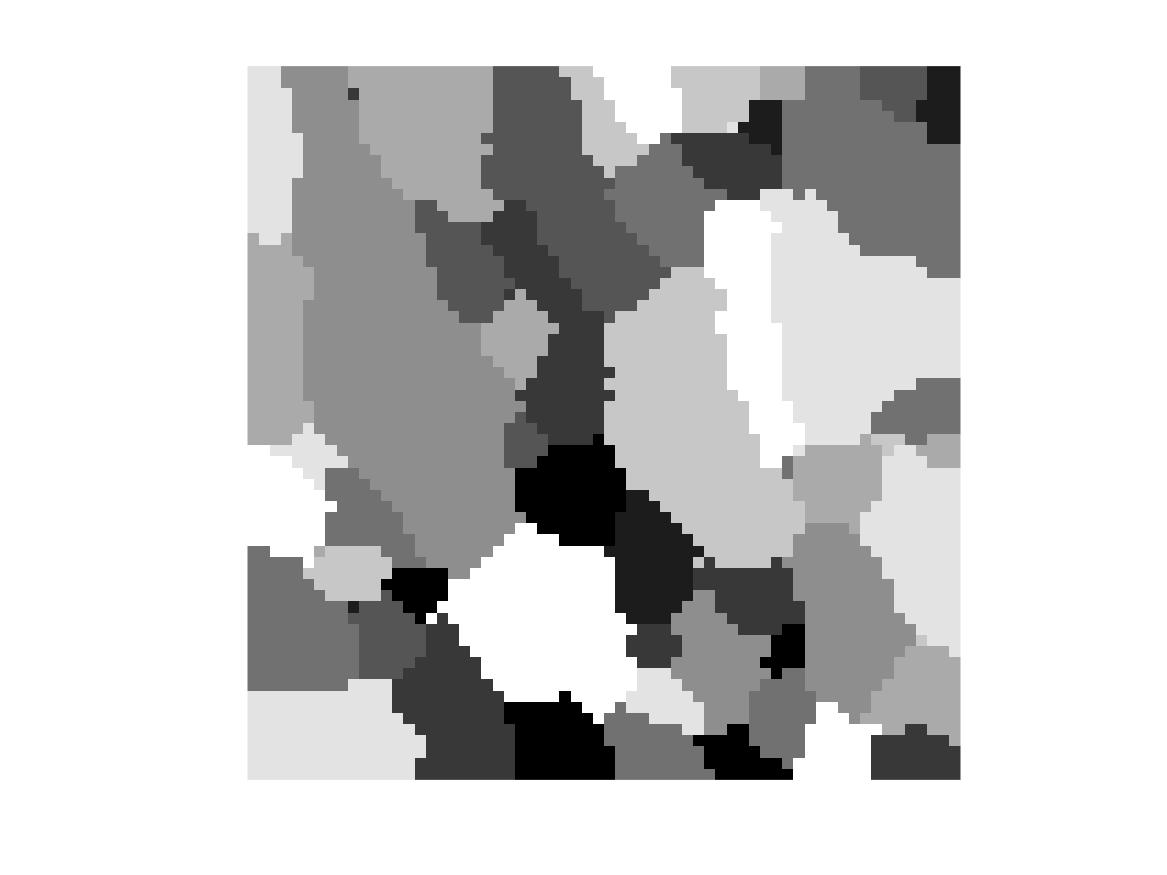}
  \label{fig:Second Final VEM Estimate X6}
  }}
\end{minipage}
\end{figure*}

To better understand the role variational Bayes plays, and the source 
of the improved performance, figures 
\protect\ref{fig:fig2}-\protect\ref{fig:fig4} present the results 
for the experiment with $K=10, M=40$ by implementing \initialI\space with 
coordinate-ascent, \initialII\space with coordinate-ascent and 
\initialII\space with variational Bayes. In the interest of space, 
we show results for subjects 1, 2 and 3 (whose subject maps $Y_1$, $Y_2$ 
and $Y_3$ are shown in figure~\ref{fig:fig2}).  
Figure~\ref{fig:fig3} shows the true and estimated binary masks for these 
three subjects for \initialI, as well as \initialII\space with variational 
Bayes. We see that the latter accurately recovers the truth, to which 
the latter bears little resemblance.
Figure~\ref{fig:fig4} shows how sensitivity to initialization is an 
important factor at play. It compares the true group-representative map 
$X$ to the estimated ones for the three schemes for two different 
initializations, random and greedy. In the latter, if a voxel is on for 
any of the subject maps, the corresponding voxel in the group-representative 
map is set to one.
We see that \initialII\ with variational Bayes is relatively insensitive 
to initialization, accurately recovering the true map in both cases.
In this example (though not always), the other methods do well for 
greedy initialization but poorly for random.


\subsection{Synthetic data from \initialII }\label{412} 
\begin{figure*}
   
  \centerline{
  \subfloat[]{
  \includegraphics[width=0.34\textwidth]{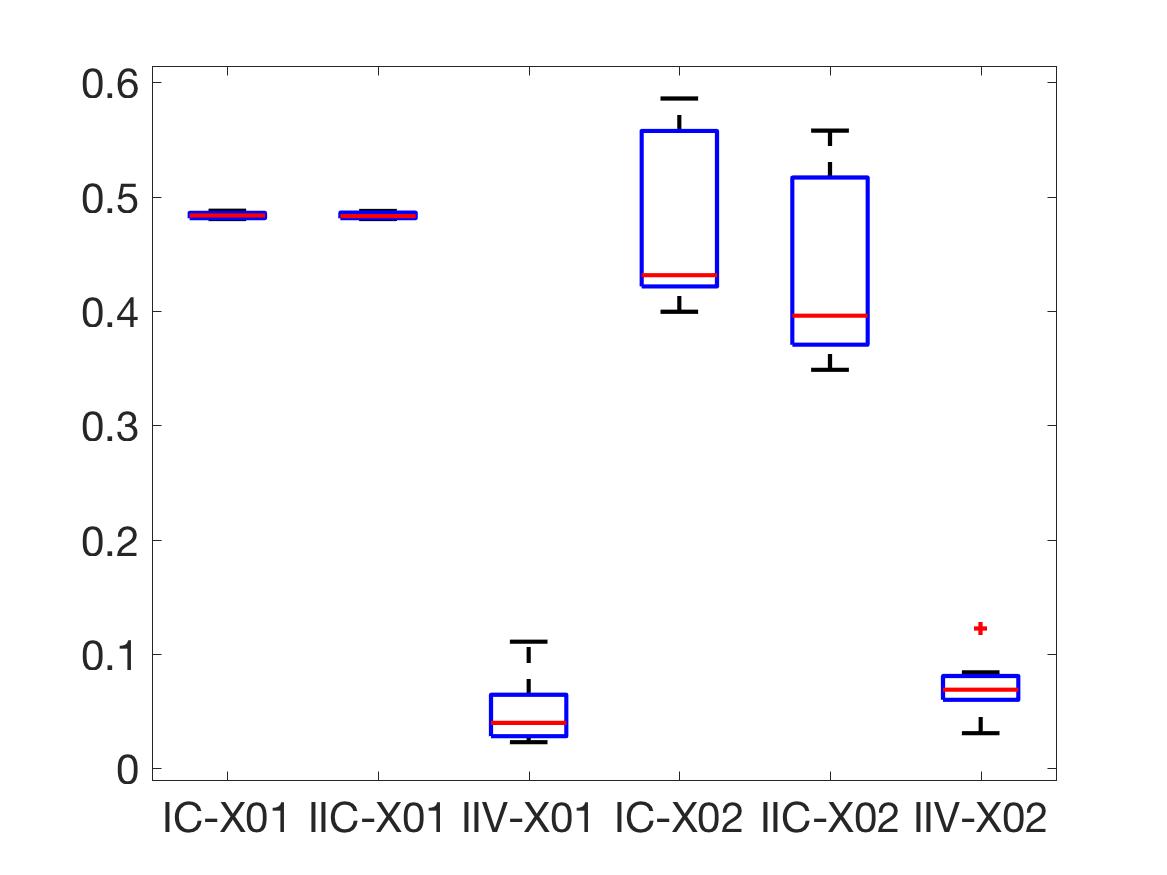}
  \label{fig:VM10K2}
  }   
  \subfloat[]{
  \includegraphics[width=0.34\textwidth]{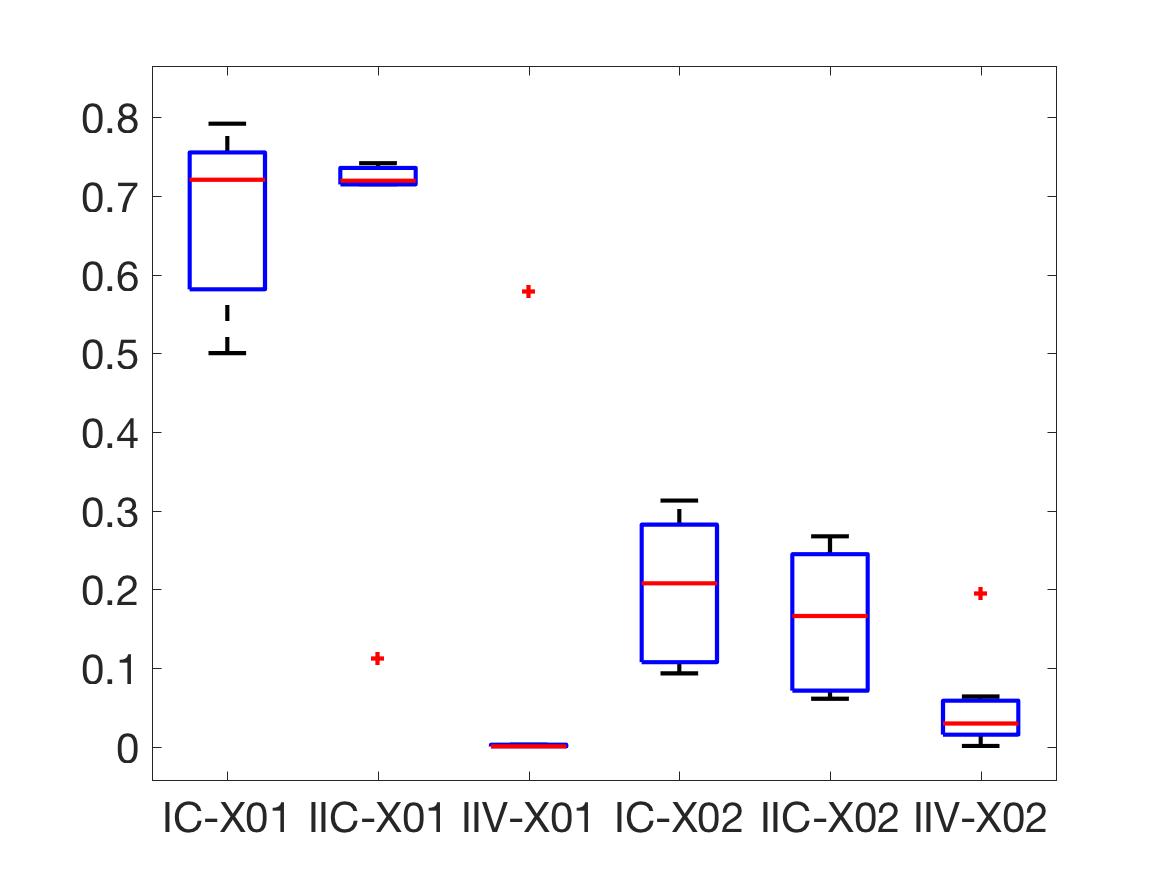}
  \label{fig:VM10K5}
  }
  \subfloat[]{
  \includegraphics[width=0.34\textwidth]{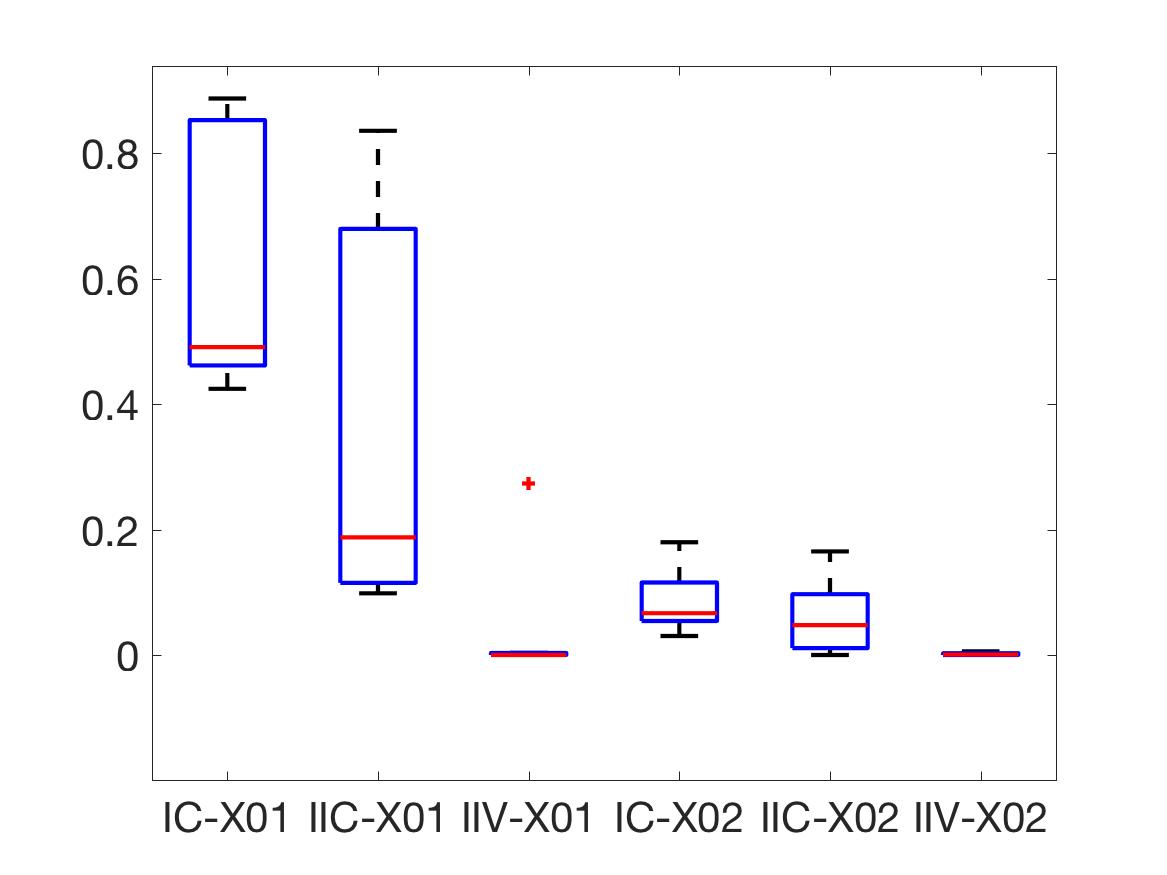}
  \label{fig:VM10K10}
  }
  }
 
  \centerline{
  \subfloat[]{
  \includegraphics[width=0.34\textwidth]{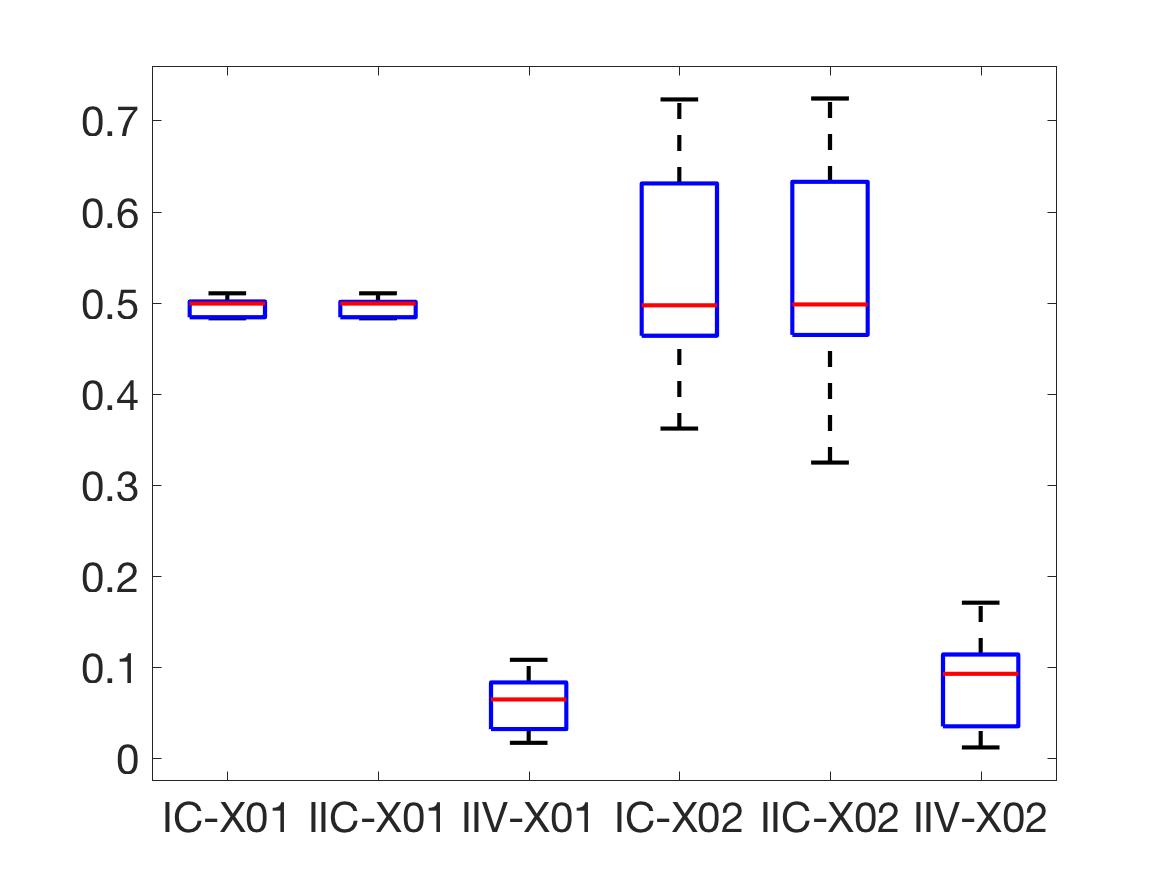}
  \label{fig:VM20K2}
  }   
  \subfloat[]{
  \includegraphics[width=0.34\textwidth]{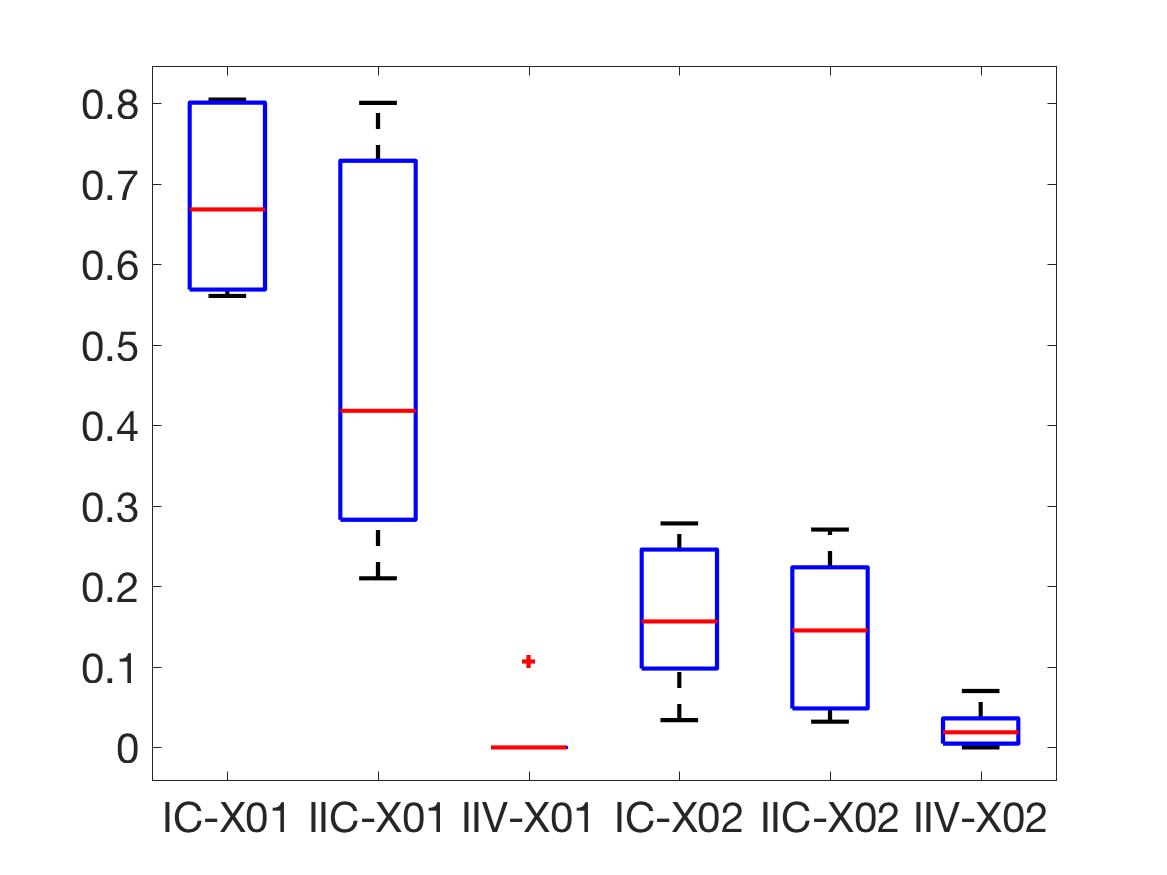}
  \label{fig:VM20K5}
  }
  \subfloat[]{
  \includegraphics[width=0.34\textwidth]{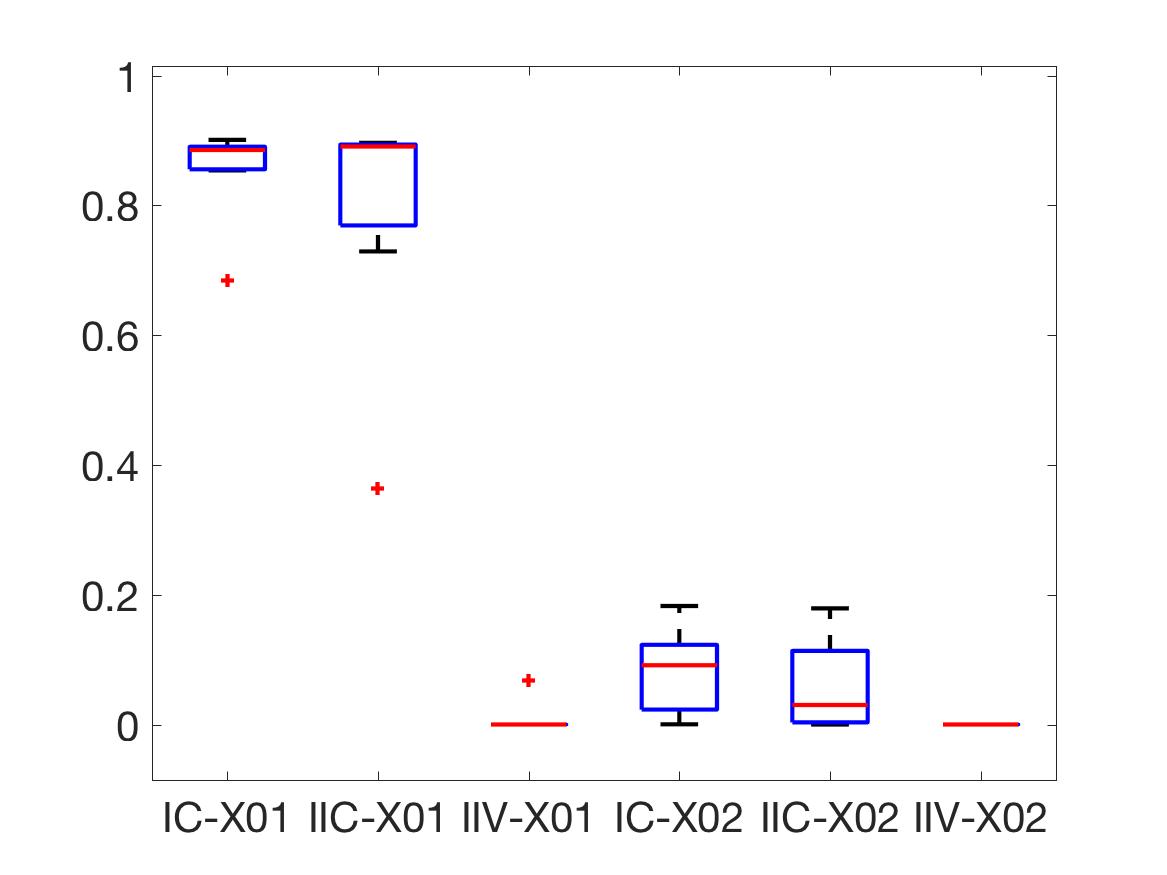}
  \label{fig:VM20K10}
  }
  }
  
  \centerline{
  \subfloat[]{
  \includegraphics[width=0.34\textwidth]{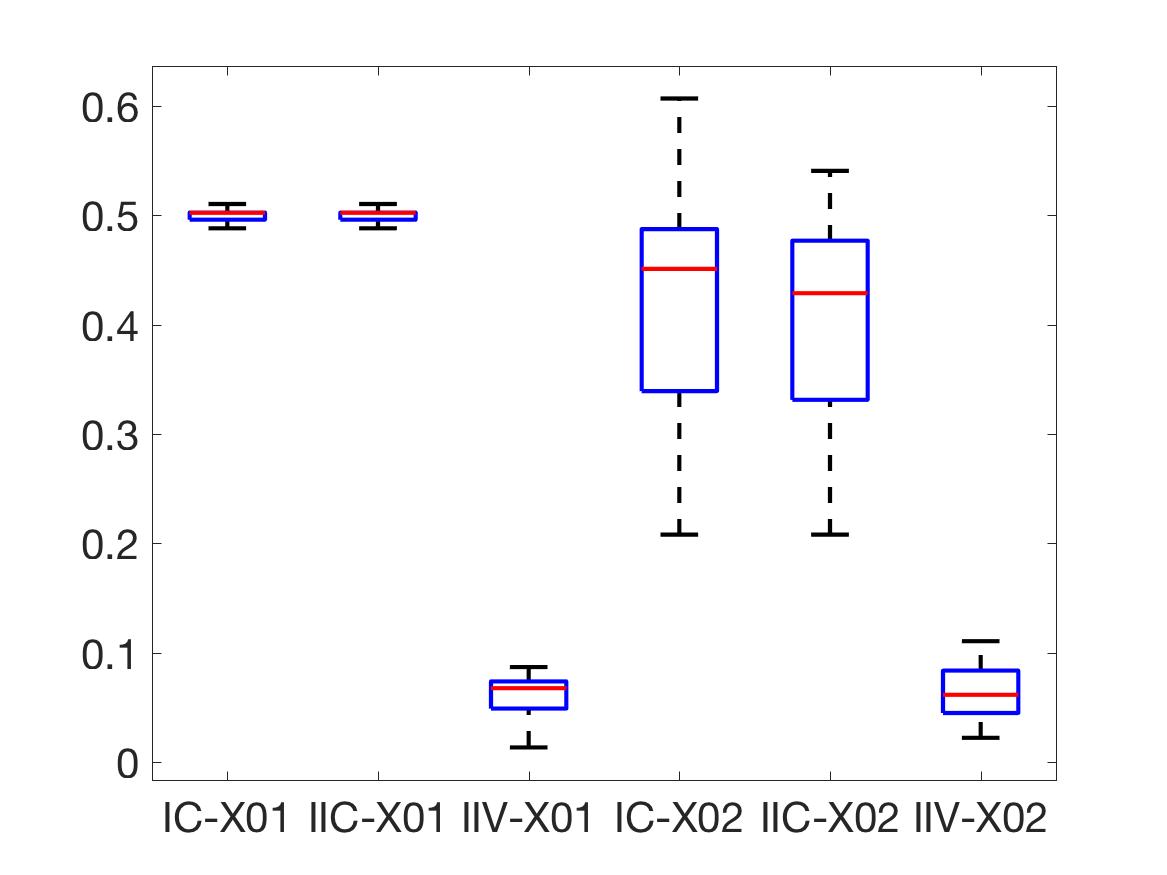}
  \label{fig:VM40K2}
  }   
  \subfloat[]{
  \includegraphics[width=0.34\textwidth]{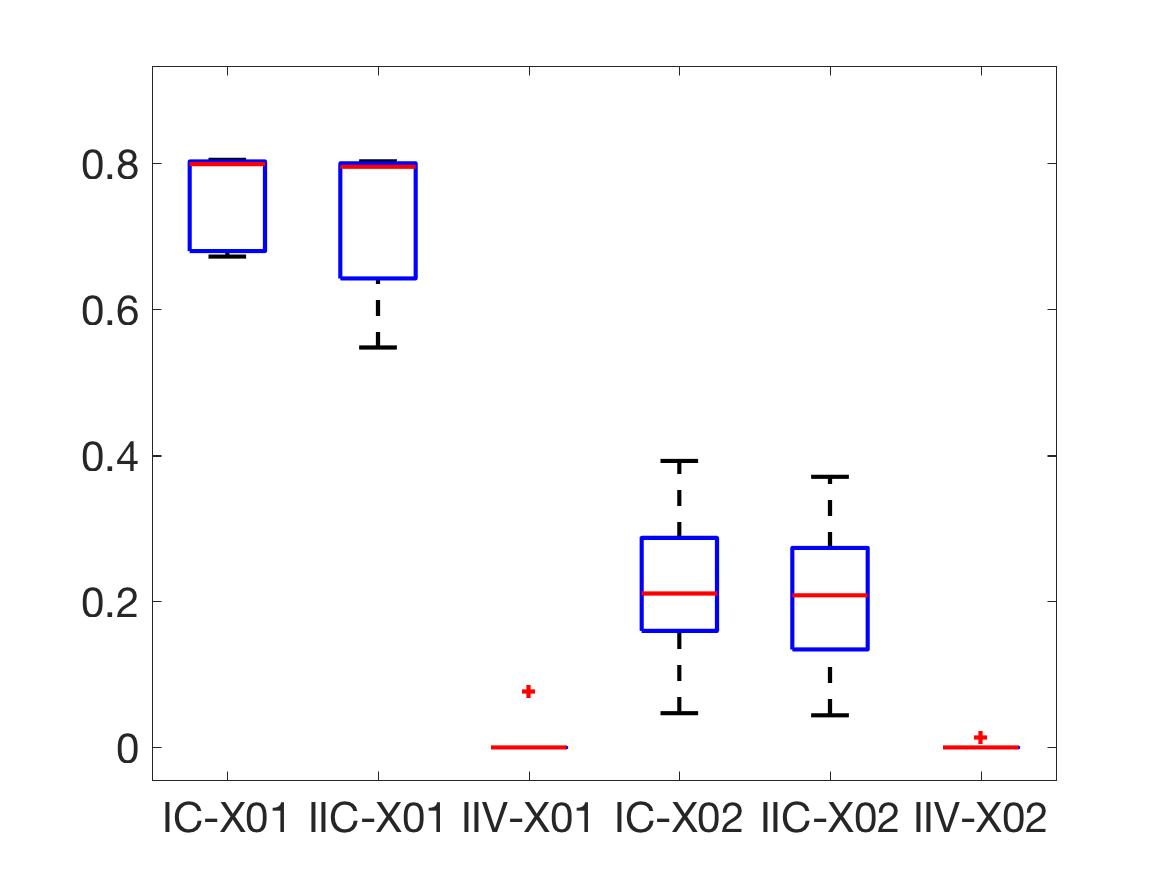}
  \label{fig:VM40K5}
  }
  \subfloat[]{
  \includegraphics[width=0.34\textwidth]{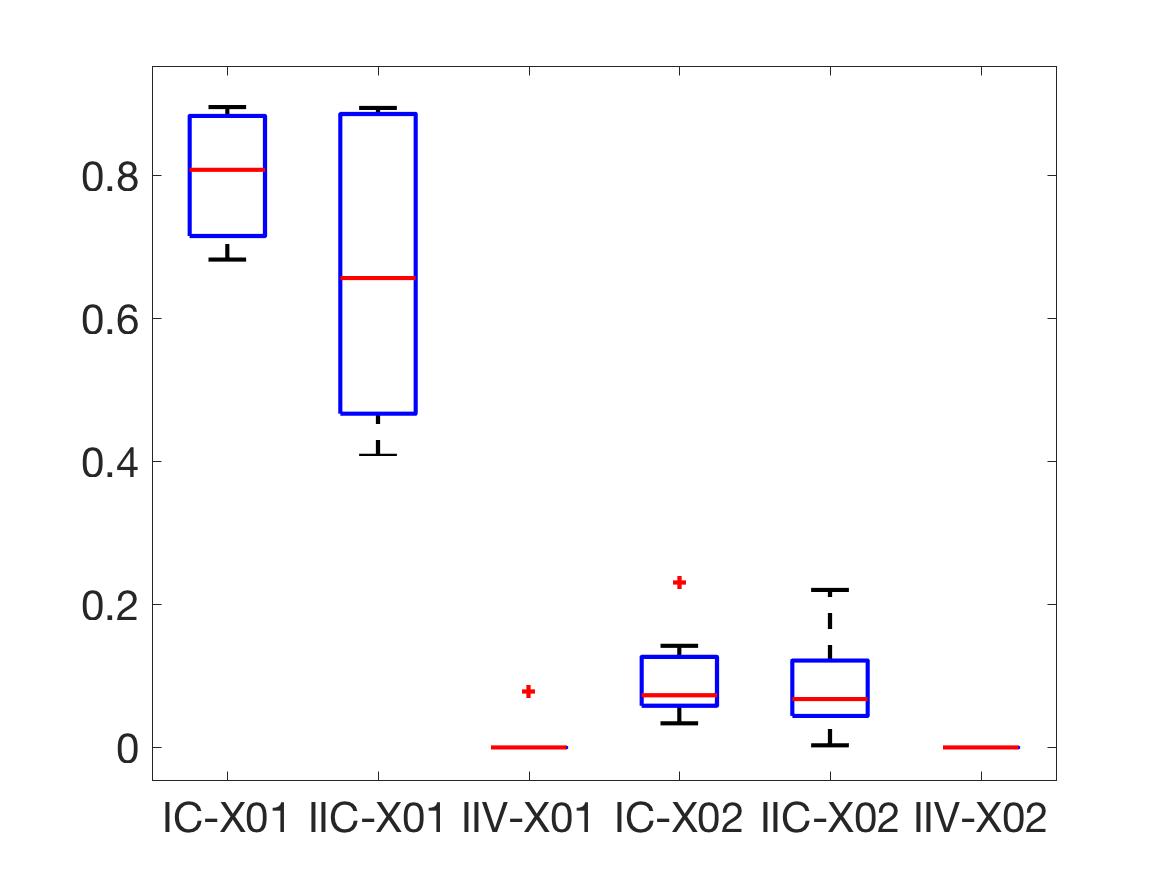}
  \label{fig:VM40K10}
  }
  }

  \caption{Results for datasets generated from forward model \initialII. 
    Columns correspond to different settings of $K$ (2, 5 and 10 
    respectively).  
    Rows correspond to different settings of $M$ (10, 20 and 40 
    respectively).  
    Boxplots within each subplot represent misclassification rates for 
    10 repeats for (from left to right): \initialI\ with coordinate 
    ascent ($IC$), \initialII\space with coordinate ascent ($IIC$), 
    \initialII\space with variational Bayes ($IIV$). These are repeated 
    twice, with random initializations ($X_{01})$, and with a greedy 
    initialization that
    contains all the nonzero components from the $M$ images
    ($X_{02}$). 
}
\label{fig:fig5}
\end{figure*}

In this experiment, we repeat the evaluation from the previous section, 
now using data with measurement noise $Z$ (see Table~\ref{table:1}). 
In generating this noise, we set the noise probability $\epsilon=0.01 $.
Figure \protect\ref{fig:fig5} shows a quantitative comparison of the 
three schemes, plotting misclassification rates of \initialI\space with 
coordinate-ascent, \initialII\space with coordinate-ascent 
and \initialII\space with variational Bayes. Once again, we consider 
two different initializations (random and greedy) of the 
group-representative map $X$. 
This problem is harder than the earlier one, and unsurprisingly, \initialI\ 
performs worst. However once again for \initialII, using variational 
Bayes results in a markedly improved performance over coordinate-ascent,
showing that even with the addition of measurement noise, coupling 
between $X$ and the $H_i$'s is sufficient to warrant a non-trivial 
algorithm.

\begin{figure*}
\begin{minipage}{.7\textwidth}
  \centerline{
  \subfloat[]{
  \includegraphics[width=0.3\textwidth]{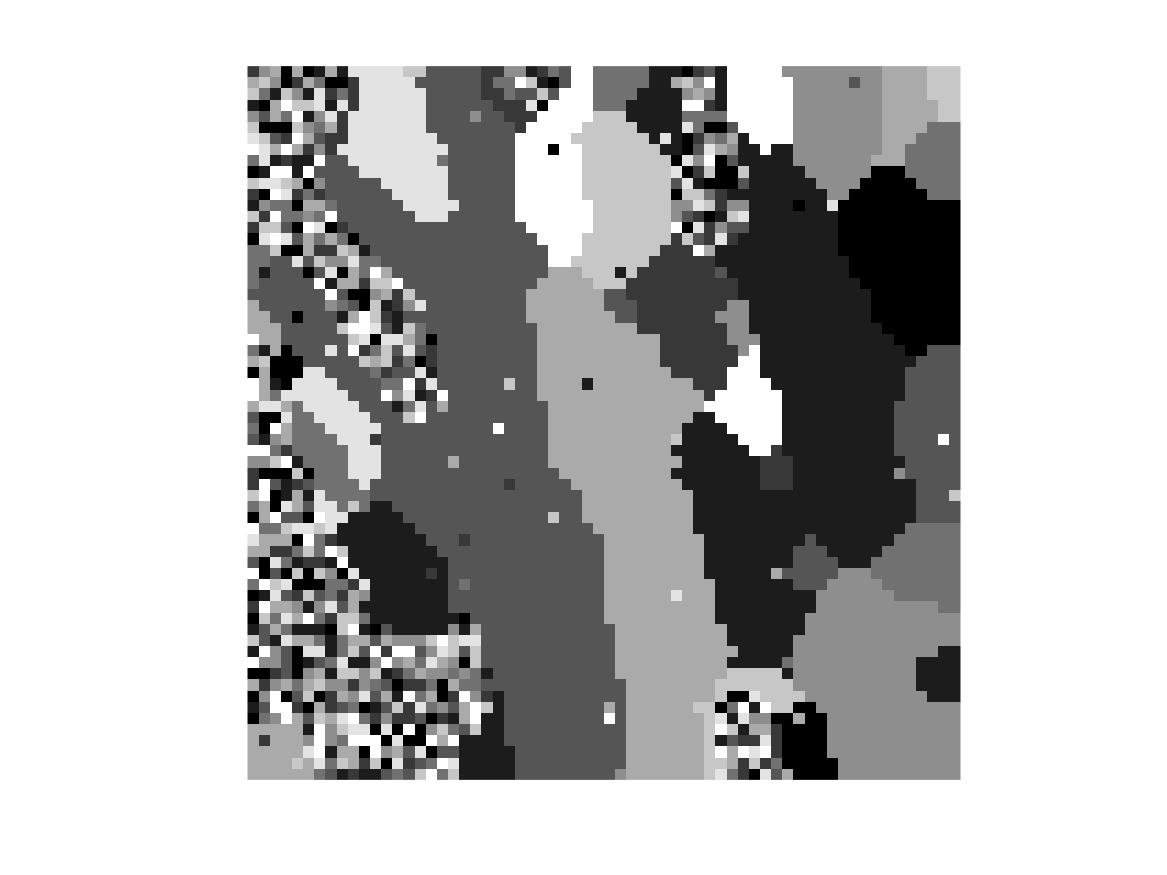}
  \label{fig:Y1}
  }   
  \subfloat[]{
  \includegraphics[width=0.3\textwidth]{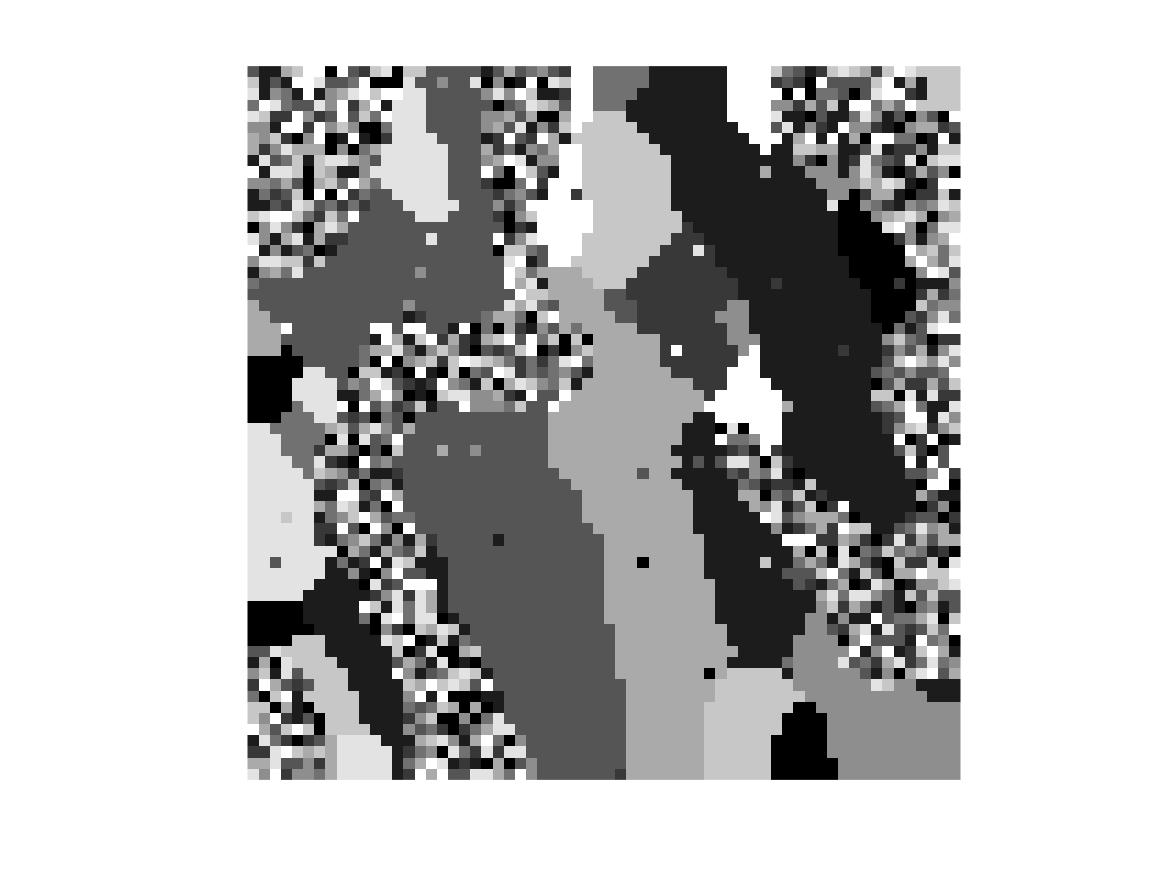}
  \label{fig:Y2}
  }
  \subfloat[]{
  \includegraphics[width=0.3\textwidth]{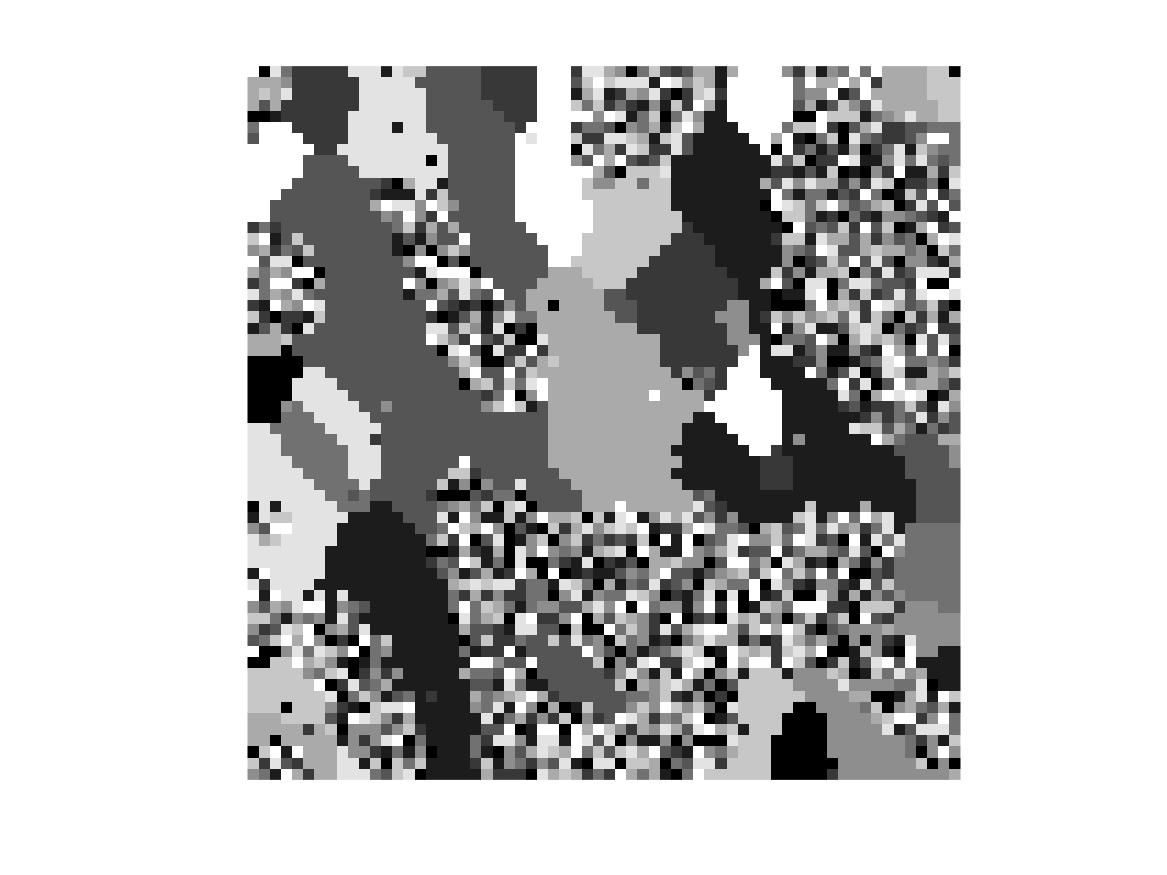}
  \label{fig:Y3}
  }
  }
\end{minipage}
\begin{minipage}{.25\textwidth}
  \caption{Figures \protect\subref{fig:Y1} -\protect\subref{fig:Y3}: 
    Individual subject maps, $Y_1$, $Y_2$, $Y_3$, respectively.
}
\label{fig:fig6}
\end{minipage}

\begin{minipage}{.7\textwidth}
  \centerline{
  \subfloat[]{
  \includegraphics[width=0.3\textwidth]{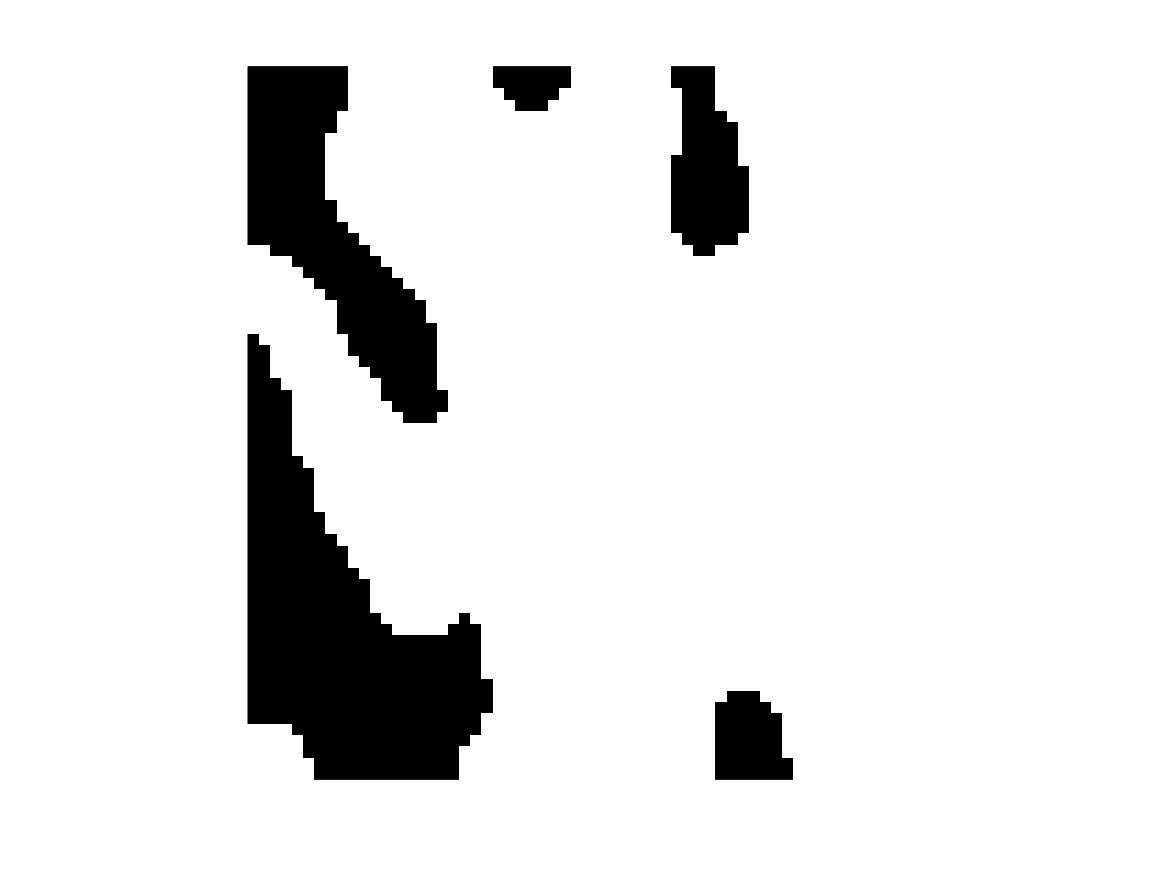}
  \label{fig:H1_gt}
  }   
  \subfloat[]{
  \includegraphics[width=0.3\textwidth]{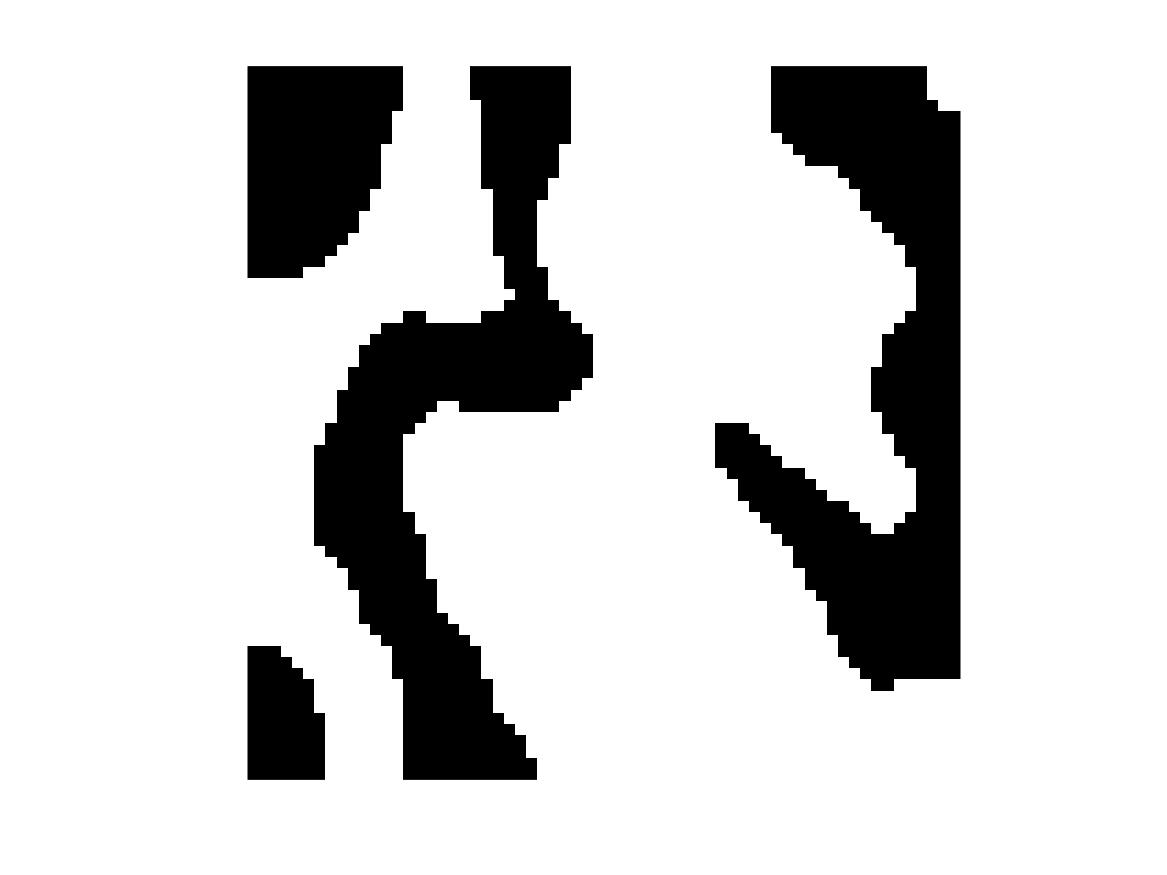}
  \label{fig:H2_gt}
  }
  \subfloat[]{
  \includegraphics[width=0.3\textwidth]{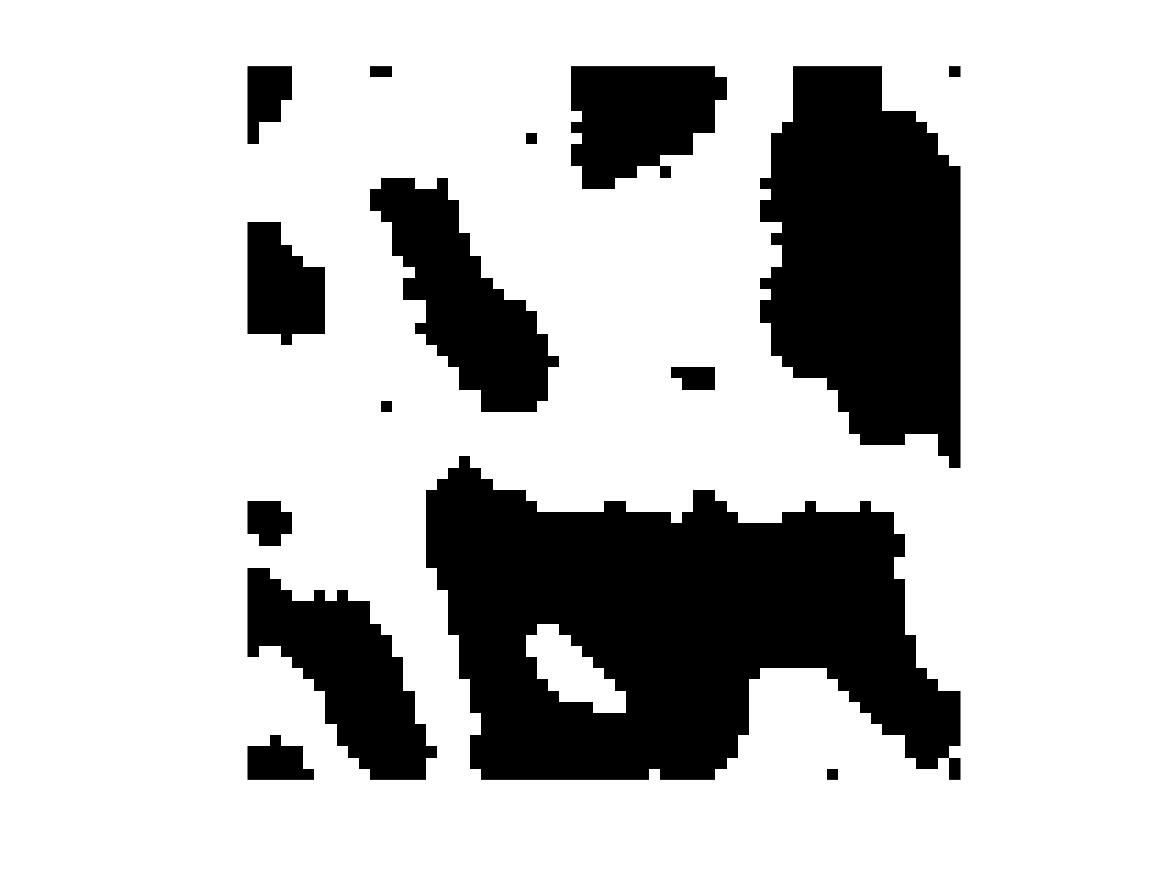}
  \label{fig:H3_gt}
  }
  }
 
  \centerline{
  \subfloat[]{
  \includegraphics[width=0.3\textwidth]{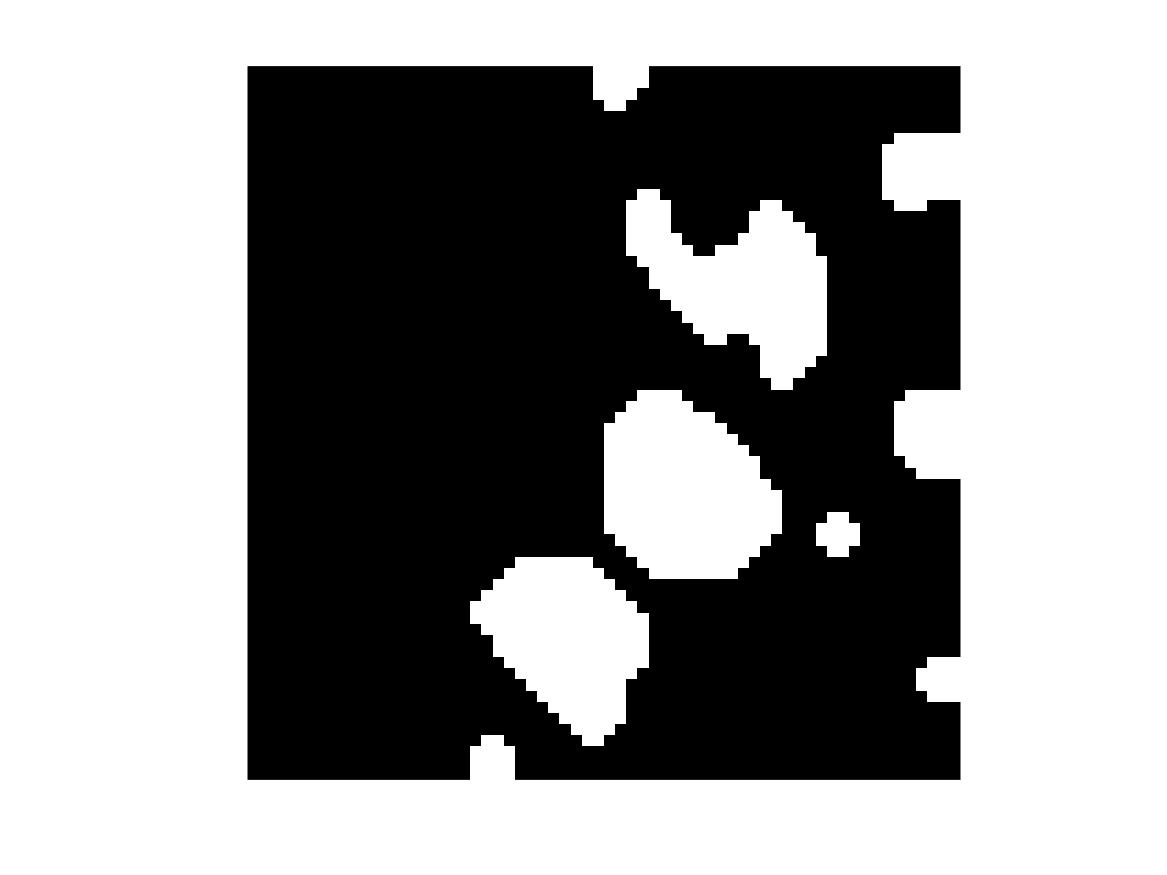}
  \label{fig:Hout1IIrandom}
  }   
  \subfloat[]{
  \includegraphics[width=0.3\textwidth]{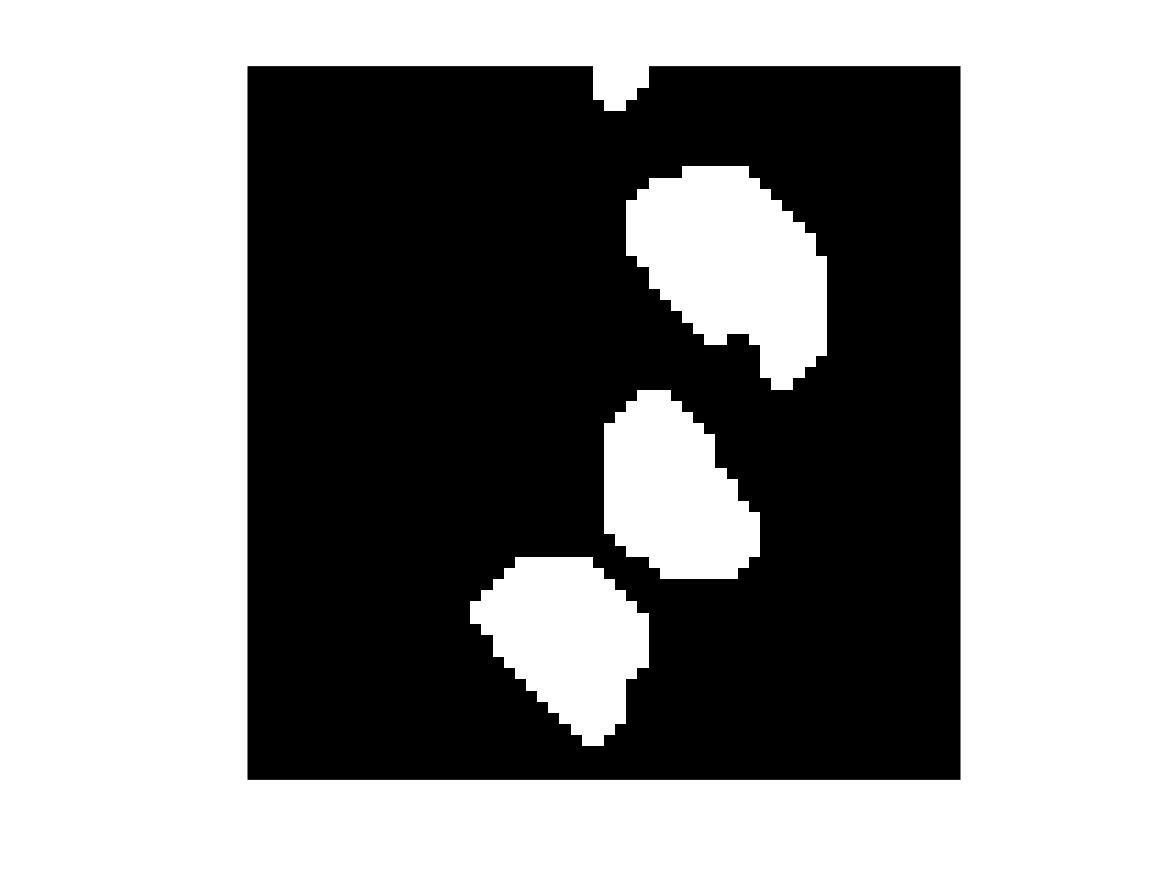}
  \label{fig:Hout2IIrandom}
  }
  \subfloat[]{
  \includegraphics[width=0.3\textwidth]{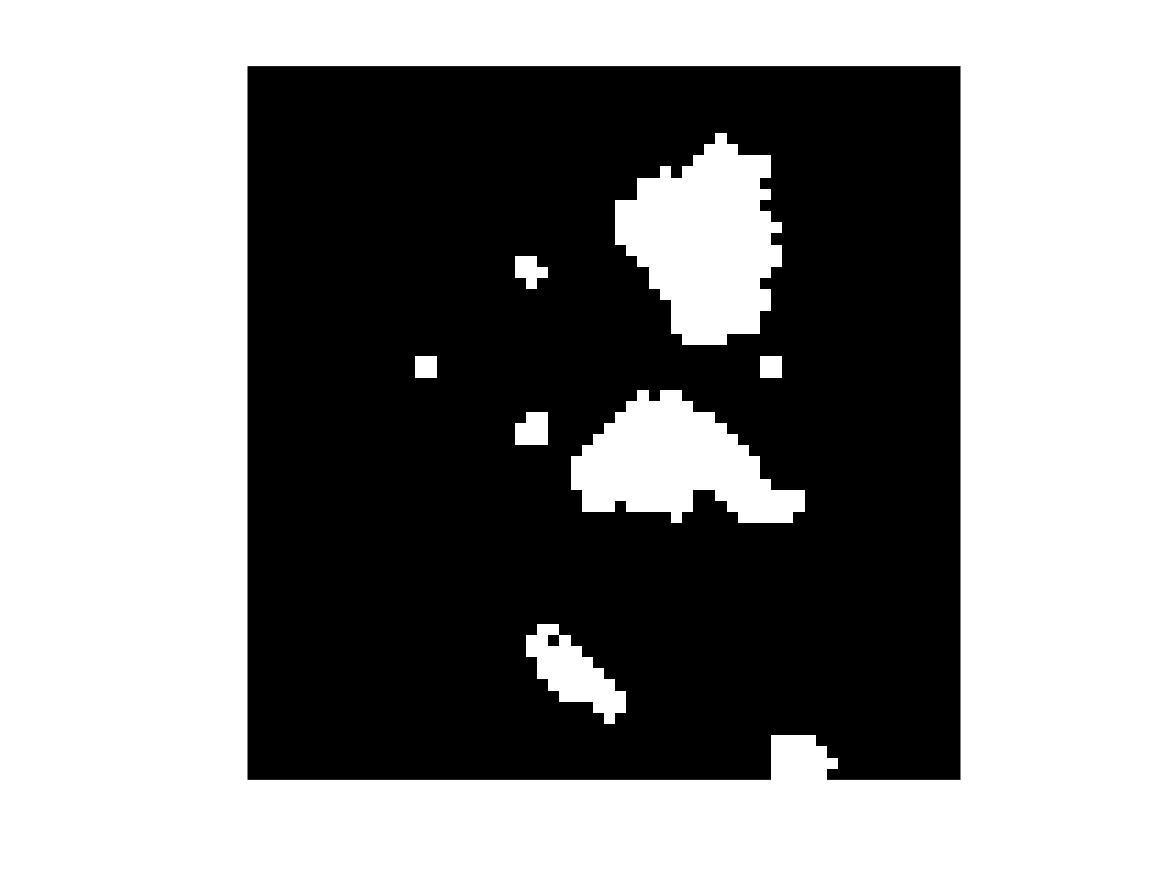}
  \label{fig:Hout3IIrandom}
  }
  }
  \centerline{
  \subfloat[]{
  \includegraphics[width=0.3\textwidth]{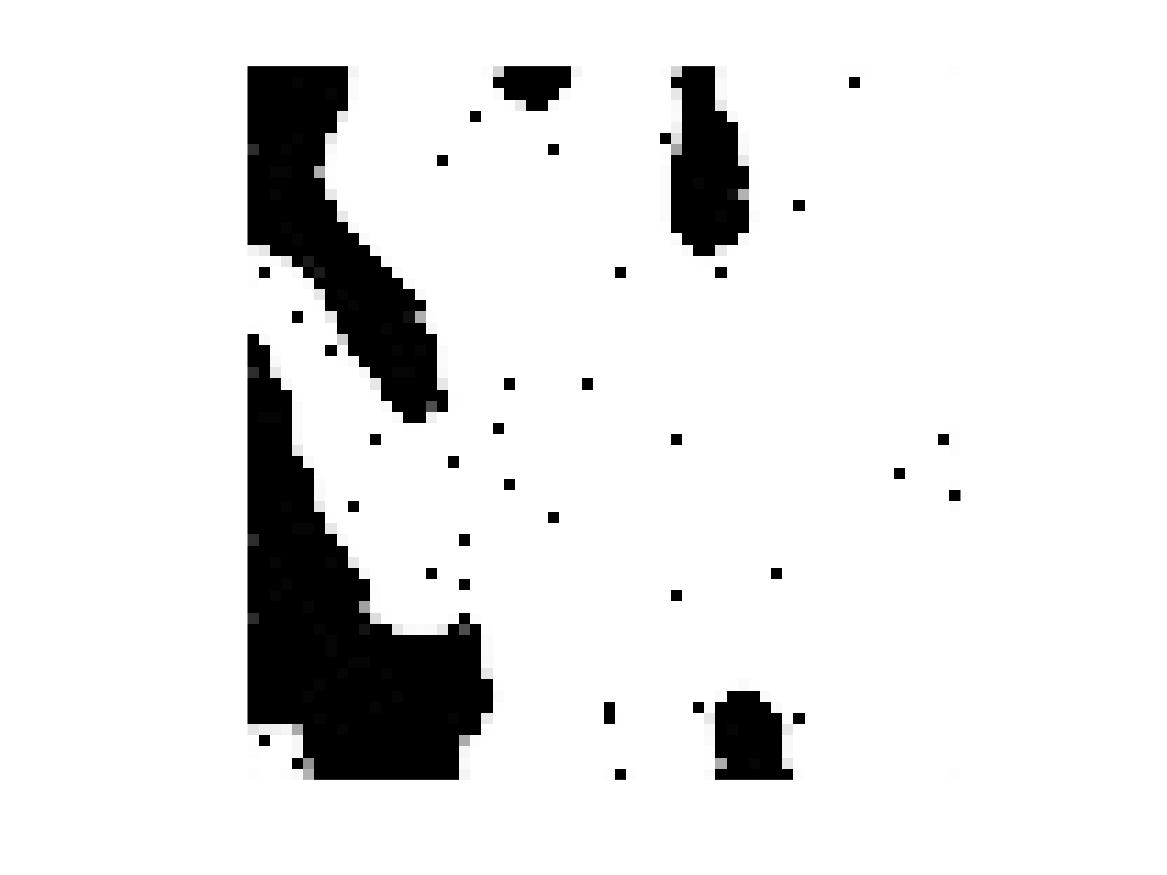}
  \label{fig:Qout1random}
  }   
  \subfloat[]{
  \includegraphics[width=0.3\textwidth]{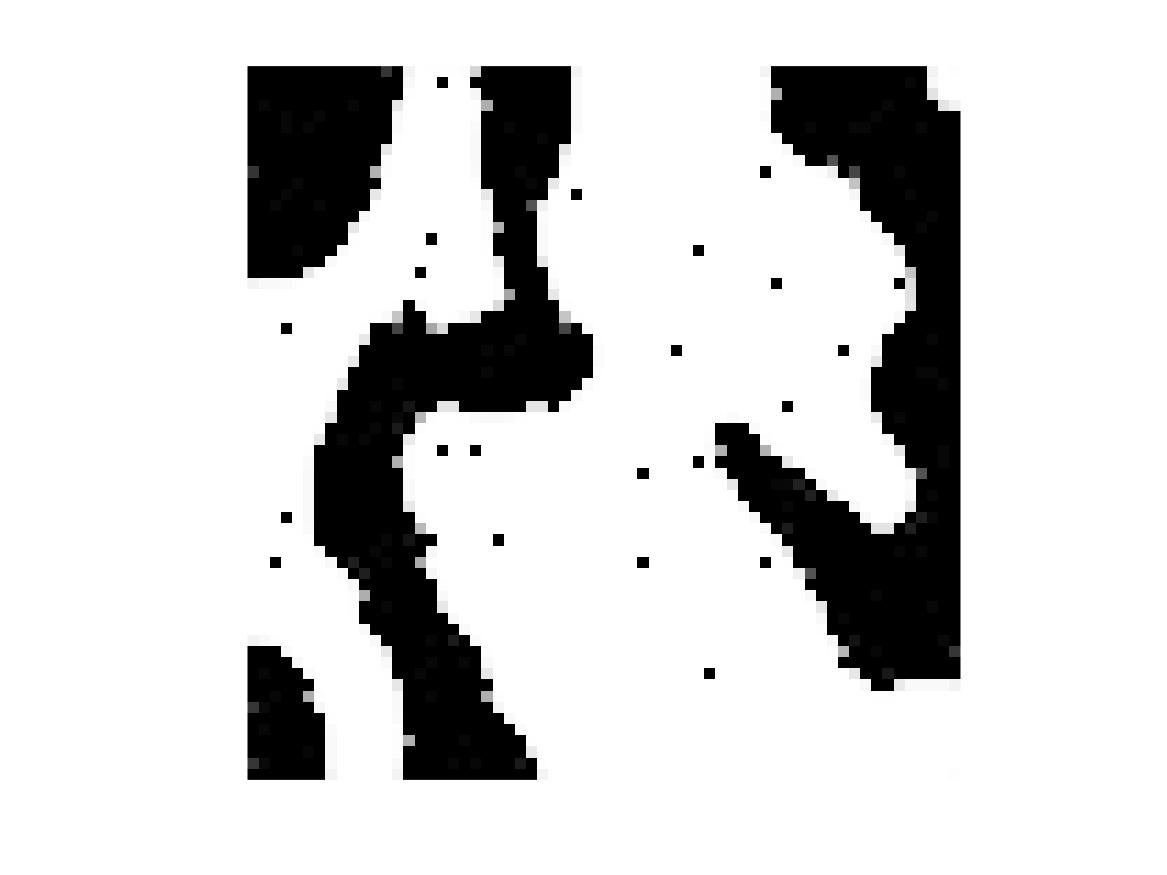}
  \label{fig:Qout2random}
  }
  \subfloat[]{
  \includegraphics[width=0.3\textwidth]{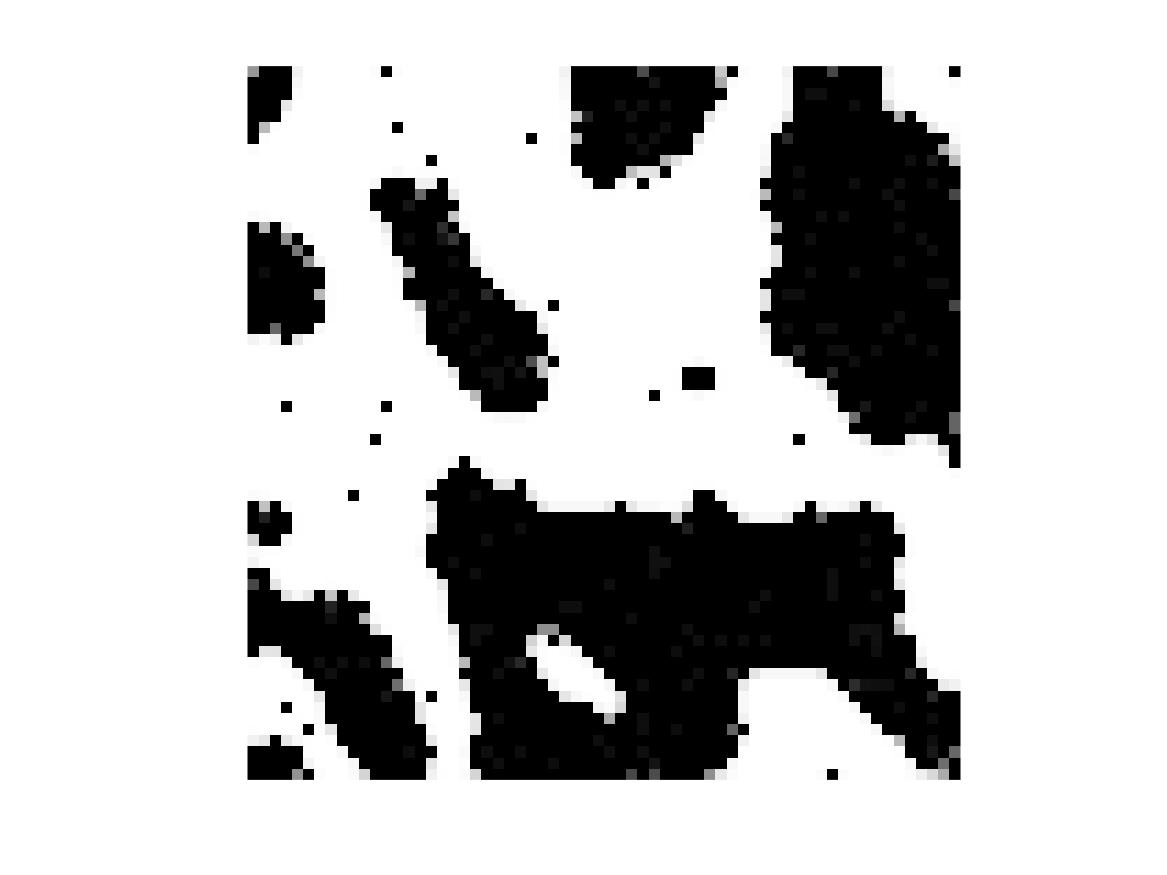}
  \label{fig:Qout3random}
  }
  }
\end{minipage}
\begin{minipage}{.25\textwidth}
  \caption{Figures \protect\subref{fig:H1_gt} -\protect\subref{fig:H3_gt}: 
    the ground truth binary masks, $H_1$, $H_2$, $H_3$, respectively. Figures \protect\subref{fig:Hout1IIrandom} -\protect\subref{fig:Hout3IIrandom} the final estimates of the three ground truth binary masks based on \initialII\space with coordinate ascent and $X_{01}$, $\hat{H}_{II11}$, $\hat{H}_{II21}$, $\hat{H}_{II31}$, respectively. Figures \protect\subref{fig:Qout1random} -\protect\subref{fig:Qout3random} the final estimated probability matrices of the three ground truth binary masks based on \initialII\space with variational Bayes and $X_{01}$, $\hat{q}_{11}$, $\hat{q}_{21}$, $\hat{q}_{31}$, respectively. 
}
\label{fig:fig7}
\end{minipage}
\end{figure*}

\begin{figure*}
\begin{minipage}{.34\textwidth}
  \centerline{
  \subfloat[]{
  \includegraphics[width=0.8\textwidth]{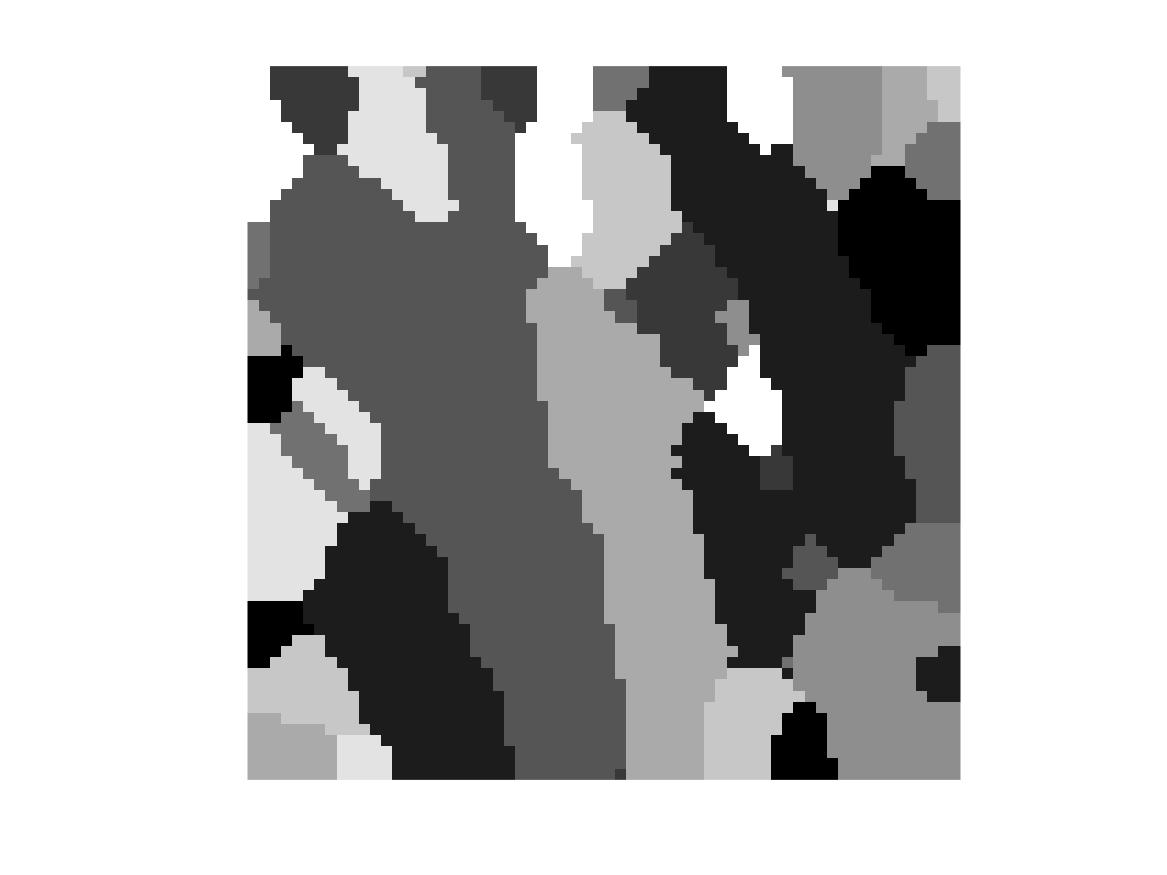}
  \label{fig:Ground truth group-representative map X}
  }}
\end{minipage}
\begin{minipage}{.5\textwidth}
  \caption{\protect\subref{fig:Ground truth group-representative map X} the true group-representative map $X$. 
    The leftmost column in the two rows below show two different 
    initializations. The remaining columns show corresponding estimates 
    of the group-representative map produced by (from left to right) 
\initialI\space with coordinate ascent, \initialII\space with coordinate ascent, and 
 \initialII\space with variational Bayes.  
}
\label{fig:fig8}
\end{minipage}
\begin{minipage}{.98\textwidth}
  \centerline{
  \subfloat[]{
  \includegraphics[width=0.23\textwidth]{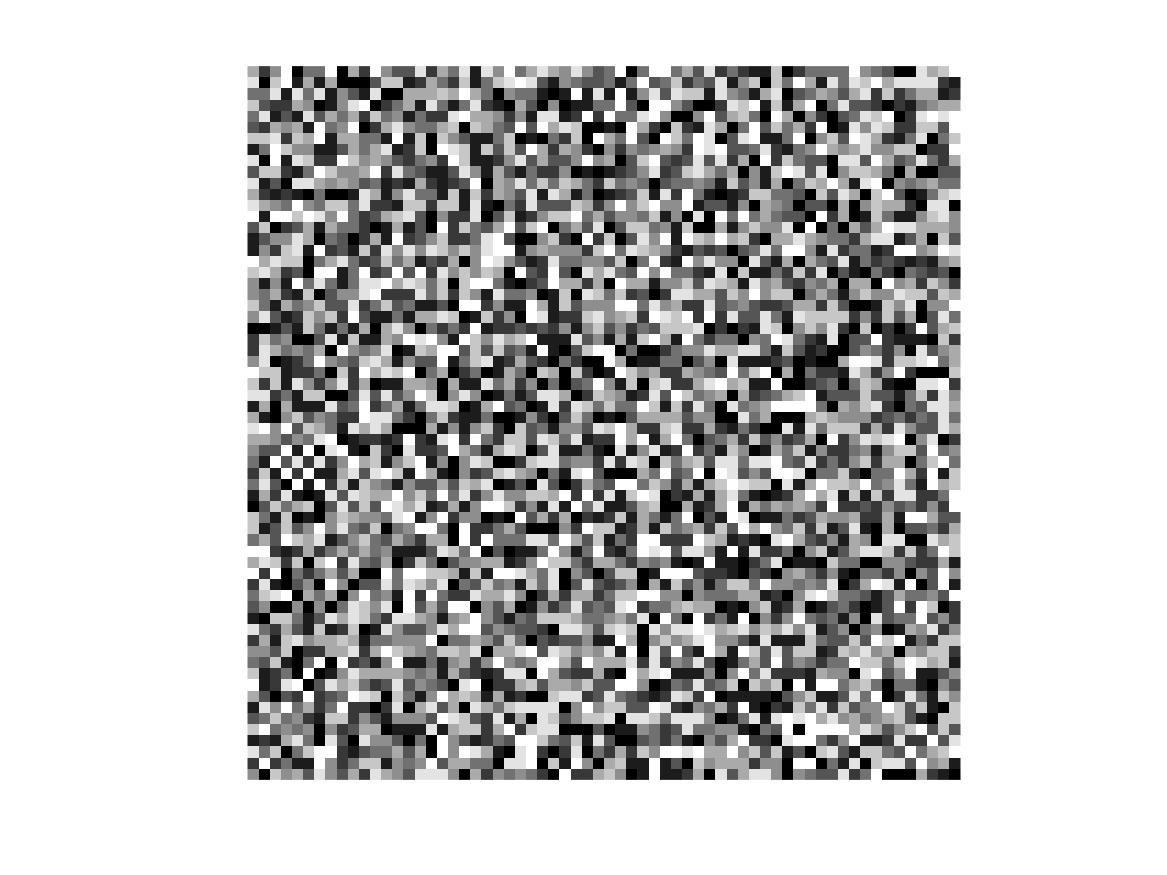}
  \label{fig:First Initial I Estimate X01}
  }   
  \subfloat[]{
  \includegraphics[width=0.23\textwidth]{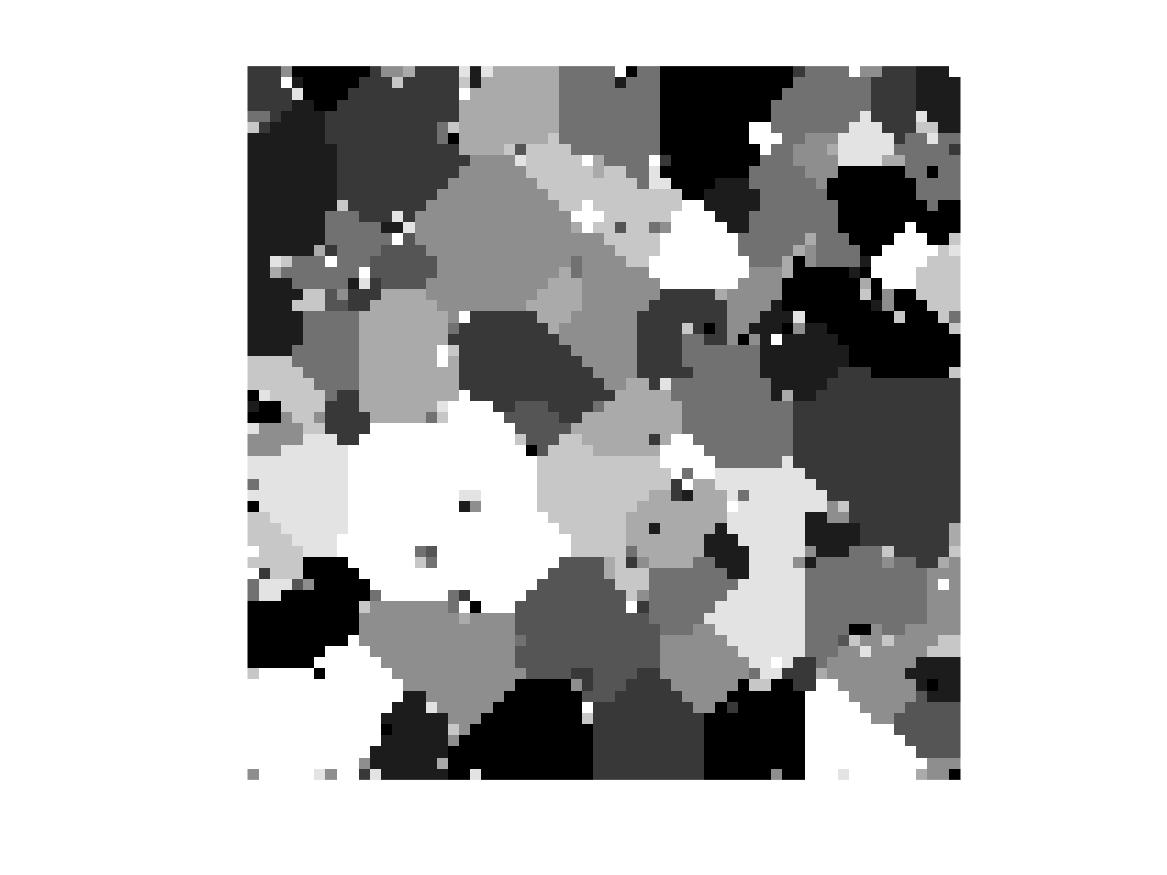}
  \label{fig:First Final I Estimate X1}
  }
  \subfloat[]{
  \includegraphics[width=0.23\textwidth]{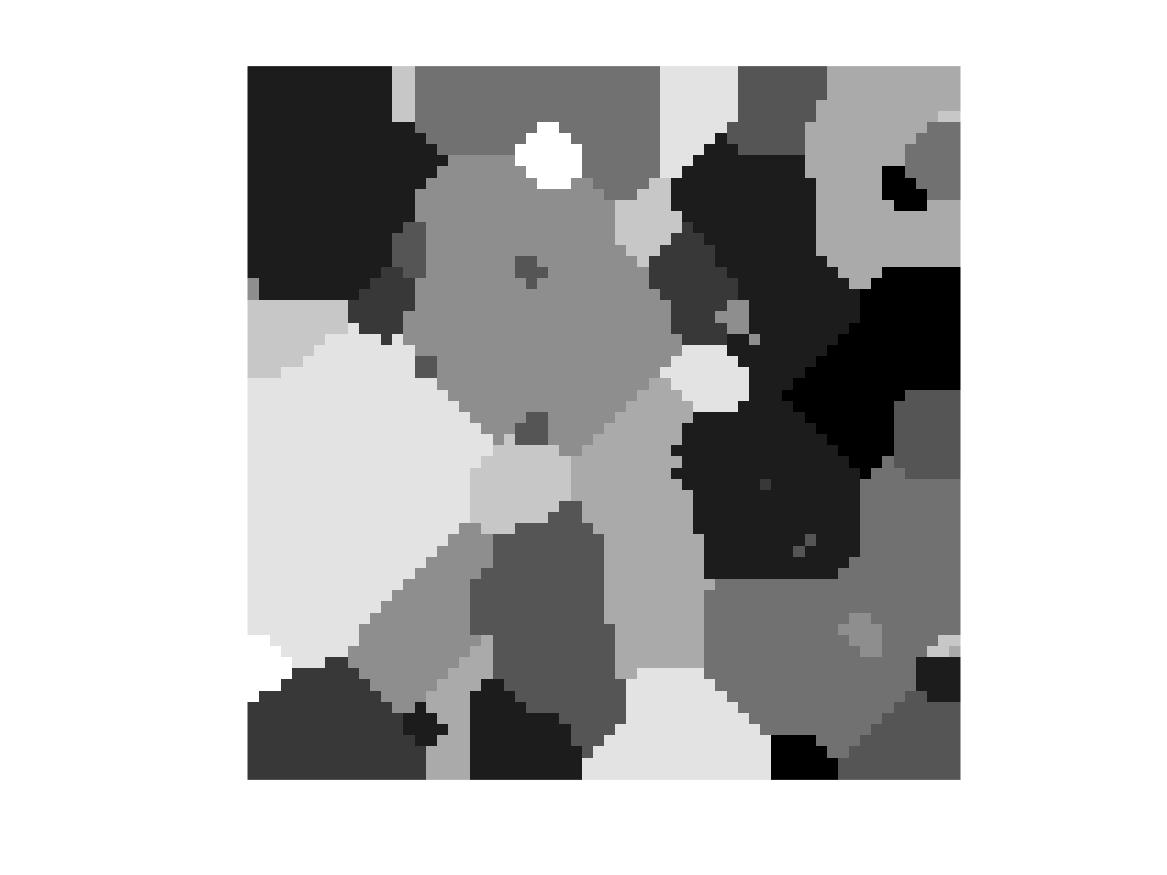}
  \label{fig:First Final II Estimate X2}
  }
  \subfloat[]{
  \includegraphics[width=0.23\textwidth]{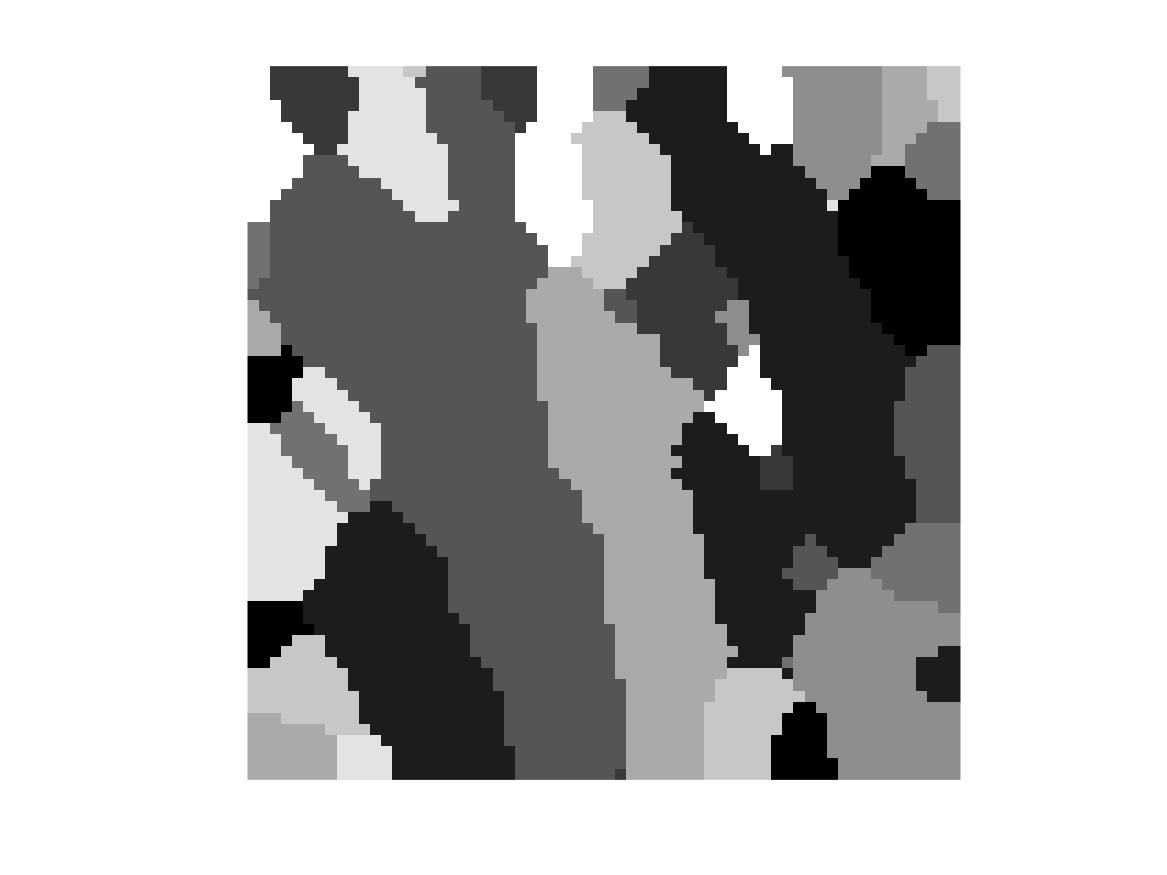}
  \label{fig:First Final VEM Estimate X3}
  }}  
  \centerline{
   \subfloat[]{
  \includegraphics[width=0.23\textwidth]{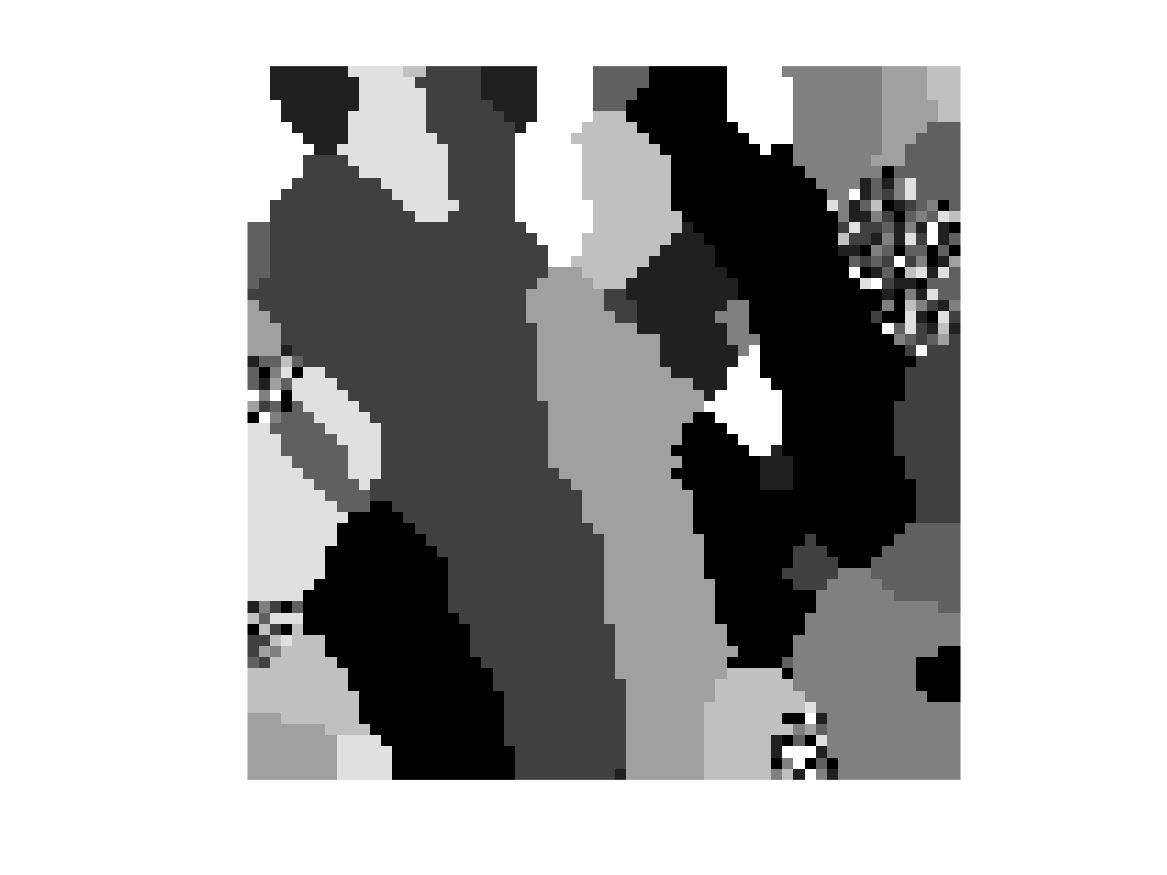}
  \label{fig:Second Initial I Estimate X02}
  }   
  \subfloat[]{
  \includegraphics[width=0.23\textwidth]{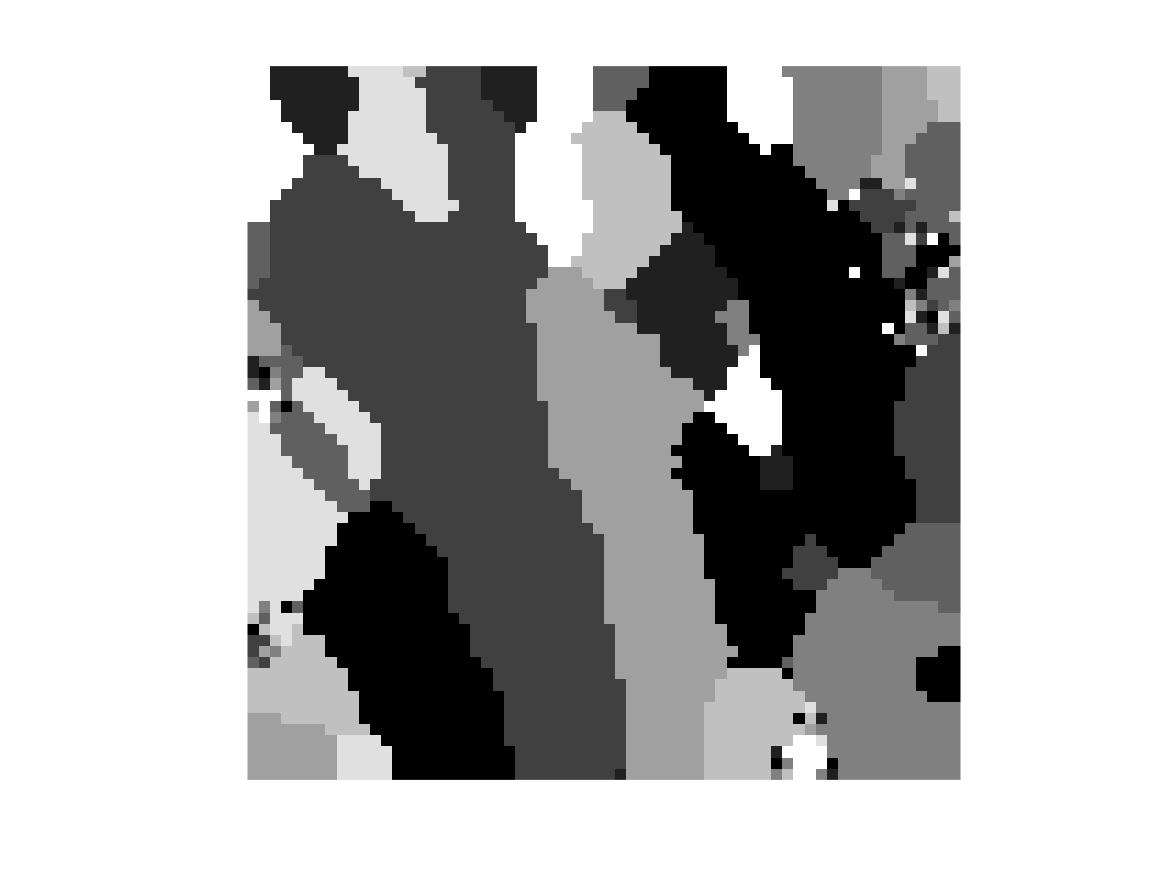}
  \label{fig:Second Final I Estimate X4}
  }
  \subfloat[]{
  \includegraphics[width=0.23\textwidth]{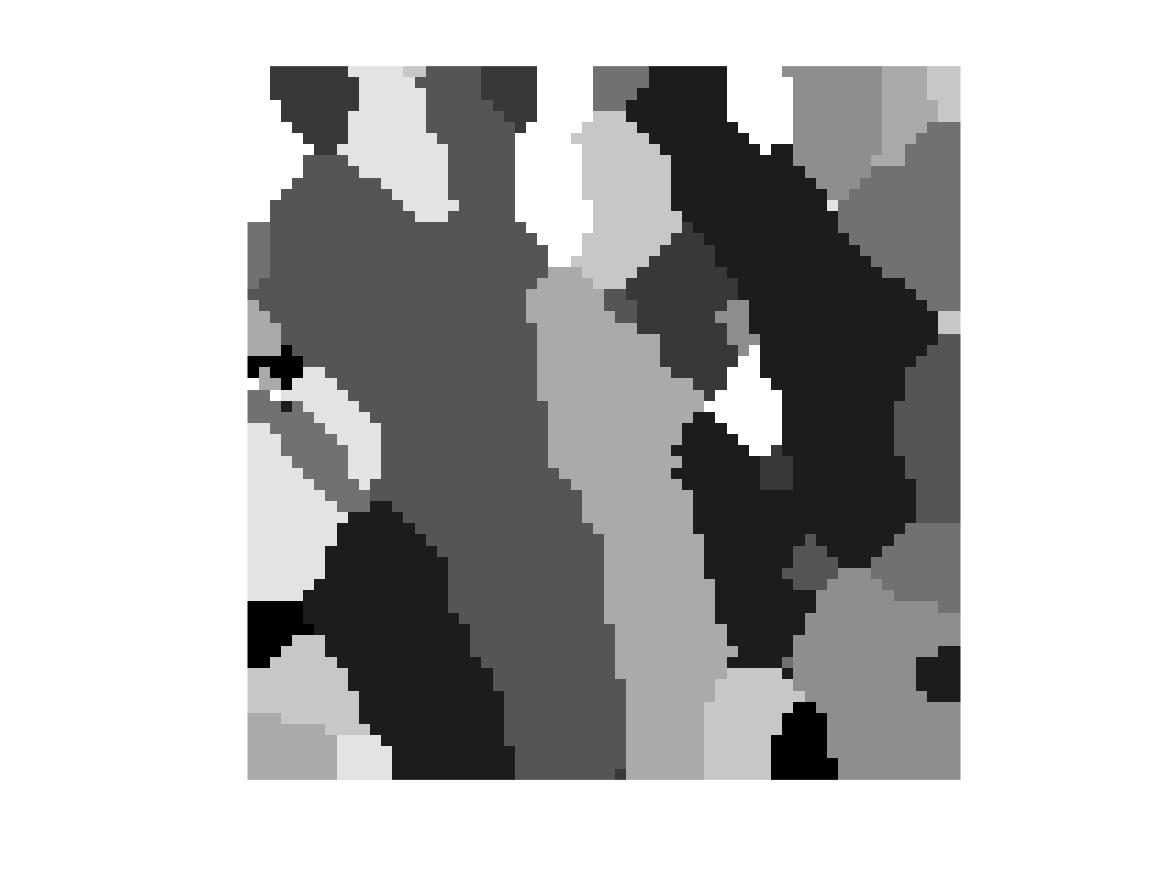}
  \label{fig:Second Final II Estimate X5}
  }
  \subfloat[]{
  \includegraphics[width=0.23\textwidth]{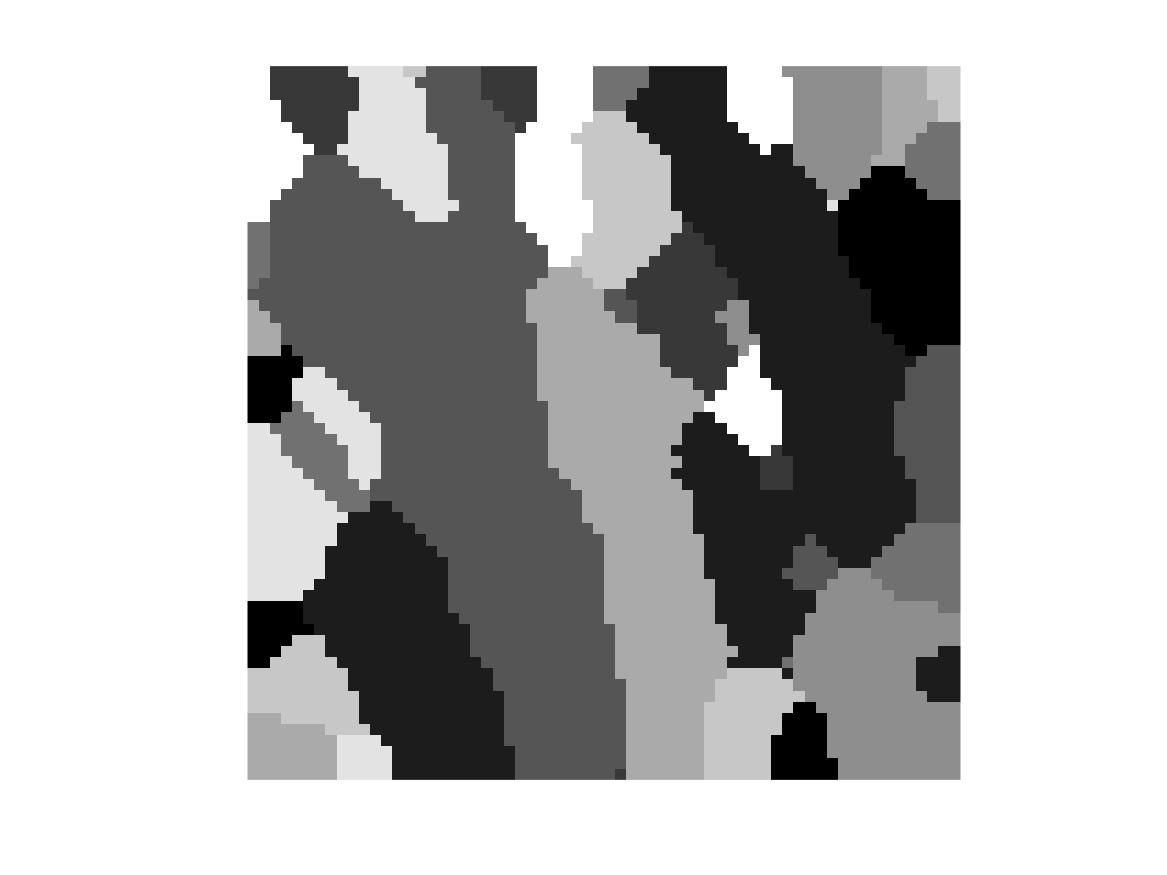}
  \label{fig:Second Final VEM Estimate X6}
  }}
\end{minipage}
\end{figure*}

In figures \protect\ref{fig:fig6}-\protect\ref{fig:fig8}, we present a 
qualitative analysis of the effect of initialization, repeating the 
steps from the corresponding plots in the previous section.
As before, we set $K=10$ and $M=40$. Figure~\ref{fig:fig6} shows three of 
the images presented to the algorithms, and figures~\ref{fig:fig7} and 
\ref{fig:fig8} show results for $H$ and $X$ respectively. 
Again, we see improved performance for variational Bayes in terms of 
its ability to avoid local optima that trap coordinate-ascent.

\subsection{Simulated fMRI datasets}\label{413}
Next we test our method on simulated fMRI data generated using the SimTB 
toolbox for MATLAB{\texttrademark}. SimTB facilitates the flexible 
generation of fMRI datasets under a model of spatio-temporal separability 
\citep{37}. Note that this generative model is not the forward model we have 
proposed, and to which our algorithms correspond. In keeping with the sample sizes of previous fMRI-based 
connectivity studies \citep{38}, we simulate $M$=30 subjects. The 
synthesized scans have a repetition time of $3$s/sample, 
with slices of size $64\times64$ at $T$=150 time points. To maintain a 
reasonable computation time, we set the number of components at 30. 
Of these components, not all are uniformly present in all the subjects. 
We instead consider a subset of 17 components of interest, for each of 
which there is a 90\% probability of occurrence in every subject. In 
addition, we assign to each of the remaining components, a 30\% 
probability of occurrence. To model the spatial variability in the regions 
of activity under each component  across the subjects, we incorporate 
independent normal translation, rotation, and spread. Activation centers 
are translated vertically and horizontally with a standard deviation of 
0.3 voxels, rotated by a deviation of 1 degree, and their spatial extent 
(compression or expansion) is determined following the normal distribution 
$N(1,0.3)$. See Appendix-\ref{sec:simtb} for more details.

\begin{figure*}
\begin{minipage}{.7\textwidth}
  \centerline{
  \subfloat[]{
  \includegraphics[width=0.24\textwidth]{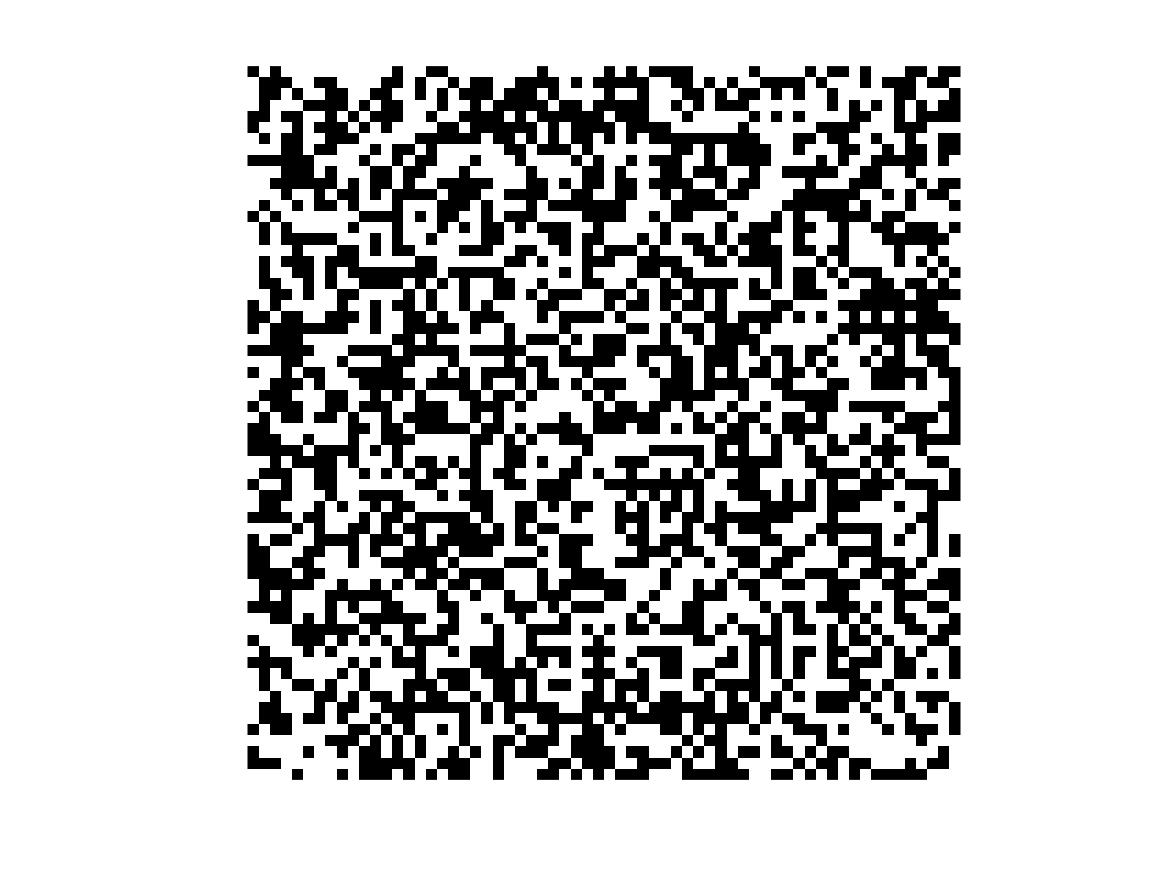}
  \label{fig:x1}
  }   
  \subfloat[]{
  \includegraphics[width=0.24\textwidth]{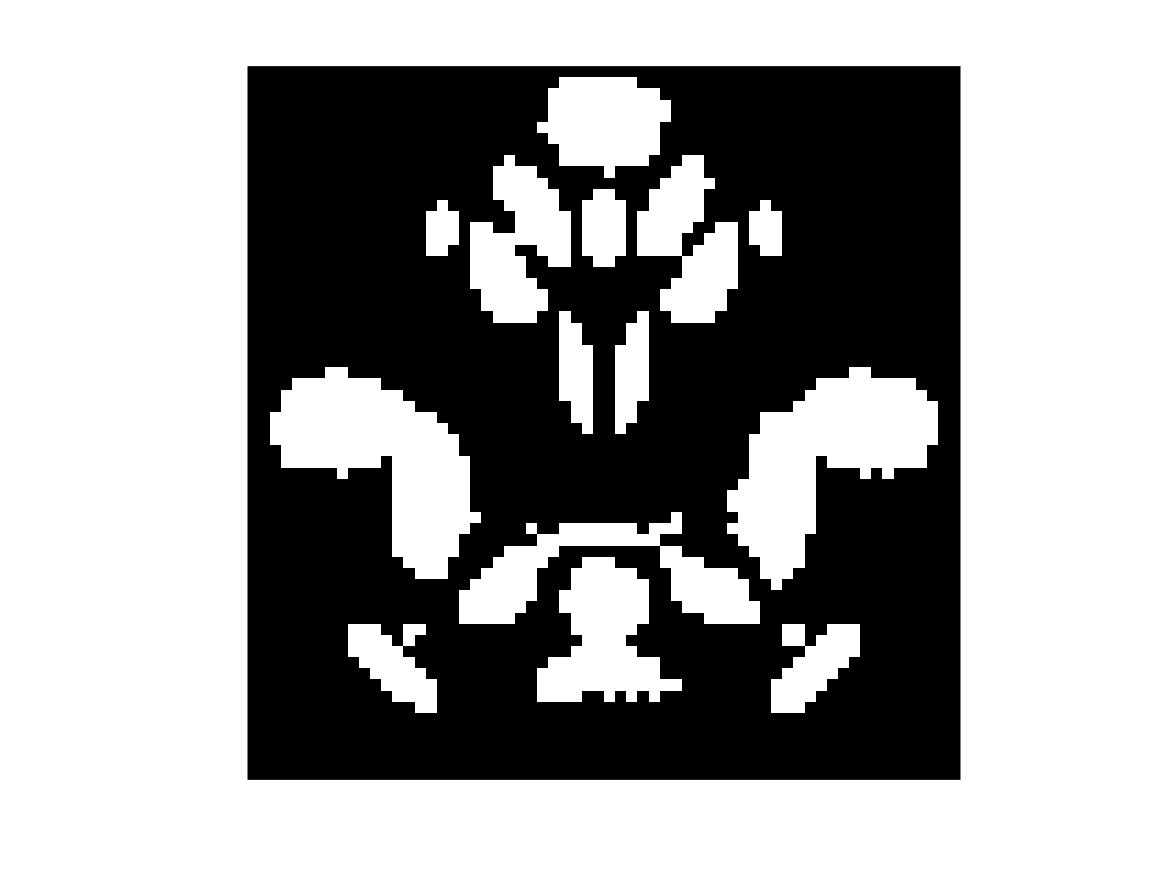}
  \label{fig:x30}
  }
  \subfloat[]{
  \includegraphics[width=0.24\textwidth]{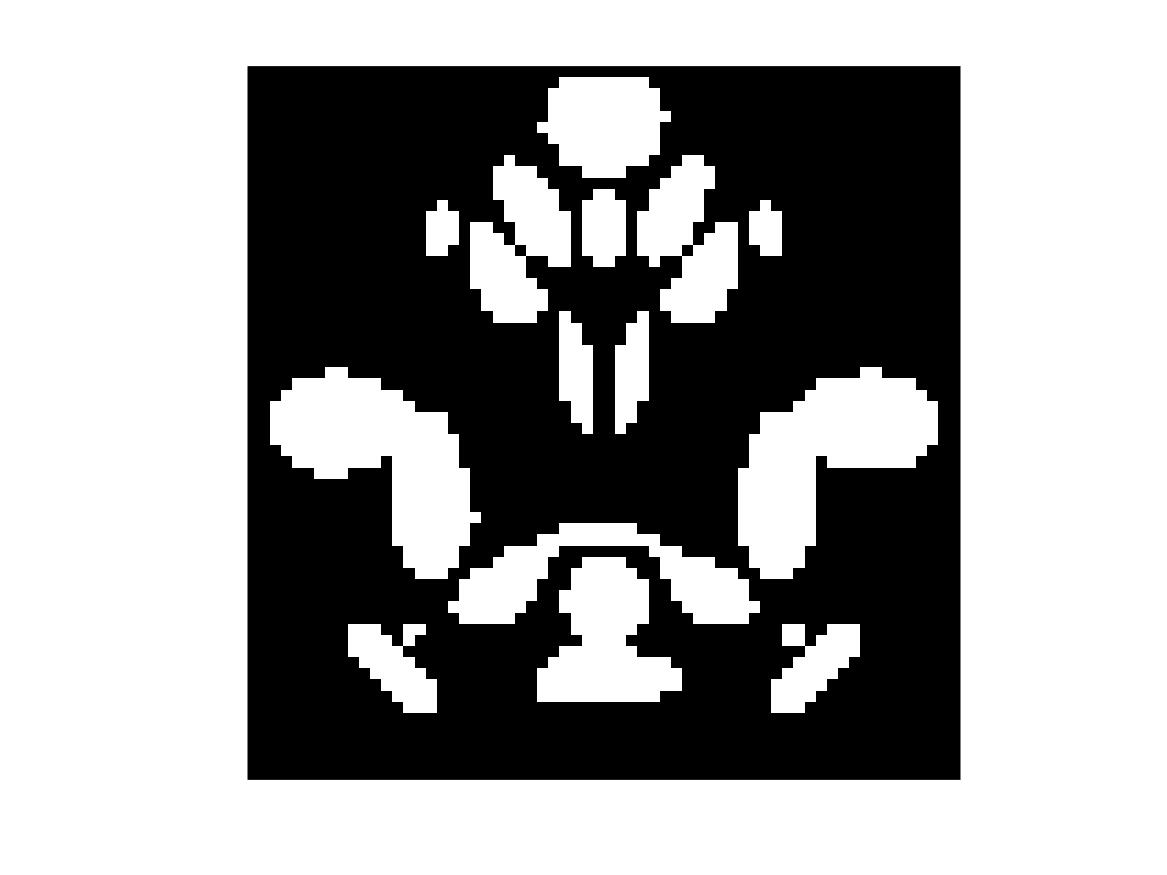}
  \label{fig:x60}
  }
  \subfloat[]{
  \includegraphics[width=0.24\textwidth]{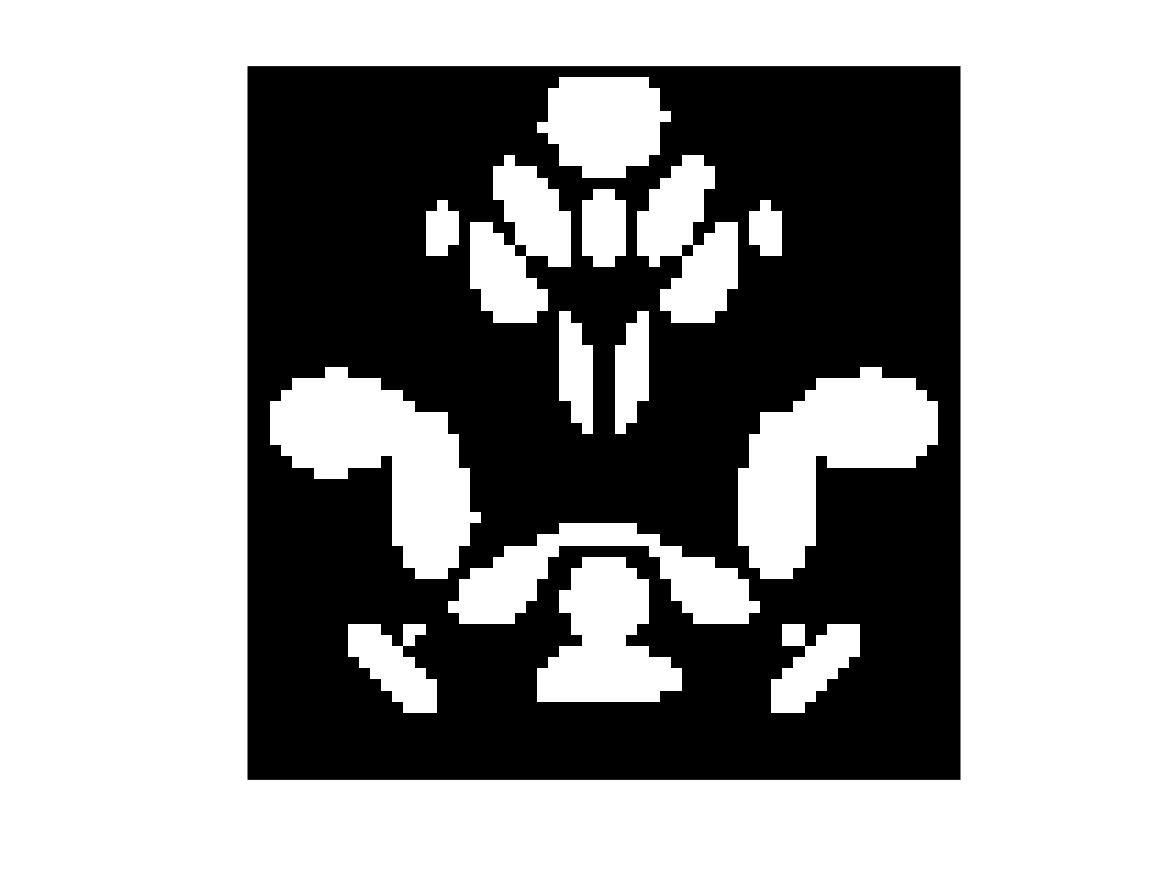}
  \label{fig:x80}
  }
  }
\end{minipage}
\begin{minipage}{.28\textwidth}
  \caption{Estimates of the group map $X$ at iterations 1, 30, 60, 
    80.}
\label{fig:fig9}
\end{minipage}
\end{figure*}

\begin{figure*}
\begin{minipage}{.7\textwidth}
  \centerline{
  \subfloat[]{
  \includegraphics[width=0.3\textwidth]{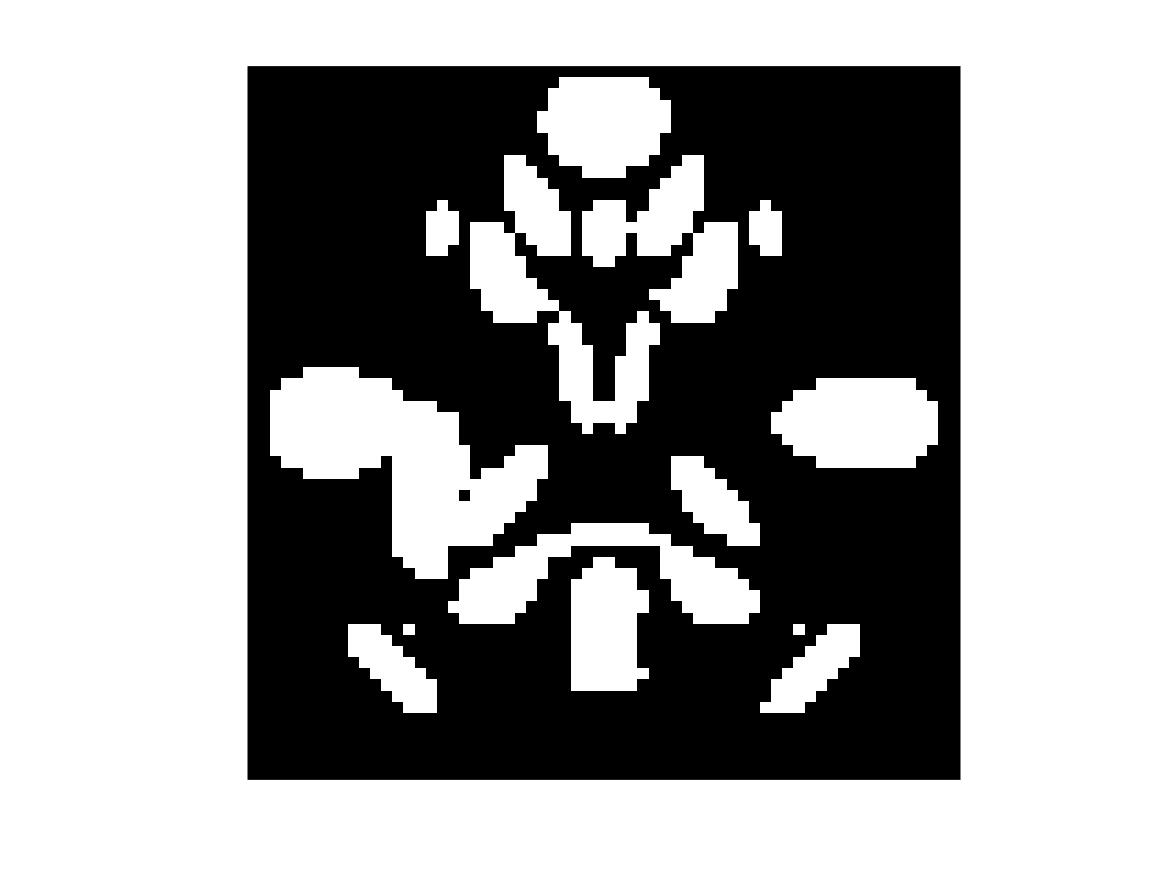}
  \label{fig:Y1}
  }   
  \subfloat[]{
  \includegraphics[width=0.3\textwidth]{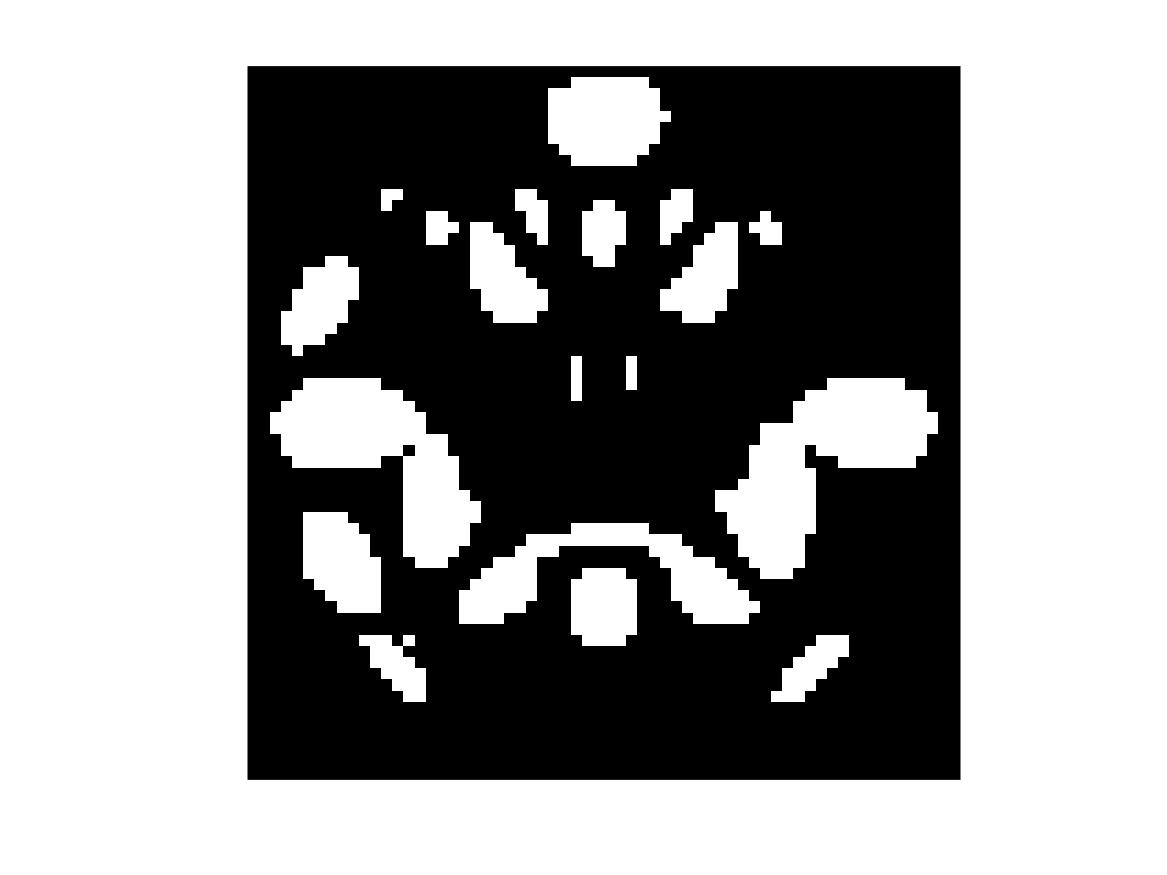}
  \label{fig:Y2}
  }
  \subfloat[]{
  \includegraphics[width=0.3\textwidth]{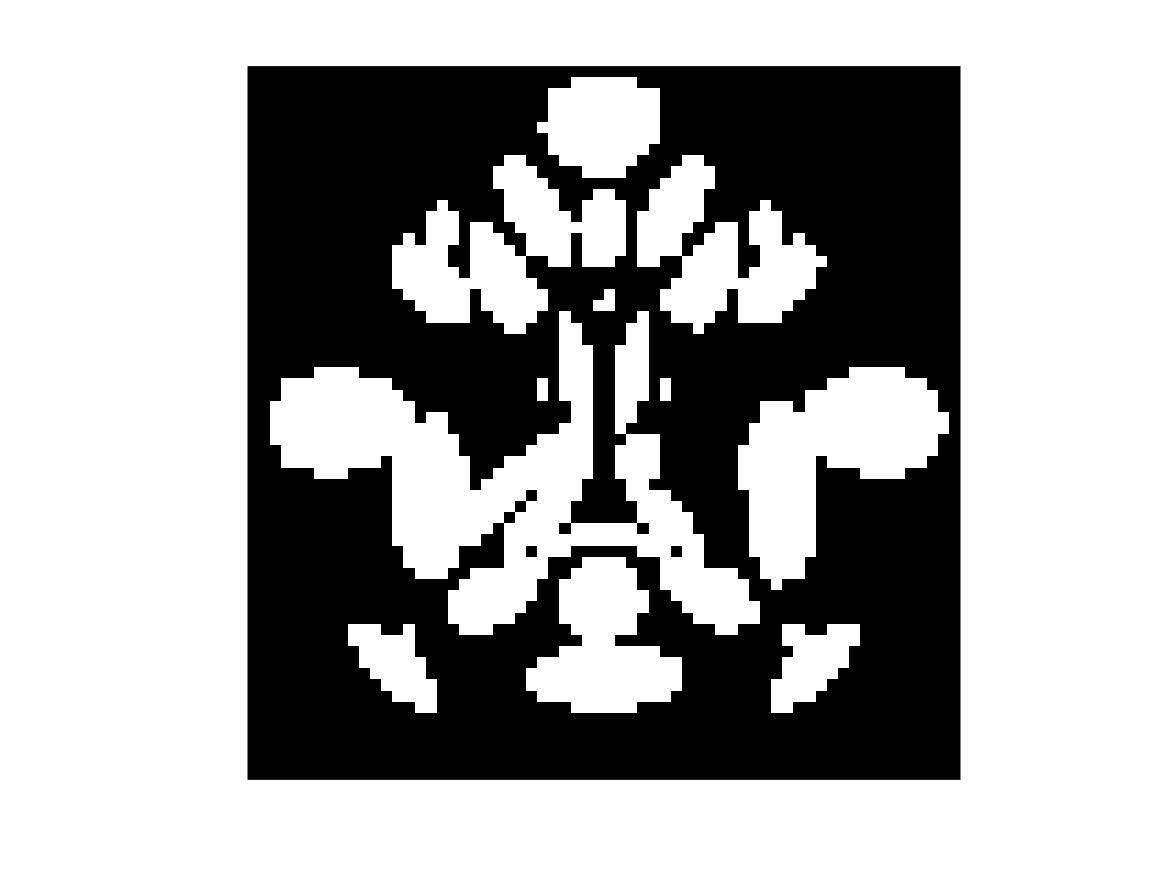}
  \label{fig:Y3}
  }
  }
\end{minipage}
\begin{minipage}{.28\textwidth}
  \caption{Estimated subject maps $Y_1$, $Y_2$, and $Y_3$ for individuals 
    $1,2$ and $3$.
}
\label{fig:fig10}
\end{minipage}
\end{figure*}

\begin{figure*}
  \centerline{
  \subfloat[]{
  \includegraphics[width=0.2\textwidth]{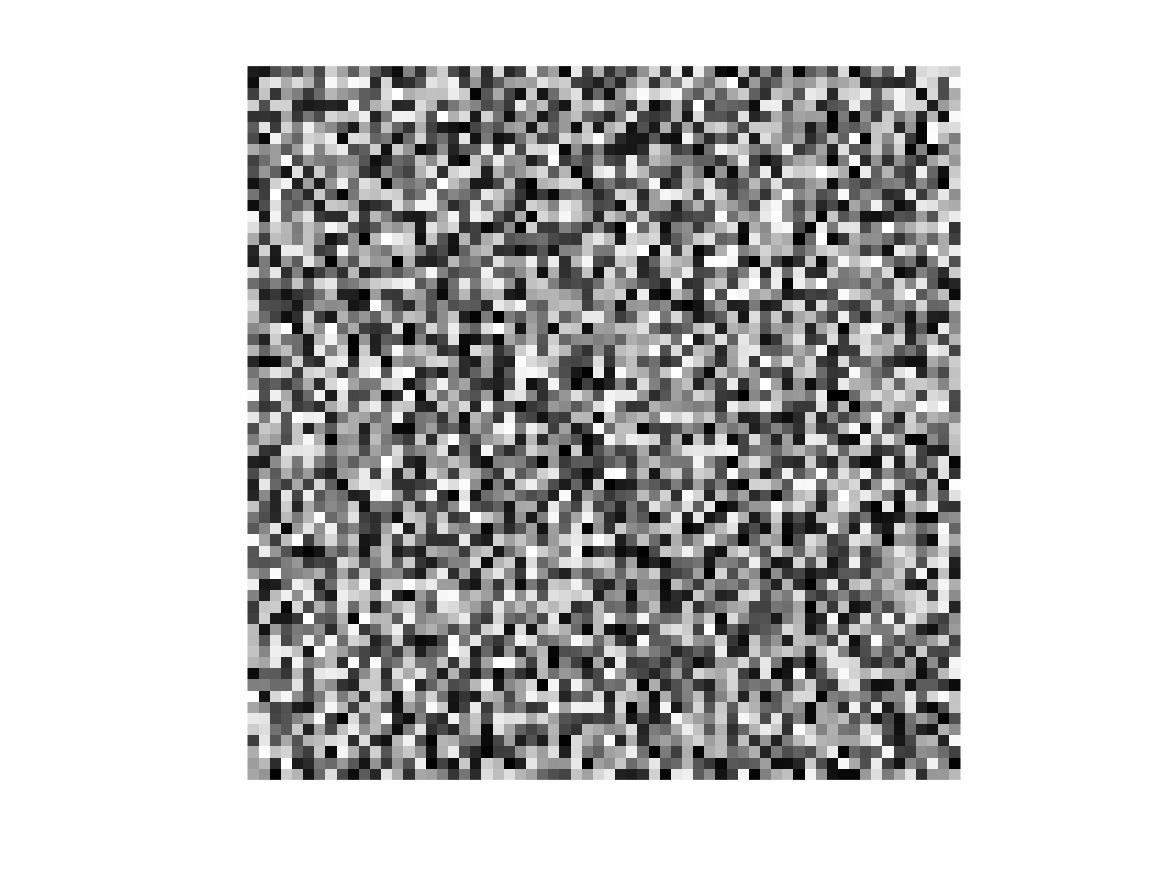}
  \label{fig:Q0}
  }   
  \subfloat[]{
  \includegraphics[width=0.2\textwidth]{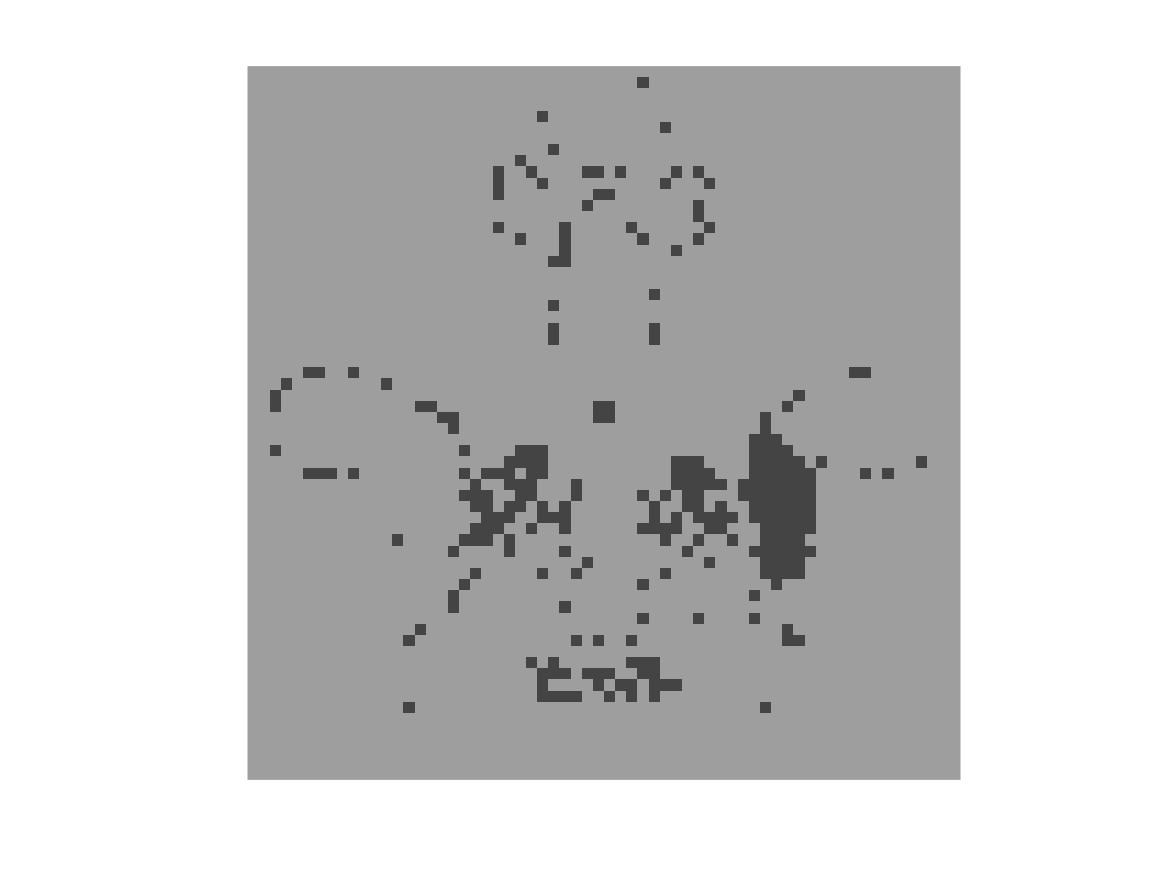}
  \label{fig:Q10}
  }
  \subfloat[]{
  \includegraphics[width=0.2\textwidth]{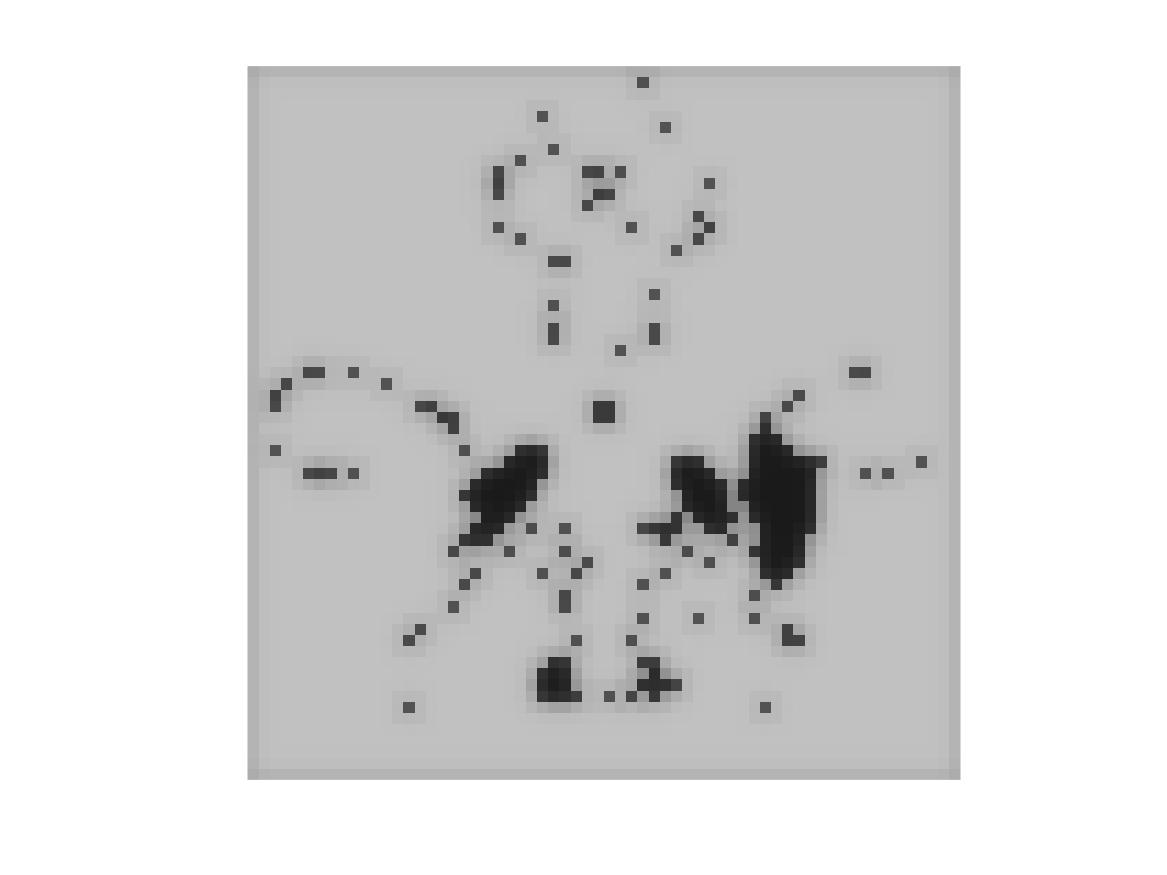}
  \label{fig:Q30}
  }
  \subfloat[]{
  \includegraphics[width=0.2\textwidth]{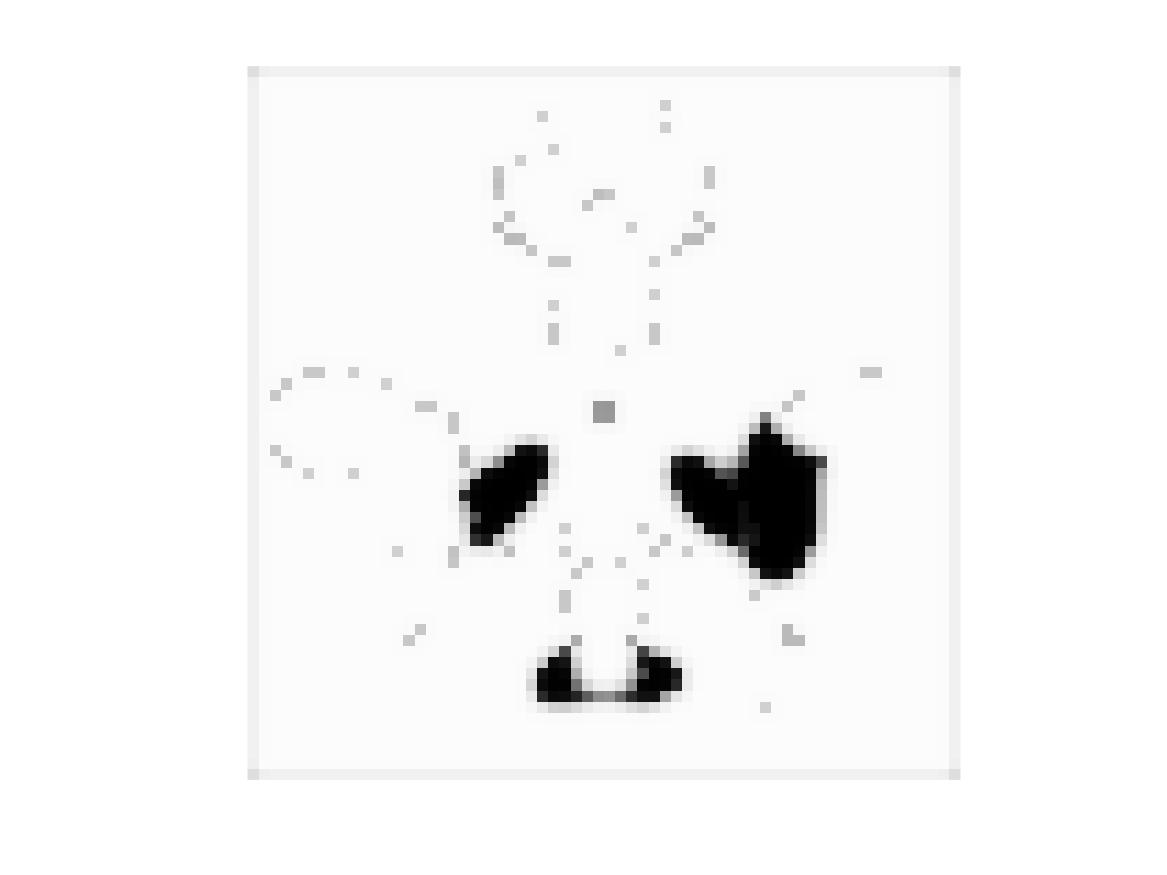}
  \label{fig:Q70}
  }
  \subfloat[]{
  \includegraphics[width=0.2\textwidth]{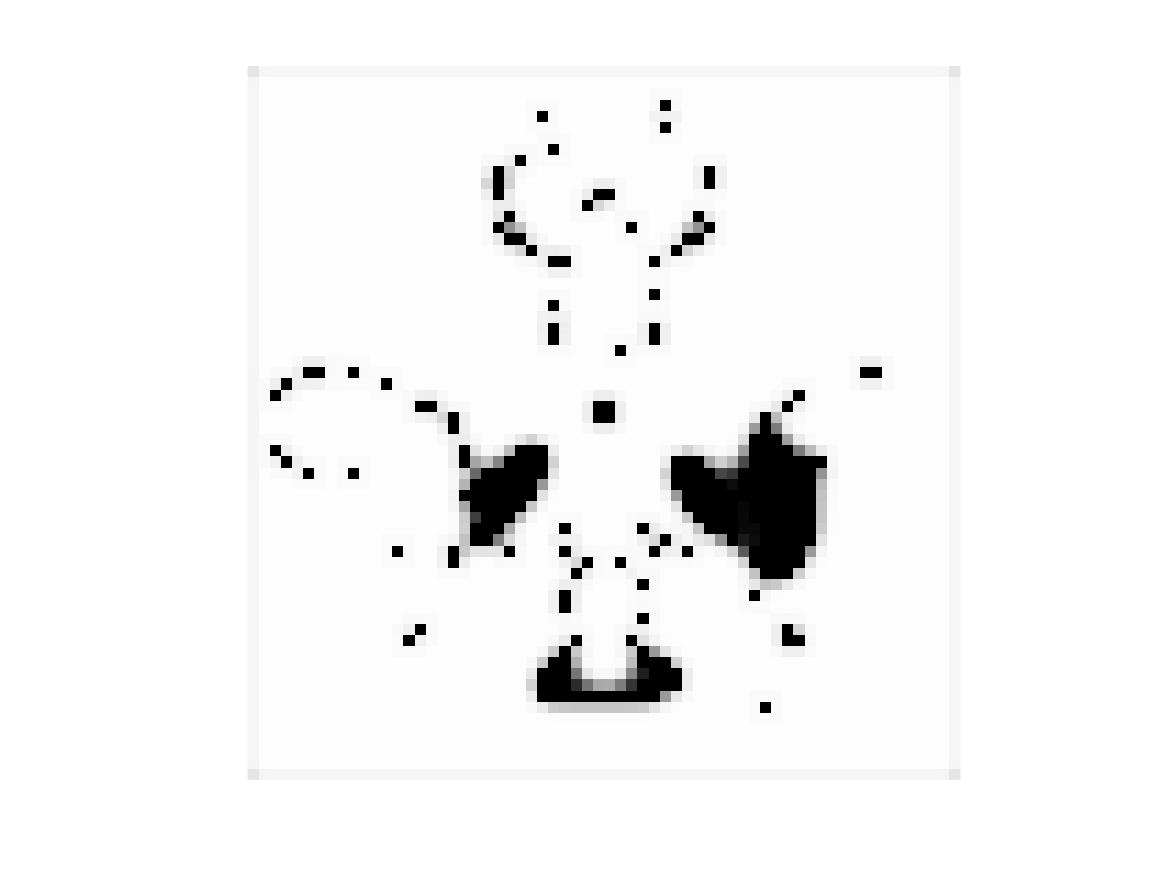}
  \label{fig:Q100}
  }
  }
  \caption{Estimates of the variational posterior $q_1$ for the mask 
  $H_1$ of subject $1$ at iterations 1, 30, 60, 80, 100} 
\label{fig:fig11}
\end{figure*}

We apply the pre-processing from section~\ref{sec:preproc} to this 
data to generate individual subject maps to input to the variational 
Bayes algorithm. We use this algorithm with random initialization to 
estimate $X_0$, the group-representative map. 
Figure \ref{fig:fig9} below plots the evolution of the estimate of $X$ over 
iterations of the variational Bayes algorithm until convergence, at which 
point, the underlying $X$ is correctly identified. 
In figure~\ref{fig:fig10} we plot the estimated subject maps for three 
subjects, while figure~\ref{fig:fig11} plots the evolution of the 
variational posterior $q_1(H_1)$ of the first subject. All these 
estimates are clearly reasonable, indicating our modeling and computational 
assumptions are appropriate for fMRI images according to the standardized 
SimTB toolbox.

\begin{figure*}
\begin{minipage}{.78\textwidth}
  \centerline{
  \subfloat[]{
  \includegraphics[width=0.23\textwidth]{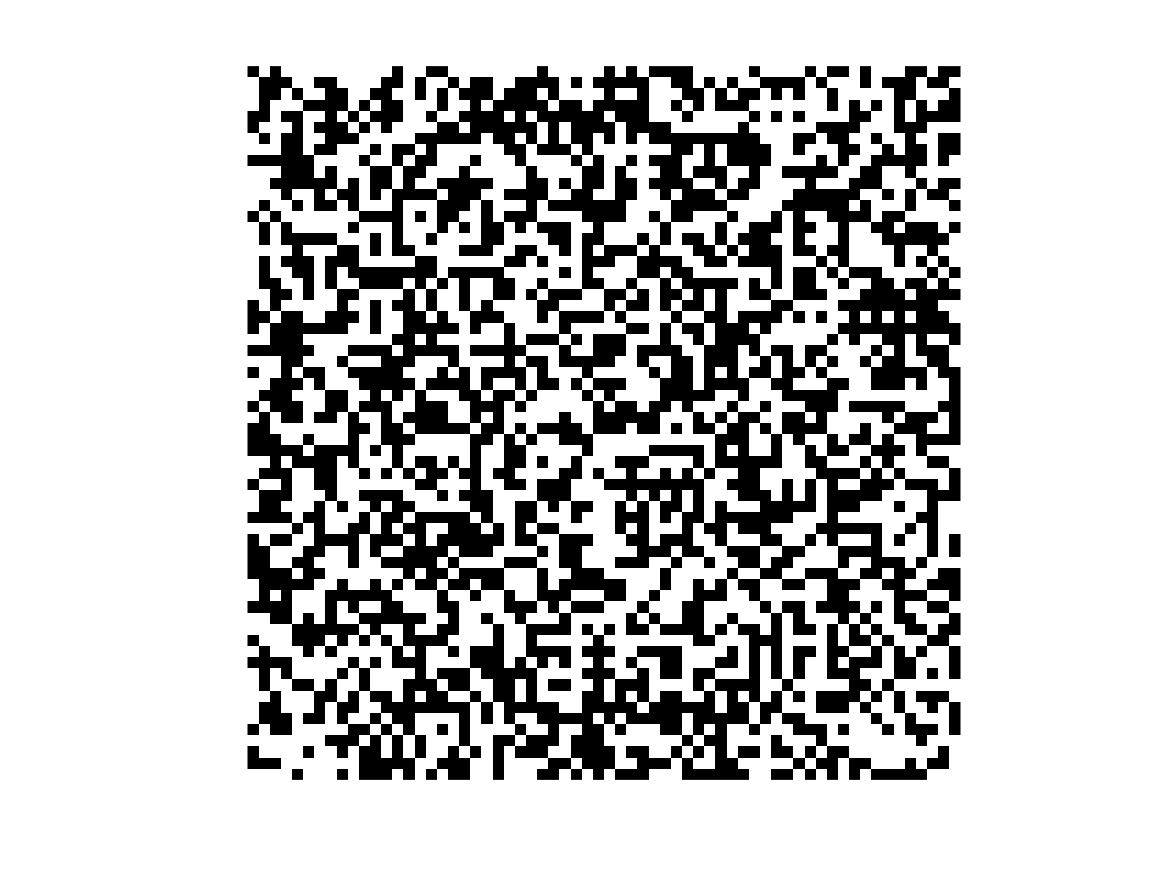}
  \label{fig:X01}
  }   
  \subfloat[]{
  \includegraphics[width=0.23\textwidth]{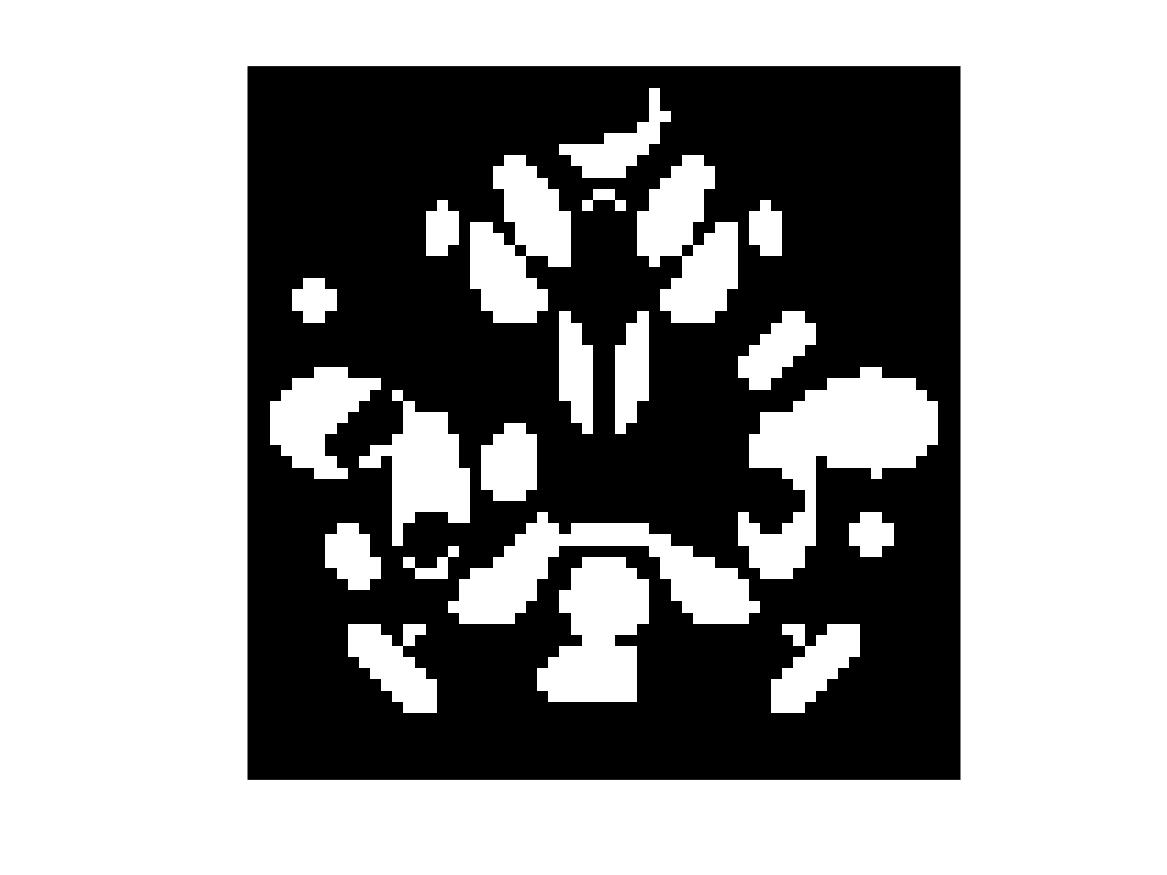}
  \label{fig:Xout1}
  }
  \subfloat[]{
  \includegraphics[width=0.23\textwidth]{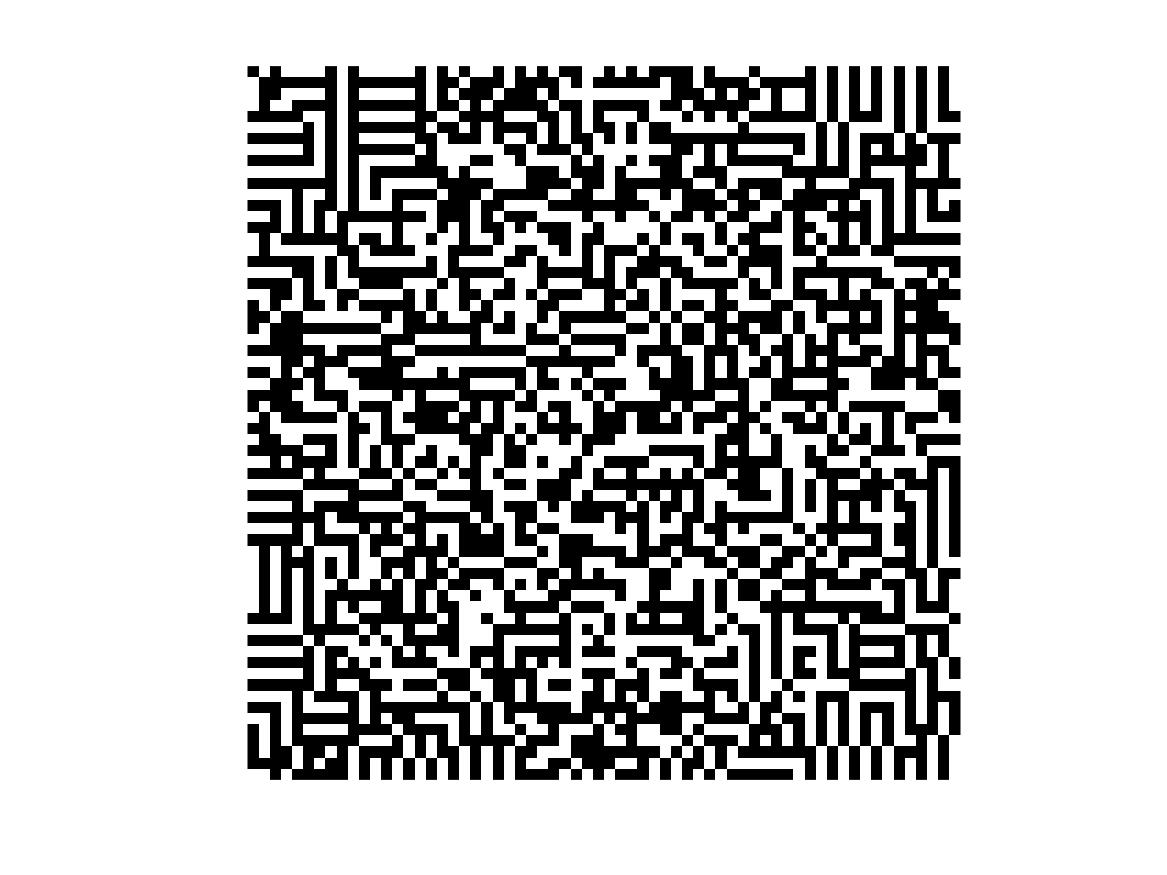}
  \label{fig:initialIXout1}
  }
  \subfloat[]{
  \includegraphics[width=0.23\textwidth]{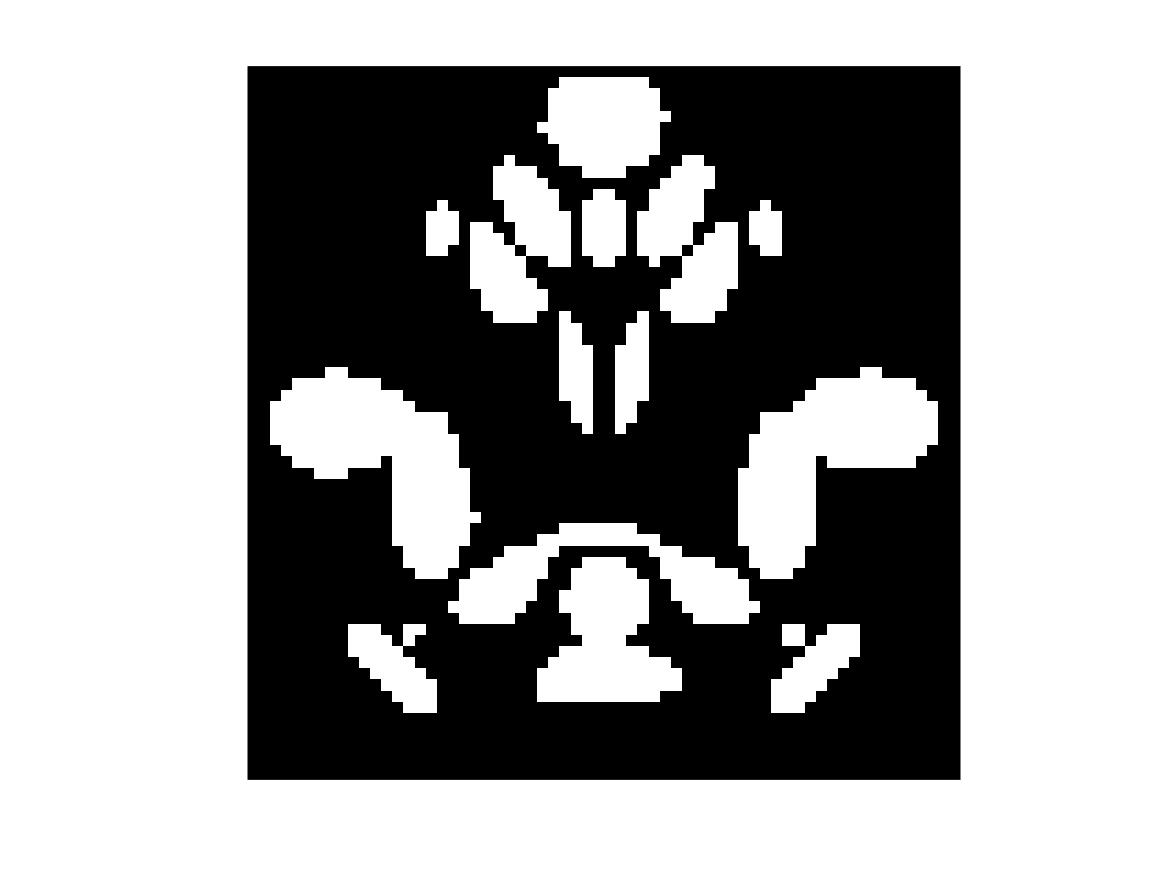}
  \label{fig:initialIIXout1}
  }}

  \centerline{
  \subfloat[]{
  \includegraphics[width=0.23\textwidth]{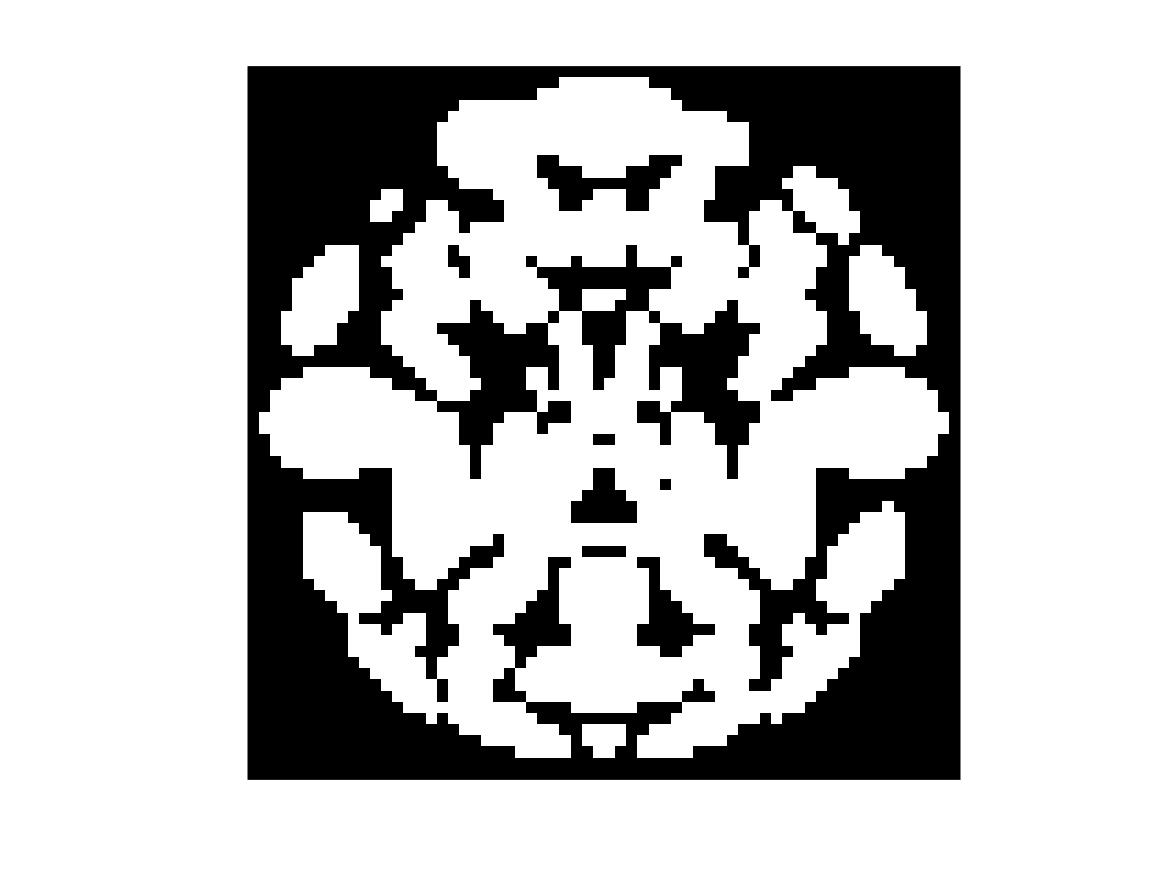}
  \label{fig:X02}
  }   
  \subfloat[]{
  \includegraphics[width=0.23\textwidth]{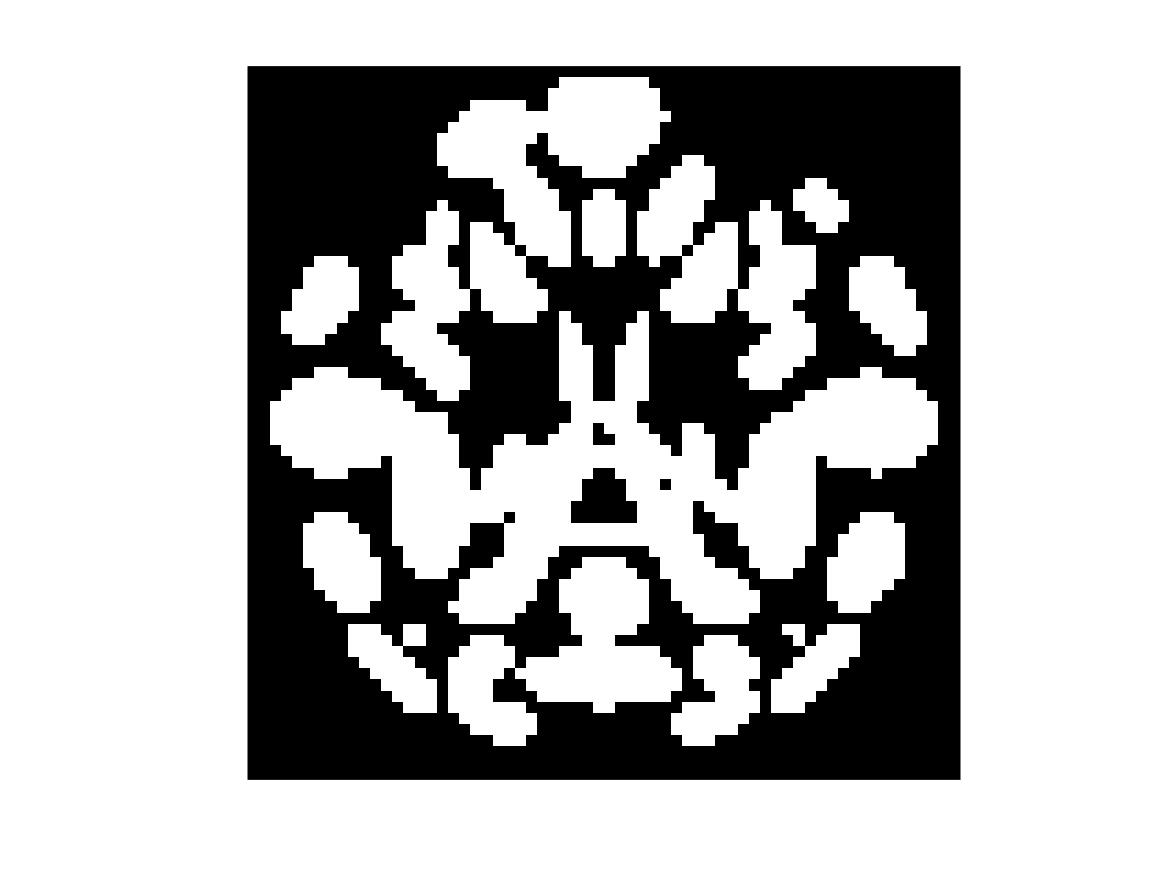}
  \label{fig:Xout2}
  } 
  \subfloat[]{
  \includegraphics[width=0.23\textwidth]{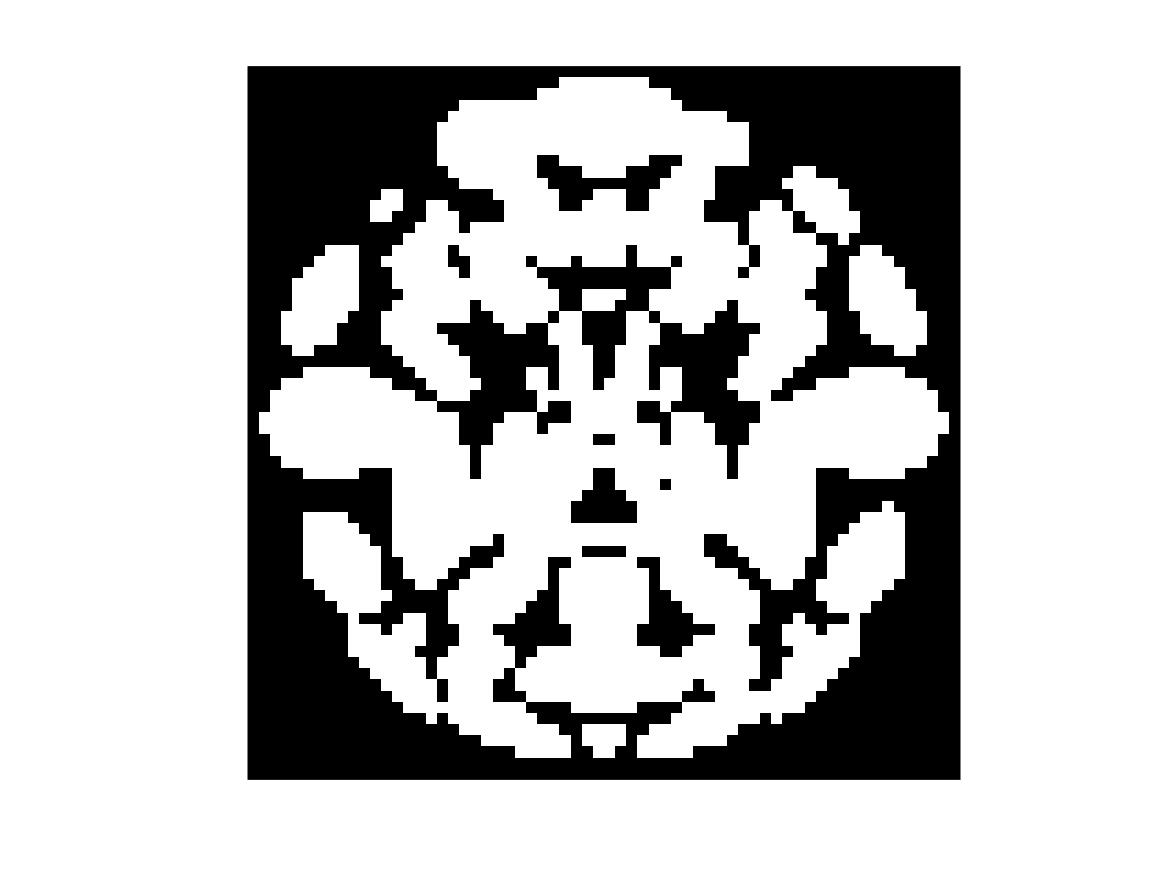}
  \label{fig:initialIXout2}
  }
  \subfloat[]{
  \includegraphics[width=0.23\textwidth]{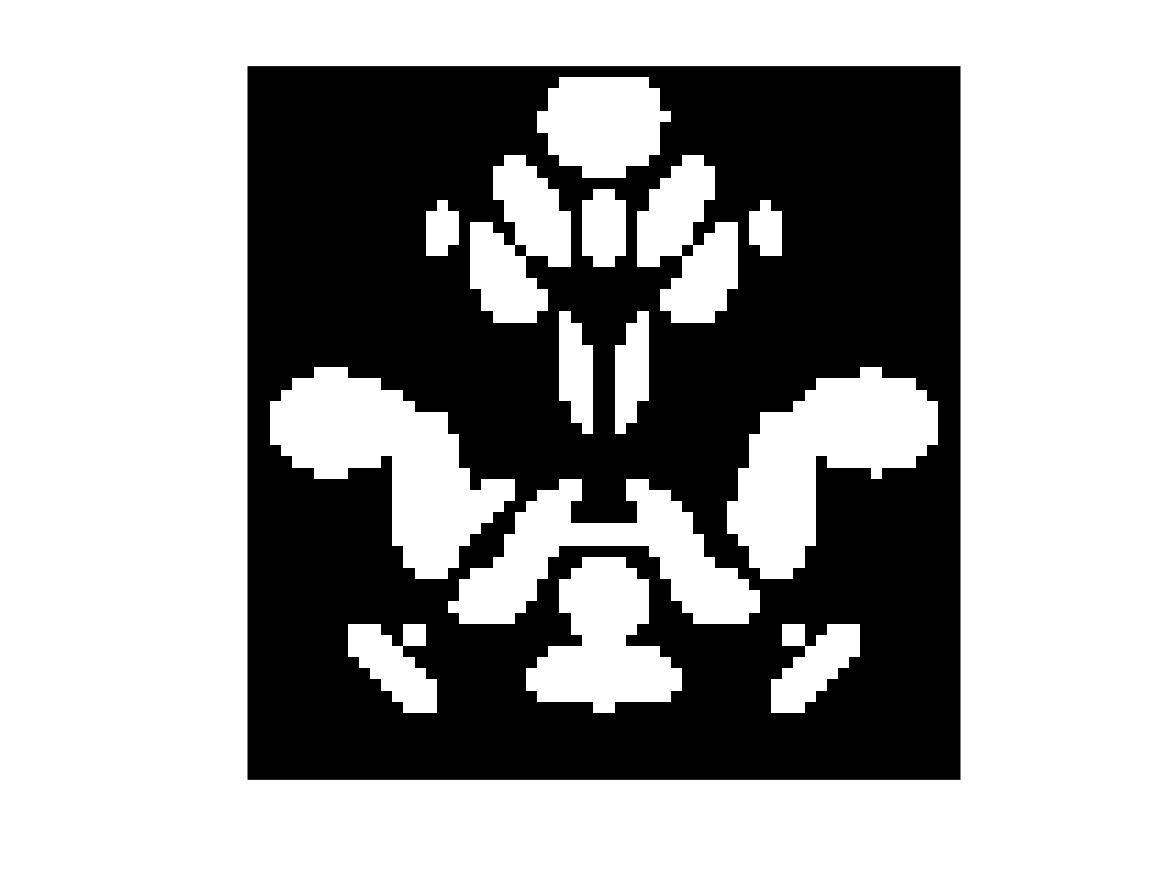}
  \label{fig:initialIIXout2}
  }}
\end{minipage} 
\begin{minipage}{.2\textwidth}
  \caption{ The leftmost column shows two different 
    initializations (random and greedy). The remaining columns show 
    corresponding estimates of the group-representative map produced by 
    (from left to right) \initialI\space with coordinate-ascent, \initialII\space with coordinate-ascent, and 
 \initialII\space with variational Bayes.  
}
\label{fig:fig12}
\end{minipage} 
\end{figure*}

Next, we use the simulated fMRI dataset to compare the 
robustness of the different algorithms to initialization.
The leftmost column of figure 
\protect\ref{fig:fig12} shows two different initializations, 
random (top) and greedy (bottom). From left to right, we plot the 
corresponding estimated group-representative maps for \initialI\ and 
\initialII\  with coordinate-ascent, and \initialII\ with variational 
Bayes respectively. Clearly, the last is the only one that a) is robust 
to the initialization, and b) that recovers a solution close to the ground 
truth. We reiterate again that this not for data generated according to 
any of the models.

\section{Discussion and Conclusions}\label{sec:conc} \
We propose a novel approach to estimate group-representative 
functional connectivity maps from multi-subject fMRI data. Our overall 
contribution is a framework consisting of two steps, a pre-processing step 
and a MAP-MRF model with an associated variational Bayes algorithm.
Our preprocessing step overcome limitations of standard averaging and 
concatenation-based ICA schemes, and unlike other methods, 
does not require pre-registration, or 
the imposition of a common set of 
associated temporal signals. Our MAP-MRF framework involves a novel 
forward model that describes the generation of individual subject maps 
from an underlying group-representative map, as well as a novel 
variational . The solution to the resulting inverse problem of estimating the group-representative map is then obtained using the MAP-MRF framework.

To capture the complexity of real data, future work should explore more 
sophisticated prior and generative models, for instance, defining 
distinct inverse temperature parameters for different clique types 
(vertical, horizontal, diagonal). There are also opportunities for 
improving the generative model through modification of the noise 
distribution, so as to better capture the inter-subject variations. 


\appendix 

\section{Data pre-processing}\label{sec:app_preproc}
We represent the fMRI data of each subject $i = 1,\ldots,M$ with a 
$T \times V$ matrix $D^{(i)}$, where $T$ is the number of 
time points, and $V$ the number of voxels. The $t$\textsuperscript{th} 
row of such a matrix contains all voxels imaged at 
time $t$, and the $v$\textsuperscript{th} column contains the time 
course of the corresponding voxel.

\subsection{Outlier exclusion and dimension reduction}\label{sec222}
We first use the RV coefficient \citep{29} to exclude individual scans 
showing significant disparity from the rest of the group, due 
to factors such as undetected scanning problems and anatomical deviations 
\citep{11}. 
The RV coefficient for subjects $i$ and $j$ is:
\begin{equation}
RV (D^{(i)},D^{(j)}) = Tr(Z_i Z_j)/ \sqrt{(Tr(Z_i^2) Tr(Z_j^2))},
\end{equation}
where $Tr(\cdot)$ is the matrix trace and $Z_i$ is the covariance matrix 
of subject $i$.  
Following~\cite{30}, we calculate the $Z_i$'s as 
$Z_i = D_*^{(i)} {D_*^{(i)}}^T$, 
where subscript $*$ indicates zero-correction along the
diagonal of $D^{(i)}$. 
 A distance matrix is then 
populated using the pairwise RV coefficients and 
outliers are excluded from the following analysis. 

Next, we reduce the number of time points in the selected data matrices 
$D^{(i)}$ by performing PCA.  We elect to achieve maximum 
dimensionality reduction following a preset CPV rule of 95\% for all 
subjects \citep{31}, and the $D^{(i)}_{T \times V}$  is 
reduced to $D^{(i)}_{t_i \times V}$ for subject $i$. 
The reduced data matrices are then whitened to have unit variance. 

\subsection{ Spatial ICA }\label{sec221}
Next, we use spatial ICA to decompose the fMRI data of each 
subject into a set of 
statistically independent spatial maps and associated time courses 
\citep{24}.  Assuming $P$ independent components, this 
decomposition is given by $D^{(i)} = A_i B_i$, where the $P \times V$ 
matrix $B_i$ denotes the 
spatial maps of the $P$ independent components and the $t_i \times P$ 
matrix $A_i$ denotes the time course of the mixing proportions.

To perform the above decomposition, 
we follow \cite{25} and iteratively improve an initial estimate of the de-mixing matrix $W_i$ 
(the inverse of the $A_i$ matrix) by maximizing the independence of the 
source components $\hat{B}_i = W_i D^{(i)}$. For this, we utilize FastICA Toolbox 2.1 
\citep{26} for MATLAB{\texttrademark}, which is chosen for 
its fast and robust iterative fixed-point implementation of ICA \citep{27}. 
Since the decomposition depends on the choice of the initial 
estimate, we run the FastICA algorithm multiple times with random 
initialization. 


\subsection{Component Clustering}\label{sec223}
The resulting components are clustered across runs using 
the ICASSO toolbox \citep{40}, and their centrotypes are chosen as final 
estimates of the unknown independent components.
For this clustering, we measure spatial similarity of the ICA components, 
as well as similarity of their corresponding time courses. For the 
former, we use the IMED measure \citep{32}: 
as shown in~\cite{47}, this takes into account 
both disagreement in the image intensity as well as pixel distance on 
the image lattice, and is comparatively insensitive to minor image 
misalignments.  The IMED between two vectorized images $X$ and 
$Y$, normalized by their maximum intensities, is given by 
$R_s(X,Y) = (X-Y)^T G (X-Y)$,
where $G = [g_{ij}]_{N \times N}$ is a symmetric, positive-definite 
weight matrix, and $g_{ij} = 1/2 \pi\sigma^2 \exp\{-(i-j)^2 /2 \sigma^2 \}$ 
for pixel coordinates $i=(i_1,i_2)$ and $j=(j_1,j_2)$. 
We normalize the IMED by dividing by the size of the image. 

To measure similarity of component time courses 
(which can have different durations), 
we use the DTW distance of \citep{35},
which stretches or compresses sequences locally to obtain the 
best possible alignment of any pair of sequences. Specifically, for two 
time courses $X: = [x_1,x_2,\ldots ,x_n]$ and $Y:= [y_1,y_2,\ldots ,y_m]$ of 
lengths $n$ and $m$ respectively, we define matrix $M$ to be their 
point-to-point Euclidean distance matrix, in which element $M_{i,j}$ is 
the distance $d(x_i,y_j)$ between $x_i$ and $y_j$. The alignment of $X$ 
and $Y$ may then be represented by a warping path $W = <w_1,w_2,\ldots ,
w_K>$, with $\max(m,n)$ ${\leq}$ $K < m+n-1$. The total cost along such 
a path $W$ is given by $C_W(X,Y) = \sum_{k=1}^{K} d_k$, where $d_k = 
d(x_i,y_j)$ and $k=1,2,\ldots,K$. The DTW distance then corresponds to 
the lowest-cost warping path between $X$ and $Y$, namely $R_t(X,Y) = 
\min_{W}C_W(X,Y)$, where $W$ is a candidate warping path. 

\

Having calculated spatial and temporal dissimilarities between components,
the overall inter-component distance for each pair of subjects is then 
computed as the average of the two. Components are clustered based on 
these distances using the Average Link method of hierarchical 
clustering \citep{41}.
These spatial patterns are then rescaled to \textit{z}-scores and thresholded using FDR-corrected \textit{p}-values, to highlight the voxels that are active under each of the components.

\section{Additional results of Synthetic Data}
These are presented in Table~\ref{table:2}.
   
\begin{table*}
\footnotesize
\begin{center}
 \begin{tabular}
{
  @{\kern-.5\arrayrulewidth}
  |p{\dimexpr1.5cm-2\tabcolsep-.5\arrayrulewidth}
  |p{\dimexpr0.7cm-2\tabcolsep-.5\arrayrulewidth}
  |p{\dimexpr0.7cm-2\tabcolsep-.5\arrayrulewidth}
  |p{\dimexpr2.1cm-2\tabcolsep-.5\arrayrulewidth}
  |p{\dimexpr2cm-2\tabcolsep-.5\arrayrulewidth}
  |p{\dimexpr2cm-2\tabcolsep-.5\arrayrulewidth}
  |p{\dimexpr2cm-2\tabcolsep-.5\arrayrulewidth}
  |p{\dimexpr2cm-2\tabcolsep-.5\arrayrulewidth}
  |p{\dimexpr2cm-2\tabcolsep-.5\arrayrulewidth}
  |@{\kern-.5\arrayrulewidth}
}
 \hline
 Dataset & M & K & \initialI\space with coordinate ascent and $X_{01}$ & \initialII\space with coordinate ascent and $X_{01}$ & \initialII\space with variational Bayes and $X_{01}$ & \initialI\space with coordinate ascent and $X_{02}$& \initialII\space with coordinate ascent and $X_{02}$ & \initialII\space with variational Bayes and $X_{02}$\\ [0.5ex]
 \hline\hline
 1 & 10 & 2 &  0.4788& 0.4813& 0.0287&0.4510&0.3988 &0.0348 \\ 
 \hline
 2 & 10 & 5 & 0.6478 & 0.5261& 0.0229&0.2735& 0.1936&0.1174\\
 \hline
 3 & 10 & 10 & 0.7179 & 0.6556& 0.0103&0.1355&0.0771 &0.0092\\
 \hline
 4 & 20 & 2 & 0.4981 & 0.4979&0.0266&0.5005&0.4870 &0.1090\\
 \hline
 5 & 20 & 5 & 0.7297 & 0.6298&0&0.2010& 0.1445&0.0126\\
 \hline
 6 & 20 & 10 & 0.7550 & 0.6535&0&0.0653&0.0487 &0.0017\\
 \hline
 7 & 40 & 2 & 0.5001 & 0.5001&0.0144&0.4537& 0.4493&0.0939\\
 \hline
 8 & 40 & 5 & 0.7506 & 0.7300&0.0065 &0.2698&0.2572 &0\\
 \hline
 9 & 40 & 10 & 0.8477 &0.6953 & 0.0071&0.0986& 0.0903&0\\ [1ex] 
 \hline
 \hline\hline
 1 & 10 & 2 & 0.4837  & 0.4837&0.0512&0.4822&0.4388&0.0717\\ 
 \hline  
 2 & 10 & 5 & 0.6794 & 0.6389&0.0834&0.1952&0.1571&0.0522\\
 \hline
 3 & 10 & 10 & 0.6325 & 0.3466&0.0398&0.0854&0.0596&0.0018\\
 \hline
 4 & 20 & 2 & 0.4947 & 0.4947&0.0613&0.5332&0.5290&0.0829\\
 \hline
 5 & 20 & 5 & 0.6827 & 0.4747&0.0152&0.1652&0.1391&0.0236\\
 \hline
 6 & 20 & 10 & 0.8517 &0.7936 &0.0096&0.0803&0.0608&0\\
 \hline
 7 & 40 & 2 & 0.4997 & 0.4997&0.0599&0.4254&0.4022&0.0646 \\
 \hline
 8 & 40 & 5 & 0.7493 & 0.7259&0.0108&0.2192&0.2034&0.0018\\
 \hline
 9 & 40 & 10 & 0.8023 & 0.6809&0.0111&0.0978&0.0856&0\\ [1ex] 
 \hline
\end{tabular}
\end{center}
\caption{The average misclassification rates of 7 datasets generated from forward model 
\initialI\space (top) and \initialII\space (bottom). $M$ corresponds to different number of subjects and $K$ corresponds to different number of labels. $X_{01}$ is a random initialization, and $X_{02}$ is a greedy initialization that contains all the nonzero components from $Y_1, Y_2,\ldots, Y_M$, whose nonzero label at a given voxel s is the most frequent label among  $Y_1(s), Y_2(s),\ldots, Y_M(s)$.}
\label{table:2}
\end{table*}

\section{Generation of the simulated fMRI dataset} \label{sec:simtb}
We use the SimTB toolbox for MATLAB{\texttrademark} to generate the 
simulated fMRI dataset 
\citep{37}. 
As with \citep{38}, we simulate $M$=30 subjects in this experiment, with a repetition time $TR = 3$s/sample, 
with slices of size $64\times64$ at $T$=150 time points. We set the number of components at $C$ = 30. Of these components, not all are uniformly present in all the subjects. We instead consider a subset of 17 components of interest, for each of which there is a 90\% probability of occurrence in every subject. In addition, we assign to each of the remaining components, a 30\% probability of occurrence. To model the spatial variability in the regions of activity under each component  across the subjects, we incorporate independent normal translation, rotation, and spread. Activation centers are translated vertically and horizontally with a standard deviation of 0.3 voxels, rotated by a deviation of 1 degree, and their spatial extent (compression or expansion) is determined following the normal distribution $N(1,0.3)$.

Following \citep{38}, we set the baseline component activation amplitude at 800 and draw the peak-to-peak percentage signal change from a Gaussian distribution with mean 3 and standard deviation 0.3. By default, the SimTB toolbox defines four different tissue types representing white matter, gray matter, sinus signal dropout, and cerebrospinal fluid (CSF). We set the corresponding tissue modifiers at 0.8 for white matter, 1.2 for CSF, 0.3 for the sinus signal, and 1.15 for frontal white matter, relative to the global mean intensity of 1, to approximate the statistical moments of real data, as reported in \citep{37}. To generate component time courses, we select the spike model for CSF, and obtain the remaining component time courses through convolution with the haemodynamic response function (HRF). Following the event-related experimental design, we set the amplitudes for component time courses to be consistent across subjects. Additionally, we simulate head motion through independent random translation and rotation, following $N(0,1)$. This distribution assumes random head motion between imaging instants, with a central position being more likely than the extremes \citep{37}. Finally, we add Rician noise to the generated data to simulate typical CNR levels, i.e., uniformly distributed from 0.65 to 2 \citep{39}. 


To generate individual subject maps, we whiten each subject's synthetic fMRI data matrix and reduce dimensionality, retaining 95\% variance. We then examine the resulting data for irregularity using the RV coefficient, which is computed between subjects as a measure of mutual ``distance''. Next, we compute the average distance of each subject's data matrix from the rest. Those over one standard deviations away from the average distance are considered outliers and excluded from the group-estimation framework. Of the 30 subjects in our experiment, we identify eight as atypical, which are then omitted from further analysis. These subjects are observed to have high noise levels, head motion, or spatial translation of regions of activation. 
 
Next, we decompose the synthetic data for each subject into a set of spatially independent components and associated time courses through ICA, over 20 runs with random initialization using the ICASSO toolbox \citep{40}. We identify the centrotypes of the clusters of components generated over the 20 runs, which form the final estimates of the independent spatial components. We then apply average-link clustering to retain those components that are present consistently across the group of subjects. In our experiment, we observe 17 consistent components. We obtain the individual subject maps through back-projection of these consistent components. Finally, we rescale the component maps to \textit{z}-scores and determine the active regions by thresholding with an FDR-corrected \textit{p}-value of 0.05. 

After observing that some components of the individual subjects are of the same shape but different colors, we decide to assign 1 
to all previous nonzero labels as the group-representative map is characteristic of the subjects in our experiment, and summarizes their shared patterns of functional connectivity.



\bibliographystyle{elsarticle-harv} 
\bibliography{Aditi_paper}

\end{document}